\documentclass[12pt,a4paper]{article}
\pdfoutput=1
\usepackage{jheppub}
\usepackage{amsthm,amsbsy,amsfonts,mathrsfs,enumerate,float,wrapfig,amsmath,amssymb}
\usepackage[colorinlistoftodos,prependcaption,textsize=scriptsize]{todonotes}
\newcommand{\nn}{\nonumber}

\newcommand{\ft}{\mathfrak{t}}
\newcommand{\fq}{\mathfrak{q}}
\newcommand{\calN}{\mathcal{N}}
\newcommand{\calR}{\mathcal{R}}
\newcommand{\calM}{\mathcal{M}}

\newcommand{\calZ}{\mathcal{Z}}
\newcommand{\calF}{\mathcal{F}}
\newcommand{\calO}{\mathcal{O}}

\allowdisplaybreaks
\usepackage{tikz}
\title{Refined topological vertex with ON-planes}

\author{Sung-Soo Kim,}
\author{Xing-Yue Wei}

\affiliation{School of Physics, University of Electronic Science and Technology of China,\\
No. 2006 Xiyuan Ave, West Hi-Tech Zone, Chengdu, Sichuan 611731, China}

\emailAdd{sungsoo.kim@uestc.edu.cn}
\emailAdd{xingyue\_wei@std.uestc.edu.cn}

\abstract{
We propose refined topological vertex formalism for 5-brane systems with ON-planes by introducing a new vertex associated with reflection over an ON-plane, which gives rise to new vertex and edge factors. We test our proposal against various 5d $\mathcal{N}=1$ gauge theories which can be realized as 5-brane webs with ON-planes, which include $D$-type quiver theories. In particular, we compute the refined partition functions for 6d E-string theory on a circle as well as 5d SU(3) theory at the Chern-Simons level 9, which can be realized as 5-brane webs with two ON-planes. Our results completely agree with the known results. 
}

\begin{document}
\maketitle

\section{Introduction}\label{sec:intro}

String theory and M-theory have provided various useful tools for studying supersymmetric field theories. For example, 5-brane webs in Type IIB string theory \cite{Aharony:1997bh,Leung:1997tw} and M-theory compactified on a Calabi-Yau threefold  \cite{Morrison:1996xf,Douglas:1996xp,Intriligator:1997pq} have shed light on uncovering various noble aspects for supersymmetric theories of eight supercharges in five and six dimensions (5d/6d). In particular, 5-brane webs have been a useful tool for better understanding of 5d superconformal theories qualitatively as well as  quantitatively. Different gauge theories can be described by different brane systems with or without introduction of orientifold planes. Their interplay through Hanany-Witten transitions, resolutions of an orientifold O7$^-$-plane to two 7-branes, and S-duality has revealed various gauge theory descriptions.  

Many 6d theories on a circle with or without a twist are represented on a 5-brane web as a 5d KK theory. 5d SU(2) gauge theory with 8 hypermultiplets in the fundamental representation  is an interesting example of a KK theory for 6d E-string theory on a circle \cite{Seiberg:1996bd, Witten:1995gx, Ganor:1996mu}. As discussed in \cite{Kim:2015jba}, the corresponding 5-brane web has a repeated spiral configuration called Tao web diagram, which gives rise to a constant period related to the radius of the compactification circle. This hence provides a diagrammatic characteristic of KK spectrum of the theory. The E-string theory can also be described by an affine $D_4$ quiver which consists of an SU(2) node in the middle and four ``SU(1)'' nodes at each quadrivalent leg. This is a realization of 6d $(D_4, D_4)$ conformal matter theory. Such quiver theories are described as a 5-brane web with two ON-planes. An ON-plane is the S-dual orientifold of an O5-plane \cite{Sen:1998rg,Sen:1998ii,Kapustin:1998fa,Hanany:1999sj}. ON-planes  are even more useful than just describing a $D$-type quiver theory, as SU(3) gauge theory at higher Chern-Simons levels can be constructed with ON-planes \cite{Hayashi:2018lyv}.  

With such various 5-brane systems, one can compute their BPS spectrum. In fact, there has been various progress on computing partition functions of 5d/6d $\mathcal{N}=1$ supersymmetric gauge theories. For instance, the ADHM method \cite{Atiyah:1978ri,Nekrasov:2002qd,Nekrasov:2003rj,Marino:2004cn,Nekrasov:2004vw,Fucito:2004gi,Hwang:2014uwa},  the Ding-Iohara-Miki algebra \cite{Ding:1996mq,doi:10.1063/1.2823979,Awata:2011ce,Bourgine:2017jsi}, 
topological vertex \cite{Aganagic:2002qg,
Aganagic:2003db,Iqbal:2007ii,Awata:2008ed,Aganagic:2012hs} and blowup equation \cite{Nakajima:2003pg,Gu:2017ccq, Huang:2017mis, Gu:2018gmy,Gu:2019dan,Gu:2019pqj,Gu:2020fem,Kim:2019uqw,  Kim:2020hhh,Kim:2021gyj}. In particular, the blowup method can be applied to theories without 5-brane configurations or any Lagrangian descriptions \cite{Kim:2020hhh,Kim:2021gyj}.

In this paper, we attempt to develop and generalize the  topological vertex which is a computation tool based on 5-brane webs as they provide intuitive pictures, and we confirm our result with known results obtained by other methods. Topological vertex works very well for toric 5-brane webs but it has some challenges for non-toric webs with orientifolds. For O7$^-$ cases, one can resolve an O7$^-$-plane into two 7-branes \cite{Sen:1996vd} and hence it leads to a non-toric 5-brane web without the orientifold plane. In such case, we can obtain the partition function as the Higgsing \cite{Hayashi:2013qwa,Hayashi:2016jak,Cheng:2018wll}. 
For the O5 case \cite{Hanany:1999sj,Zafrir:2015ftn,Hayashi:2015vhy,Zafrir:2016jpu,Hayashi:2018lyv,Hayashi:2019yxj}, on the other hand, the application of topological vertex was first proposed in \cite{Kim:2017jqn} and further developed in \cite{Hayashi:2020hhb,Li:2021rqr,Kim:2021cua, Nawata:2021dlk}. The 5-brane transition on an O5-plane where two 5-branes  intersect gives rise to a phase where they can be smoothly connected to the mirror images reflected due to the O5-plane. But these are only available for unrefined cases, where the sum of two Omega deformation parameters is set to zero,  $\epsilon_1+\epsilon_2=0$. Generalization toward the refined topological vertex is still a difficult task. The main difficulty comes from the Omega deformation parameters assigned to the original web and the reflected web which are not compatible with topological vertex formulation with conventional choice of the preferred direction which is parallel to O5-plane. 
To avoid this difficulty, we, instead, consider the ON case, which one can regard as an S-dual of O5-plane \cite{Kutasov:1995te,Sen:1996na, Hanany:1999sj}. It has a technical advantage for generalizing to the refined case, as the preferred direction is perpendicular to an ON-plane.

A 5-brane web with an ON-plane describes a $D$-type quiver. For such 5-brane configurations with ON-planes, we propose  new refined vertex and edge factors so that they account for the 5-brane system reflected over an ON-plane, as a generalization of unrefined topological vertex formalism \cite{Kim:2017jqn}.\footnote{There is an algebraic construction for a $D$-type quiver based on the DIM algebra \cite{Bourgine:2017rik,Kimura:2019gon}, which may lead to refined topological vertex. Our proposal is different from this as it is a generalization of the unrefined topological vertex to the refined one.} %
In fact, these new factors are not just a computational device, it is another set of vertex and edge factors which is equally applicable for computation of the partition function. For instance, we explicitly show that one can use our new factors to obtain the same partition function based on a 5-brane web without an ON-plane. We demonstrate our proposal with various theories where ON-planes are used for constructing the corresponding 5-brane systems, which include the SU(2) theory with 8 flavors and the SU(3) theory at the CS level 7 and 9. We also present the exact form of one- and two-instanton partition function for these theories. In particular, the refined partition function for the SU(3) theory at the CS level 9 is recently computed based on the blowup equation \cite{Kim:2020hhh} and our prescription perfectly reproduces this result.

The organization of the paper is as follows: In section \ref{sec:TV}, we start with reviewing the refined topological vertex and propose a new vertex factor that accounts the reflection due to an ON-plane. In section \ref{sec:SU2withfour}, we apply the proposed refined topological vertex to various gauge theories which can be realized with one or two ON-plane(s). We discuss  SU$(2)+4\mathbf{F}$ as an instructive example, and then compute the partition functions for the E-string on a circle and 5d SU(3) gauge theory at the Chern-Simons levels 7 and 9. We summarize our result in section \ref{sec:conclusion}. In Appendix \ref{sec:App}, we list the characters of global symmetries discussed in the main text. In Appendix \ref{app:newfactors}, we discuss how one can obtain the partition function of a 5-brane web without ON-planes by only using the new vertex and edge factors.

\bigskip
\section{Refined topological vertex with ON-planes}\label{sec:TV}%

In this section, we discuss salient feature of the refined topological vertex \cite{Iqbal:2007ii} and then we generalize it to 5-brane systems with ON-planes. 

\subsection{5-brane web}
\begin{figure}[t]
	\centering
	\includegraphics[scale=0.26]{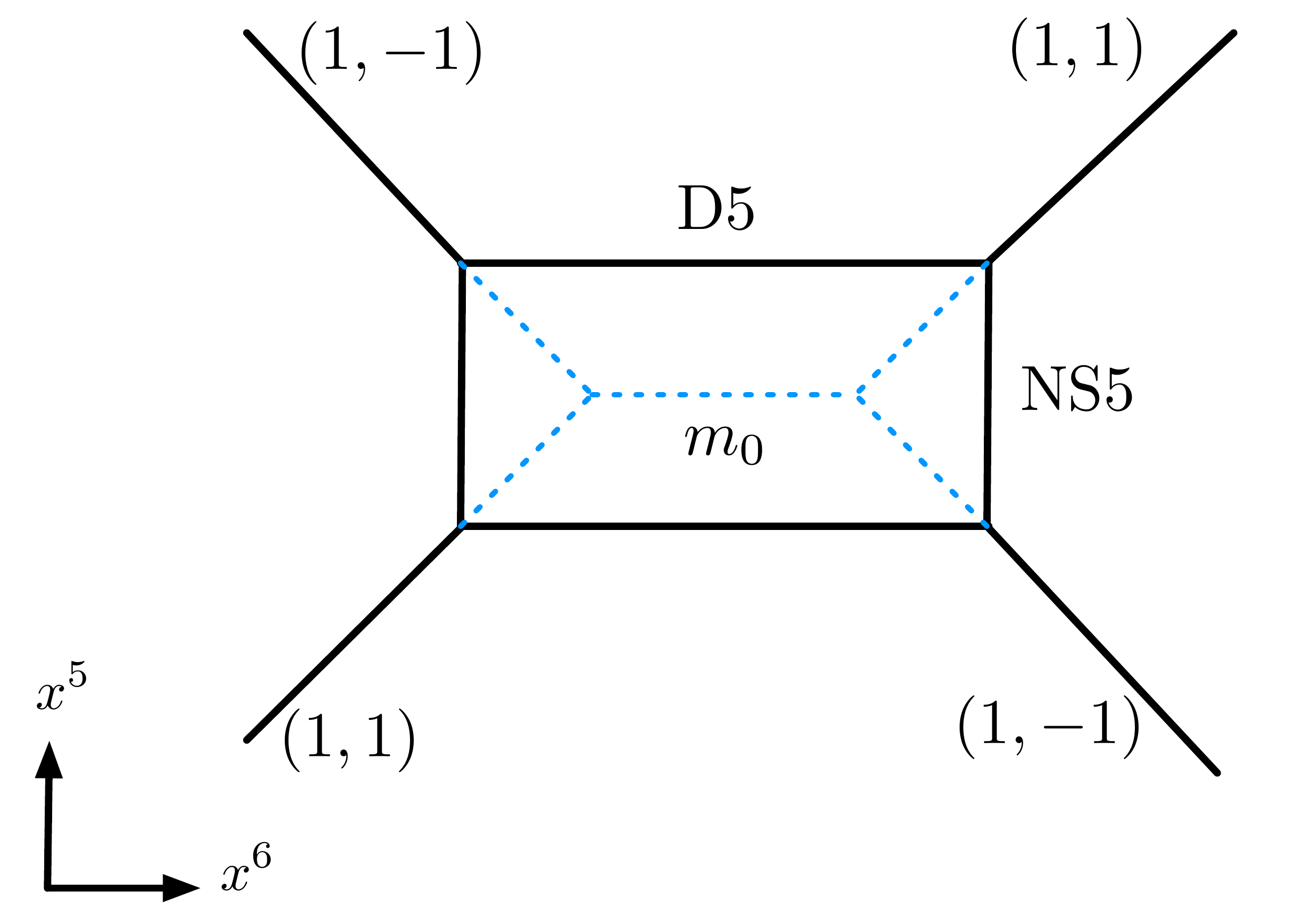}
	\caption{A 5-brane web for the SU(2)$_0$ gauge theory. }
	\label{fig:web-su2}
\end{figure}

A large class of 5d $\mathcal{N}=1$ supersymmetric theories can be realized through 5-brane webs in Type IIB string theory. With the convention that $(1,0)$ 5-brane refers to D5-brane and $(0,1)$ 5-brane refers to NS5-brane, 5d $\mathcal{N}=1$ supersymmetric field theories form a charge conserving configuration of various $(p,q)$ 5-branes. A simple web diagram is given in Figure \ref{fig:web-su2},  
where D5-branes are extended along $x^0,x^1,x^2,x^3,x^4, x^6$ while NS5-branes are extended along $x^0,x^1,x^2,x^3,x^4,x^5$ and 5-brane webs are given in a ($x^5,x^6$)-plane called the $(p,q)$-plane, as summarized in Table \ref{tab:brane cofniguration}. 
\begin{table}[H]
    \centering
    \begin{tabular}{c|ccccc|cc|ccc}
    \hline
    &0&1&2&3&4&5&6&7&8&9\\
    \hline
    NS5 / ON 
    & $\times$&$\times$&$\times$&$\times$&$\times$&$\times$ &&&&\\
    D5 / O5 
    & $\times$&$\times$&$\times$&$\times$&$\times$& &$\times$&&&\\
    $(p,q)$& $\times$&$\times$&$\times$&$\times$&$\times$&$\vartheta $&$\vartheta$&&&\\
    7-brane / O7& $\times$&$\times$&$\times$&$\times$&$\times$&&&$\times$&$\times$&$\times$\\
    \hline
\end{tabular}
\caption{Worldvolume configuration of 5-brane webs. The occupation of each brane is marked with $\times$ or $\vartheta$, where a $(p_1,q_1)$ 5-brane appears as a line on the $(p,q)$-plane with the slope given by $\vartheta=\tan^{-1}(q_1/p_1)$.  } 
\label{tab:brane cofniguration}
\end{table}
In Figure \ref{fig:web-su2}, a 5-brane web for 5d SU(2)$_0$ gauge theory of the vanishing discrete theta parameter is drawn. Here, the K\"ahler parameter along the NS5-brane connecting two D5-branes is associated with the Coulomb branch moduli, while the K\"ahler parameter between two NS5-branes is associated with a product of the instanton factor and the Coulomb branch parameter. More precisely, the instanton factor is given by $u=e^{-\beta m_0}$, where $m_0$ is the inverse gauge coupling squared and corresponds to the distance between two NS5 branes at trivial K\"ahler parameter for the Coulomb branch moduli as shown in Figure \ref{fig:web-su2}. In a similar fashion, the instanton factor for gauge theories of higher rank is defined. One can introduce 7-branes such that the external 5-branes are attached. With these 7-branes, one can explicitly show various dualities by the Hanany-Witten moves. 5-brane webs also provide a way to compute the BPS spectrum of the theory through topological vertex \cite{Aganagic:2003db,Iqbal:2007ii}, which we shall discuss in the next subsection.

\subsection{Topological vertex}
The refined topological vertex is a powerful way of computing the partition functions of 5d supersymmetric gauge theories on the $\Omega$ background, $\mathbb{R}^4_{\epsilon_1,\epsilon_2} \times S^1$.  With the fugacity of $\Omega$-deformation parameters $\fq= e^{-\beta\epsilon_1}$ and $\ft= e^{\beta\epsilon_2}$, the partition function $Z$ is given as a sum of the Young diagrams $\lambda,\mu,\nu,\cdots$ along edge factors and vertex factors for a given 5-brane web,
\begin{align}\label{eq:topological Z as Edge and Vertex}
Z = \sum_{\lambda,\mu,\nu,\cdots} \Big(\prod \text{edge factor}\Big) \Big( \prod \text{vertex factor}\Big)\ .
\end{align}
Here, following the convention used in \cite{Bao:2013pwa}, the edge factor and vertex factor are defined as follows.
For all the internal edges of a 5-brane web, we associate each edge with a K\"ahler parameter $Q$, an arrow, and a Young diagram (or integer partition) $\mu=(\mu_1, \mu_2,\cdots, \mu_{\ell(\mu)})$, where flopping the arrow corresponds to the transpose $\mu^t$ of the associated Young diagram $\mu$. The edge factor is then given as
\begin{align}\label{eq:edge factor}
	(-Q)^{|\mu|}\mathfrak{f}^{\mathfrak{n}}_\mu\ ,
\end{align}
where the framing factor function $\mathfrak{f}_\mu$ takes the form: along the preferred direction,  
\begin{align}\label{eq:framing factor}
\mathfrak{f}_\mu\rightarrow f_\mu(\ft,\fq)&=(-1)^{|\mu|}\ft^{\frac{||\mu^t||^2}{2}}\fq^{-\frac{||\mu||^2}{2}} = f_{\mu^t}(\fq,\ft)^{-1},
\end{align}
and along the non-preferred directions,
\begin{align}
\mathfrak{f}_\mu\rightarrow\tilde{f}_{\mu}(\ft,\fq)&=(-1)^{|\mu|}\ft^{\frac{||\mu^t||^2+|\mu|}{2}}\fq^{-\frac{||\mu||^2+|\mu|}{2}}= \tilde f_{\mu^t}(\fq,\ft)^{-1}\ ,
\end{align}
where 
\begin{align}
	|\mu| =\sum^{\ell(\mu)}_{i=1} \mu_i, \qquad
	||\mu||^2 =\sum^{\ell(\mu)}_{i=1} \mu_i^2 \ , 
\end{align}
and the power $\mathfrak{n}$ is defined as  $\mathfrak{n}= v_\text{in}\wedge v_\text{out}= p_1q_2-p_2q_1$ for a pair of charges $v_\text{in}=(p_1, q_1)$ and $v_\text{out}=(p_2, q_2)$ which are connected to the edge of $Q$, where $v_{\text{in}}$ and $v_{\text{out}}$ are 
two dimensional vectors $(\pm p,\pm q)$ associated with the $(p,q)$ charge of the corresponding brane with $\pm$ signs chosen to be compatible with the directions of the vectors.

The vertex factor is assigned to each vertex of three out-going edges with three Young diagrams $\lambda, \mu, \nu $ in clockwise direction where the last Young diagram $\nu$ is reserved for Young diagram on the edge of the preferred direction, and it is defined as
\begin{align}
C_{\lambda\mu\nu}(\ft,\!\fq)=\fq^{\frac{||\mu||^2+||\nu||^2}{2}}\ft^{-\frac{||\mu^t||^2}{2}}\tilde{Z}_\nu(\ft,\!\fq)\!\sum_{\eta}\!\Big(\frac{\fq}{\ft}\Big)^{\!\!\frac{|\eta|+|\lambda|-|\mu|}{2}}\!\!\! s_{\lambda^t/\eta}(\ft^{-\rho}\fq^{-\nu})s_{\mu/\eta}(\fq^{-\rho}\ft^{-\nu^t}),
\label{eq:C for refined}	
\end{align}
where $\ft$ and $\fq$ are assigned to the edge associated with $\lambda$ and $\mu$, respectively, and 
\begin{align}
\tilde{Z}_{\nu}(\ft,\fq)= \prod_{i=1}^{\ell(\nu)}\prod_{j=1}^{\nu_i}\big(1-\ft^{\nu^t_j-i+1}\fq^{\nu_i-j}\big)^{-1},
\end{align}
and $s_{\sigma/\eta}(\bf{x})$ are the skew-Schur functions of an infinite vector $\bf{x}$, {\it e.g.}, 
$\ft^{-\rho}\fq^{-\nu} = (\ft^\frac{1}{2} \fq^{-\nu_1},\ft^{\frac32}\fq^{-\nu_2}, \cdots)$. 

We list some special functions that are useful in actual computation:
\begin{align}
\calR_{\lambda\mu}(Q;\ft,\fq)&:=\! \prod_{i,j=1}^{\infty}\!\!\Big(\!1\!-\!Q \ft^{i-\frac{1}{2}-\lambda_j}\fq^{j-\frac{1}{2}-\mu_i}\!\Big)\!=\!\calM\big(Q\sqrt{\tfrac{\ft}{\fq}};\ft,\fq\big)^{-1}\calN_{\lambda^t\mu}\big(Q\sqrt{\tfrac{\ft}{\fq}};\ft,\fq\big), \label{eq:RMN}\\
\calM(Q;\ft,\fq)&:=\!\prod_{i,j=1}^{\infty}\big(1-Q \ft^{i-1}\fq^{j}\big)^{-1},\\
\calN_{\lambda\mu}(Q;\ft,\fq)&:=\! \prod_{i,j=1}^{\infty}\frac{1-Q\ft^{i-1-\lambda_j^t}\fq^{j-\mu_i}}{1-Q\ft^{i-1}\fq^j}\nonumber \\
&=\!\prod_{(i,j)\in\lambda}(1-Q\ft^{\mu_j^t-i}\fq^{\lambda_i-j+1})\prod_{(i,j)\in\mu}(1-Q\ft^{-\lambda_j^t+i-1}\fq^{-\mu_i+j})\ .
\end{align}
To evaluate the Young diagram sums along non-preferred directions, one needs to repeatedly use the Cauchy identities\footnote{These identities are slightly modified from the ones in \cite{Bao:2013pwa}.} 
\begin{align}
	&\sum_{\lambda}Q^{|\lambda|}s_{\lambda/\mu_1}(Q_1\ft^{-\rho}\fq^{-\nu_1})s_{\lambda^t/\mu_2}(Q_2\fq^{-\rho}\ft^{-\nu_2})\nonumber\\
	&=\calR_{\nu_2\nu_1}(-QQ_1Q_2)\sum_{\lambda}Q^{|\lambda|}s_{\mu_2^t/\lambda}(QQ_1\ft^{-\rho}\fq^{-\nu_1})s_{\mu_1^t/\lambda^t}(QQ_2\fq^{-\rho}\ft^{-\nu_2})\ .\\
	\nonumber\\
	&\sum_{\lambda}Q^{|\lambda|}s_{\lambda/\mu_1}(Q_1\ft^{-\rho}\fq^{-\nu_1})s_{\lambda/\mu_2}(Q_2\fq^{-\rho}\ft^{-\nu_2})\nonumber\\
	&=\calR_{\nu_2\nu_1}(QQ_1Q_2)^{-1}\sum_{\lambda}Q^{|\lambda|}s_{\mu_2/\lambda}(QQ_1\ft^{-\rho}\fq^{-\nu_1})s_{\mu_1/\lambda}(QQ_2\fq^{-\rho}\ft^{-\nu_2})\ .
\end{align}

\begin{figure}[t]
		\centering 
		\includegraphics[scale=1]{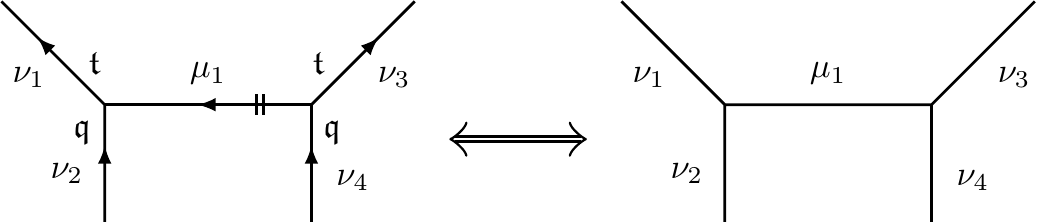}
		\caption{Convention for assignment of the arrows and the $\Omega$-deformation parameters throughout the paper.}
		\label{fig:convention}
	\end{figure}

\paragraph{Conventions.}
In the rest of the paper except for the beginning of section \ref{sub:refined_ON}, to avoid the cluttering of the 5-brane diagrams, we use the following convention when assigning $\ft$, $\fq$, arrows, and Young diagrams: 
\begin{enumerate}
	\item [(1)]The preferred direction is always along the horizontal edges and the arrows associated with it 
	are chosen to point toward the left. 
	The arrows along  non-preferred directions are chosen to point upward.
	\item[(2)]The $\Omega$-deformation parameters $\fq = e^{-\beta\epsilon_1},\ft=e^{\beta\epsilon_2}$ are assigned such that $\ft$'s are always placed above the edges associated with the preferred direction, while $\fq$'s are placed below the edges associated with the preferred direction. 
	\item[(3)]We use the Greek letters $\mu, \nu,\cdots$ to denote the Young diagrams. If an edge is labelled by a Young diagram, say $\alpha$, then we also call the edge as edge $\alpha$ or brane $\alpha$ for simplicity.
\end{enumerate}
	  For instance, the 5-brane web on the right-hand side of Figure \ref{fig:convention} should be understood as the 5-brane web with the assignments given on the left-hand side.
	
	We also define
	\begin{align}
		\calR_{\lambda\mu}(Q)\equiv\calR_{\lambda\mu}(Q;\ft,\fq),\quad
		\calN_{\lambda\mu}(Q)\equiv\calN_{\lambda\mu}(Q;\ft,\fq),\quad
		\calM(Q)\equiv\calM(Q;\ft,\fq)
	\end{align}
	for simplicity. The products of several K\"ahler parameters are expressed as a shorthand notation, for example, $Q_{i,j^2,k}\equiv Q_iQ_j^2Q_k$. On the other hand, $\mu_{i,j,k}$ means $\mu_i,\mu_j,\mu_k$ for short which we will use occasionally.
	
\subsection{Refined topological vertex with ON-planes}\label{sub:refined_ON}
The refined topological vertex formalism discussed in the previous section works well with those 5-brane webs which do not involve orientifold planes like O5-, O7${}^+$-, or ON-planes. It is also applicable for 5-brane web systems with O7${}^-$-planes, as an O7${}^-$-plane can be resolved into a pair of two 7-branes, $[1, -1], [1, 1]$ 7-branes or  $[2, -1], [0, 1]$ 7-branes \cite{Sen:1996vd}. Such 5-brane configuration with the  resolved 7-branes can be understood as the Higgsing \cite{Benini:2009gi,Hayashi:2013qwa}. 
In this subsection, we attempt to generalize the refined topological vertex to be applicable for 5-brane webs with orientifold planes.  
As discussed, unrefined topological vertex formalism with O5/ON-planes was introduced.  Let us first discuss some of relevant features of this formalism and generalize them to refined topological vertex with an orientifold plane.

To this end, we first recall the unrefined case with an O5-plane (or an ON-plane) \cite{Kim:2017jqn}. For the unrefined case, $\mathfrak{t}=\mathfrak{q}=g$, one introduces the vertex factor for the reflected image due to the orientifold plane where the Young diagrams of the vertex factor for the reflected image are all transposed. The vertex factor for the reflected images satisfies the following reflection identity
\begin{align}
	C_{\mu^t\lambda^t\nu^t} = (-1)^{|\lambda|+|\mu|+|\nu|}f_{\lambda}(g)
	f_{\mu}(g)f_{\nu}(g)C_{\lambda\mu\nu}\ .
	\label{eq:unrefrelation}
\end{align}
This is an important identity as it gives rise to the relation between the vertex factor for the reflected image of 5-branes due to an O5/ON-plane and that defined on 5-brane web without O5/ON-plane.
It is therefore natural to generalize the reflection identity \eqref{eq:unrefrelation} to the refined case. 
\bigskip

\begin{figure}[t]
	\centering
	\includegraphics[scale=1]{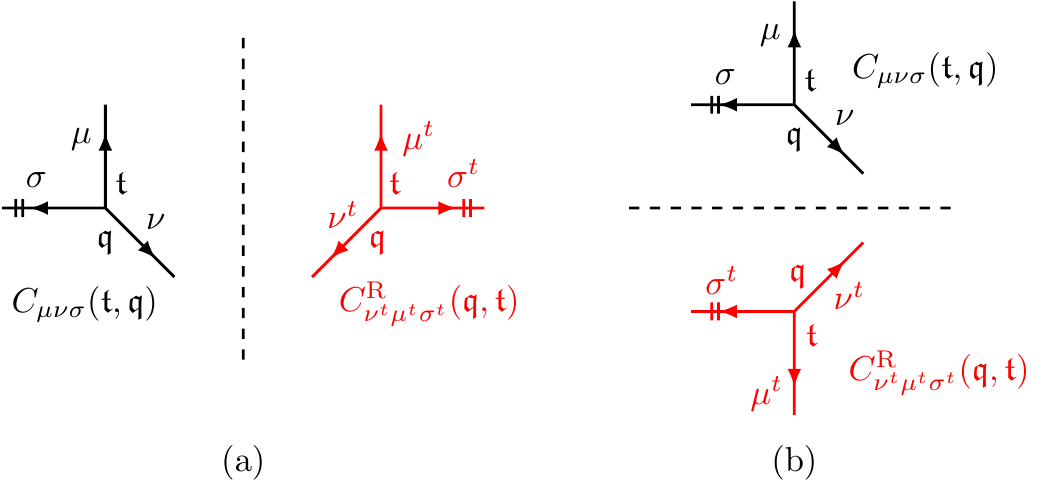}
	\caption{The usual vertex factor $C$ and the reflected vertex factor $C^{\text{R}}$.  (a) is for the case of ON-plane reflection and (b) is for the case of O5-plane reflection.}
	\label{fig:CRONO5}
	\end{figure}

\paragraph{New vertex factor.}
We introduce the following refined vertex factor for reflected image of a 5-brane web with an ON-plane or an O5-plane:  
\begin{equation}
	C^{\text{R}}_{\lambda\mu\nu}(\ft,\fq)=\fq^{\frac{||\mu||^2+||\nu||^2}{2}}\ft^{-\frac{||\mu^t||^2}{2}}\tilde{Z}_\nu(\ft,\fq)\!\sum_{\eta}\left(\frac{\fq}{\ft}\right)^{\frac{-|\eta|+|\lambda|-{}|\mu|}{2}}\!\!s_{\lambda^t/\eta}(\ft^{-\rho}\fq^{-\nu})s_{\mu/\eta}(\fq^{-\rho}\ft^{-\nu^t}).
	\label{eq:CR}
\end{equation}
Here, we used the superscript~${}^{\text{R}}$ to denote that this new vertex factor is associated with the reflected image and also to distinguish the conventional vertex factor $C$.   
Compared with the  vertex factor $C$ in \eqref{eq:C for refined}, $C^\text{R}$ has the opposite sign in front of $|\eta|$ appearing in the power of $\frac{\fq}{\ft}$. We note that these two kinds of vertex factors are related by the following relation
\begin{align}
	C^{\text{R}}_{\nu^t\mu^t\sigma^t}(\fq,\ft)=(-1)^{|\mu|+|\nu|+|\sigma|}f_{\mu}(\ft,\fq)f_{\nu}(\ft,\fq)f_{\sigma}(\ft,\fq)\frac{\tilde{Z}_{\sigma^t}(\fq,\ft)}{\tilde{Z}_{\sigma}(\ft,\fq)}C_{\mu\nu\sigma}(\ft,\fq) \ ,
	\label{eq:refrelation1}	
\end{align}
or equivalently, 
\begin{align}
	C_{\mu\nu\sigma}(\ft,\fq)=(-1)^{|\nu^t|+|\mu^t|+|\sigma^t|}f_{\nu^t}(\fq,\ft)f_{\mu^t}(\fq,\ft)f_{\sigma^t}(\fq,\ft)\frac{\tilde{Z}_{\sigma}(\ft,\fq)}{\tilde{Z}_{\sigma^t}(\fq,\ft)}C^{\text{R}}_{\nu^t\mu^t\sigma^t}(\fq,\ft)\ ,
	\label{eq:refrelation2}
	\end{align}
which is a refined version of the reflection identity  \eqref{eq:unrefrelation}. It is straightforward to check that in the unrefined limit, the reflected vertex $C^\text{R}$ reduces to the usual $C$, and the refined reflection identities \eqref{eq:refrelation1} and  \eqref{eq:refrelation2} become the unrefined identity \eqref{eq:unrefrelation}.
In Figure \ref{fig:CRONO5}, the assignments of Young diagrams and $\ft,\fq$ associated with the new vertex factor $C^\text{R}$ are depicted in comparison with those for the usual vertex factor $C$. As discussed earlier, in usual refined topological vertex formalism, the Young diagrams of  $C_{\mu\nu\sigma}(\ft,\fq)$ are assigned clockwise with the last one being the preferred direction. We follow the same rule for the new vertex factor $C^{\text{R}}$. Likewise, the assignments of $\ft,\fq$ in the reflected image are also given clockwise as in Figure \ref{fig:CRONO5}. So, for the vertex factor $C^{\text{R}}_{\nu^t\mu^t\sigma^t}(\fq,\ft)$ in the reflected image, Young diagrams and $\ft,\fq$ are treated in the same way as those in the usual $C$ vertex function as illustrated in Figure \ref{fig:CRONO5}.

 We note that when a Young diagram along the non-preferred direction of $C^{\text{R}}$ is of an empty set, $C^\text{R}$ is equal to $C$ 
as the vertex factor is only nonzero for $\eta=\text{\o}$. 
Unlike the unrefined case, both vertex factors $C$ and $C^\text{R}$ are necessary for the refined case. As a 5-brane configuration away from an ON-plane looks like  usual web without an ON-plane, only vertices near an ON-plane are related to $C^\text{R}$. Even more, we claim that $C^{\text{R}}$'s appear only on the strips of NS-branes that are next to an ON-plane
and all other vertices are assigned with $C$'s.

We also note that as shown in Figure \ref{fig:CRONO5}(b), the assignments of $\ft,\fq$ are exchanged in the reflected image of the vertex over an O5-plane which makes it more difficult to deal with in refined topological vertex formalism, so from now on we only consider cases with ON-planes in this paper. %
\bigskip
\paragraph{New Edge factors.}
\begin{figure}[t]
		\centering
		\includegraphics[scale=1]{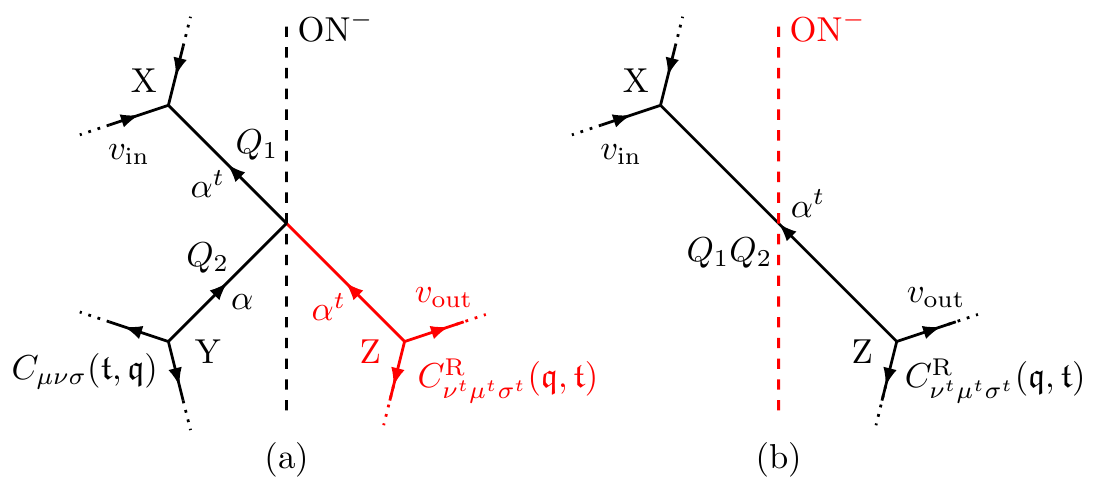}
		\caption{Reflection of the brane $\alpha$ by the ON-plane.}
		\label{fig:reflection of ref TV}
\end{figure}
Together with the new vertex factor $C^{\text{R}}$, new edge factors naturally follow in the refined topological vertex with ON-plane. Consider 5-brane configuration near an ON-plane, as depicted in Figure \ref{fig:reflection of ref TV}. Figure \ref{fig:reflection of ref TV}(a) represents typical 5-brane configuration with an ON-plane. On the other hand, Figure \ref{fig:reflection of ref TV}(b) represents a 5-brane configuration with half of 5-brane webs being reflected over an ON-plane so that the corresponding 5-brane web appears to be smoothly connected as if there is no ON-plane. Here the new vertex factor $C^\text{R}$ is introduced and the corresponding Young diagram is transposed accordingly.

To be more general, we introduce generic vertices X, Y, and Z in Figure \ref{fig:reflection of ref TV} which are associated with either usual vertex factor $C$ or the reflected vertex factor $C^\text{R}$ and denote them by $C_\text{X}, C_\text{Y}$ and $C_\text{Z}$. Let us first consider the case where $C_\text{Y}$ is the usual vertex factor $C_\text{Y}=C$, which is the case depicted in Figure \ref{fig:reflection of ref TV}(a). We will consider the case where $C_\text{Y}$ is associated with $C^\text{R}$ later. To be more specific, $C_\text{Y}$ is  $C_{\mu\nu\sigma}(\ft,\fq)$ where $\sigma$ is assumed to be along the preferred direction. We also introduce the Young diagram $\alpha$ which is one of the Young diagrams $\mu,\nu,\sigma$ associated with the vertex factor $C_{\mu\nu\sigma}(\ft,\fq)$, {\it i.e.,} $\alpha\in\{\mu,\nu,\sigma\}$. Depending on whether $\alpha$ is along the preferred direction or not, we will have different new edge factors. Note that $\text{Z}$ in Figure \ref{fig:reflection of ref TV}(b) is the reflected vertex of $\text{Y}$ in Figure \ref{fig:reflection of ref TV}(a). As we consider the case  $C_\text{Y}=C_{\mu\nu\sigma}(\ft,\fq)$, the vertex factor associated with Z is given as $C_\text{Z}=C^{\text{R}}_{\nu^t\mu^t\sigma^t}(\fq,\ft)$.

We now write the edge factor for the edge associated with the Young diagram $\alpha^t$ (of the K\"ahler parameter $Q_1Q_2$) in Figure \ref{fig:reflection of ref TV}(b), which is the edge connecting the vertices X and Z. As Z is the reflected image of Y, this would lead to how to define or construct the edge factor associated with Y. The product of the edge factor and  vertex factors associated with the edge between X and Z can be written as
\begin{align}
    & C_\text{X}\cdot(-Q_1Q_2)^{|\alpha^t|} \calF_{\alpha^t}\cdot C_\text{Z}\cr
	&= C_\text{X}\cdot(-Q_1Q_2)^{|\alpha^t|}\calF_{\alpha^t}\cdot C_{\nu^t\mu^t\sigma^t}^{\text{R}}(\fq,\ft)\nonumber\\
	& =C_\text{X}\cdot (+Q_1Q_2)^{|\alpha|}\Big(\mathcal{F}_{\alpha^t}f_{\alpha}(\ft,\fq)\Big)\left(\frac{\tilde{Z}_{\sigma^t}(\fq,\ft)}{\tilde{Z}_\sigma(\ft,\fq)}\right)^{\delta_{\alpha\sigma}}\cdot C_{\mu\nu\sigma}(\ft,\fq)\cdot(\cdots)\ ,
		\label{eq:deriveEF}
\end{align}
where we denote by $\calF_{\alpha^t}$ the framing factor associated with $Q_1Q_2$. In the second equality, we used \eqref{eq:CR} to express $C^\text{R}$ into usual vertex factor $C$. As we are only interested in finding new edge factor involving $\alpha$, we neglect irrelevant $\alpha$-independent parts by putting them into the ellipsis $(\cdots)$ in \eqref{eq:deriveEF}. If the Young diagram $\alpha$ is along the preferred diction, namely  $\alpha=\sigma$, then $\delta_{\alpha\sigma}=1$; otherwise $\delta_{\alpha\sigma}=0$. Notice that $C_{\mu\nu\sigma}(\ft,\fq)$ in \eqref{eq:deriveEF} is nothing but $C_\text{Y}$. This hence leads to the edge factor between the vertices X and Y in Figure \ref{fig:reflection of ref TV}(a). The new edge is then given by
\begin{equation}
	(+Q_1Q_2)^{|\alpha|}\left(f_{\alpha}(\ft,\fq)\mathcal{F}_{\alpha^t}\right)\left(\frac{\tilde{Z}_{\sigma^t}(\fq,\ft)}{\tilde{Z}_\sigma(\ft,\fq)}\right)^{\delta_{\alpha\sigma}},\quad\alpha\in\{\mu,\nu,\sigma\}\ .	\label{eq:newedge1}
\end{equation}	
We note that so far we consider the case where $C_\text{Z}=C^\text{R}$. In general, $C_\text{Z}$ can also be of the usual vertex factor $C$. Of course, if $C_\text{X}$ and $C_\text{Z}$ are all of $C$, then the framing factor $\calF_{\alpha^t}$ is simply $f_{\alpha^t}^{\mathfrak{n}}$ or ${\tilde{f}}^{\mathfrak{n}}_{\alpha^t}$. We will also discuss other $C_\text{X}$ and $C_\text{Z}$ cases later.

Consider also the case  $C_\text{Y}=C^\text{R}_{\mu\nu\sigma}(\ft,\fq)$, which means $C_\text{Z}=C_{\nu^t\mu^t\sigma^t}(\fq,\ft)$ accordingly. In this case, by repeating a similar calculation as \eqref{eq:deriveEF} taking into account \eqref{eq:refrelation2}, one finds that the resulting edge factor is still the same form as \eqref{eq:newedge1}. When the order of $\ft,\fq$ in the argument of Y exchanges, namely $C_\text{Y}=C^\text{R}_{...}(\fq,\ft)$, the edge factor also correspondingly changes to 
\begin{equation}
	(+Q_1Q_2)^{|\alpha|}\left(f_{\alpha}(\fq,\ft)\mathcal{F}_{\alpha^t}\right)\left(\frac{\tilde{Z}_{\sigma^t}(\ft,\fq)}{\tilde{Z}_\sigma(\fq,\ft)}\right)^{\delta_{\alpha\sigma}},\quad\alpha\in\{\mu,\nu,\sigma\}\ . \label{eq:newedge1p}
\end{equation}
We note that the analysis above is based on Figure \ref{fig:reflection of ref TV} in which the ON-plane is on the right-hand side. If an ON-plane is on the left-hand side, we can still do the similar analysis which yields that the edge factor is still given as the form of \eqref{eq:newedge1} or \eqref{eq:newedge1p}.
\begin{figure}[t]
		\centering
		\includegraphics[scale=1]{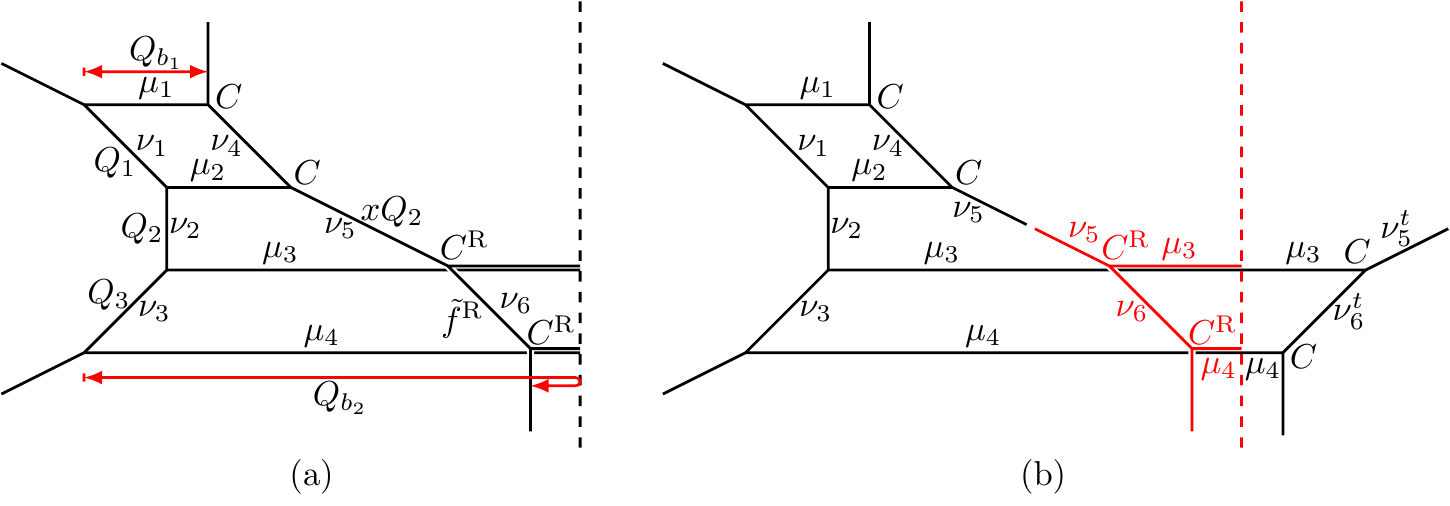}
		\caption{5-brane webs with an ON$^-$-plane for  $D_2(N_1,N_2)$ gauge theory with $N_1=2,N_2=2$.}
		\label{fig:D22,2}
	\end{figure}
	
Now that we have described how new edge factor appears when we consider the edge associated with the 5-brane being reflected by an ON-plane, we try to apply our prescription to actual computations involving an ON-plane. As a simple example, a 5-brane configuration is given in Figure \ref{fig:D22,2} which may look like a 5-brane web for a single gauge theory, though it is merely two copies of SU(2) theories without bifundamental matter. For purposes of expediency, we call such theory the ``$D_2(N_1,N_2)$" theory which is a linear sum of SU($N_1$) and SU($N_2$) gauge theories. 
	For simplicity, first we consider the case when $N_1=2$ and $N_2=2$, as given in Figure Figure \ref{fig:D22,2}. 
Here, the Young diagrams $\mu_1,\mu_2$ label the two color D5-branes for one SU$(2)$ and the Young diagrams $\mu_3,\mu_4$ label the two color D5-branes for the other SU$(2)$. As discussed earlier, a 5-brane configuration with an ON-plane like Figure \ref{fig:D22,2}(a) can be realized by folding the right part of a 5-brane configuration in Figure \ref{fig:D22,2}(b) to the left and then by gluing the edges in the middle. The 5-brane configuration before the folding is all equipped with the conventional vertex factors $C$ because the diagram contains two copies of normal SU$(2)$ webs. After the folding, on the other hand, the two vertices are reflected and they become $C^{\text{R}}$ on the left. The edge factor between this two vertices also changes due to the folding. The edge factor between the two $C$'s on the right of the ON-plane together with the two $C$'s in Figure \ref{fig:D22,2}(b) are given by 
	\begin{align}
		&C_{\nu_5^t\nu_6\mu_3}(\ft,\fq)\cdot(-Q_3)^{|\nu_6|}\tilde{f}_{\nu_6^t}^{-1}(\fq,\ft)\cdot C_{\nu_6^t{\text{\o}}\mu_4}(\ft,\fq)\nonumber\\
		&=C_{\nu_6^t\nu_5\mu_3^t}^{\text{R}}(\fq,\ft)\cdot\left(-Q_3\right)^{|\nu_6|}\left(\frac{\ft}{\fq}\right)^{\frac{|\nu_6|}{2}}f_{\nu_6}(\fq,\ft)\cdot C_{{\text{\o}}\nu_6\mu_4^t}^{\text{R}}(\fq,\ft)\cdot(\cdots)\nonumber\\
		&=C_{\nu_6^t\nu_5\mu_3^t}^{\text{R}}(\fq,\ft)\cdot\left(-Q_3\right)^{|\nu_6|}\tilde{f}_{\nu_6}^{\text{R}}(\fq,\ft)\cdot C_{{\text{\o}}\nu_6\mu_4^t}^{\text{R}}(\fq,\ft)\cdot(\cdots)
		\label{eq:intronewedge}\ .
	\end{align}
	In the second line of the above equation, \eqref{eq:refrelation2} is used and edge factors that do not involve $\nu_6$ are put into the ellipsis. In the third line, we have defined
	\begin{equation}
	\tilde{f}^{\text{R}}_{\mu}(\fq,\ft)\equiv(-1)^{|\mu|}\fq^{\frac{||\mu^t||^2-|\mu|}{2}}\ft^{-\frac{||\mu||^2-|\mu|}{2}}\ ,
	\label{eq:fR}
	\end{equation}
	which plays the role of the new framing factor function for edge that connects two $C^{\text{R}}$ vertices, and one can check that this definition is also consistent for the case when ON-plane is on the left of the web diagram. So 
	\begin{equation}
		\left(-Q_3\right)^{|\nu_6|}\tilde{f}_{\nu_6}^{\text{R}}(\fq,\ft)
		\label{eq:newedge2}
	\end{equation}
	is the edge factor of edge $\nu_6$ in Figure \ref{fig:D22,2}(a). As $\tilde{f}^\text{R}$ is the reflected correspondence of the normal $\tilde{f}$, one may think whether there is also a reflected correspondence of the preferred direction framing factor $f$, but actually one can check that the reflected correspondence is the same as the normal $f$.
	
	The edge $\nu_5$ is the glued edge which has two different kinds of vertices at its two ends. The framing factor of this edge is zero due to $\mathfrak{n}=0$, but we still need to modify the edge factor in order to get the correct partition function. It turns out that a multiplier $x$ needs to be added to the K\"ahler parameter $Q_2$ of $\nu_5$. Then we compute the partition function of the web diagram in Figure \ref{fig:D22,2}(a) with the prescription that we discussed, 
\begin{align}
		Z^{D_2(2,2)}=&~C_{\nu_1^t{\text{\o}}\mu_1^t}(\fq,\ft)C_{\nu_2^t\nu_1\mu_2^t}(\fq,\ft)C_{\nu_3^t\nu_2\mu_3^t}(\fq,\ft)C_{\text{\o}\nu_3\mu_4^t}(\fq,\ft)C_{\text{\o}\nu_4^t\mu_1}(\ft,\fq)\cr
		&\times C_{\nu_4\nu_5^t\mu_2}(\ft,\fq) C^{\text{R}}_{\nu_6^t\nu_5\mu_3^t}(\fq,\ft)C^{\text{R}}_{\text{\o}\nu_6\mu_4^t}(\fq,\ft)(-Q_1)^{|\nu_1|}\tilde{f}_{\nu_1}(\fq,\ft)(-Q_2)^{|\nu_2|}\tilde{f}_{\nu_2}(\fq,\ft)\crcr
		&\times(-Q_3)^{|\nu_3|}\tilde{f}_{\nu_3}(\fq,\ft)(-Q_1)^{|\nu_4|}\tilde{f}_{\nu_4}(\fq,\ft)^{-1}(-xQ_2)^{|\nu_5|}(-Q_3)^{|\nu_6|}\tilde{f}^{\text{R}}_{\nu_6}(\fq,\ft)\cr
		&\times(-Q_{b1})^{|\mu_1|}f_{\mu_1}(\ft,\fq)^{-1}(-Q_{b1})^{|\mu_2|}f_{\mu_2}(\ft,\fq)(+Q_{b2})^{|\mu_3|}f_{\mu_3}(\ft,\fq)^{-1}\cr
		&\times f_{\mu_3^t}(\fq,\ft)\frac{\tilde{Z}_{\mu_3}(\ft,\fq)}{\tilde{Z}_{\mu_3^t}(\fq,\ft)}(+Q_{b2})^{|\mu_4|}f_{\mu_4}(\ft,\fq)f_{\mu_4^t}(\fq,\ft)\frac{\tilde{Z}_{\mu_4}(\ft,\fq)}{\tilde{Z}_{\mu_4^t}(\fq,\ft)}\ .
	\end{align}
	Using the extended Cauchy identities repeatedly, we can sum over the non-preferred direction Young diagrams along the vertical strips, we obtain
	\begin{align}
		Z^{D_2(2,2)}&=\fq^{||\mu_1||^2+||\mu_3||^2}\ft^{||\mu_2^t||^2+||\mu_4^t||^2}Q_{b1}^{|\mu_1|+|\mu_2|}Q_{b2}^{|\mu_3|+|\mu_4|}\prod_{i=1}^4\tilde{Z}_{\mu_i}(\ft,\fq)\tilde{Z}_{\mu_i^t}(\fq,\ft)\nonumber\\
		&\times \calR_{\mu_1^t\mu_3}(xQ_1Q_2)\calR_{\mu_1^t\mu_4}(xQ_1Q_2Q_3)\calR_{\mu_2^t\mu_3}(xQ_2)\calR_{\mu_2^t\mu_4}(xQ_2Q_3)\nonumber\\
		&\times\Bigg(\calR_{\mu_1^t\mu_2}(\sqrt{\frac{\fq}{\ft}}Q_1)\calR_{\mu_1^t\mu_2}(\sqrt{\frac{\ft}{\fq}}Q_1)\calR_{\mu_1^t\mu_3}(\sqrt{\frac{\fq}{\ft}}Q_1Q_2)\calR_{\mu_1^t\mu_4}(\sqrt{\frac{\fq}{\ft}}Q_1Q_2Q_3)\nonumber\\
		&\times\calR_{\mu_2^t\mu_3}(\sqrt{\frac{\fq}{\ft}}Q_2)\calR_{\mu_2^t\mu_4}(\sqrt{\frac{\fq}{\ft}}Q_2Q_3)\calR_{\mu_3^t\mu_4}(\sqrt{\frac{\fq}{\ft}}Q_3)\calR_{\mu_3^t\mu_4}(\sqrt{\frac{\ft}{\fq}}Q_3)\Bigg)^{-1}\ .
	\end{align}
	By fixing
	\begin{equation}
		x=\sqrt{\frac{\fq}{\ft}}\ ,
		\label{newedge3}
	\end{equation}
	the factors $\calR_{\mu_i^t\mu_j}$ with $i=1,2$ and $j=3,4$ which are the bifundamental contributions cancel, and the result turns out to be the correct partition function of $D_2(2,2)$ gauge theory. So the new edge factor for the edge $\nu_5$ is
	\begin{equation}
		\left(-\sqrt{\frac{\fq}{\ft}}Q_2\right)^{|\nu_5|}\ ,
		\label{eq:newedge3}
	\end{equation}
	and we indicate this edge factor in the figures as $\sqrt{\frac{\fq}{\ft}}Q_2$ for short.
	\begin{figure}[t]
		\centering
		\includegraphics[scale=1]{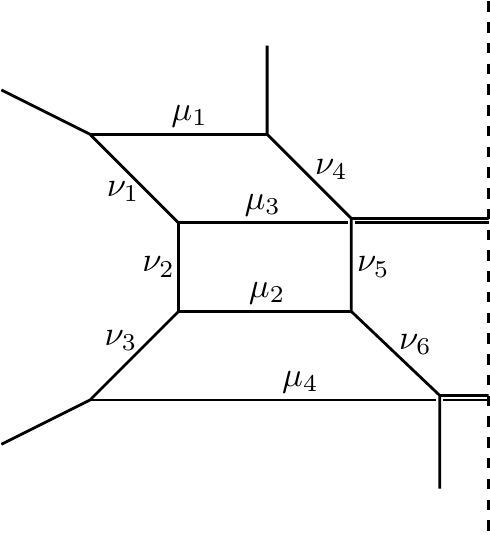}
		\caption{5-brane configuration for $D_2(2,2)$ gauge theory of overlapping sub webs.}
		\label{fig:D22,2s}	
	\end{figure}
	
	The two SU$(2)$ nodes in Figure \ref{fig:D22,2}(a) are drawn in the way that their sub webs do not overlap with each other. If we move the color brane $\mu_3$ upward passing the color brane $\mu_2$, we then end up with Figure \ref{fig:D22,2s} in which the sub webs of the two SU(2) nodes overlap. Our refined topological vertex prescription does not work well in this kind of overlapping case. If we simply assume the vertices on the strip of NS-branes next to the ON-plane to be $C$ or $C^{\text{R}}$ or some modified vertices which are obtained by changing the coefficients of $|\eta|,|\lambda|,|\mu|$ in the power of $\frac{\fq}{\ft}$ in \eqref{eq:CR} and multiply the K\"ahler parameters of $\nu_{4,5,6}$ by some $\ft,\fq$ factors as we have done in fixing the $x$, then after summing over Young diagrams in the non-preferred direction, we always end up with non-vanishing bifundamental contributions or incorrect vector contributions\footnote{The correct vector contribution should have the form of $\left(\calR_{\alpha^t\beta}(Q\sqrt{\frac{\fq}{\ft}})\calR_{\alpha^t\beta}(Q\sqrt{\frac{\ft}{\fq}})\right)^{-1}$ where $Q$ is the K\"ahler ``distance'' from the color brane $\alpha$ to $\beta$ with the vertical position of brane $\alpha$ being higher than $\beta$.}. We suspect that when $\mu_3$ is moved upward passing $\mu_2$ the two vertices $C$ and $C^{\text{R}}$ are entangled and the entangled vertices are hard to determine. So in this paper, we will always use separated sub webs which do not overlap for computation.

	\begin{figure}[t]
		\centering
		\includegraphics[scale=1]{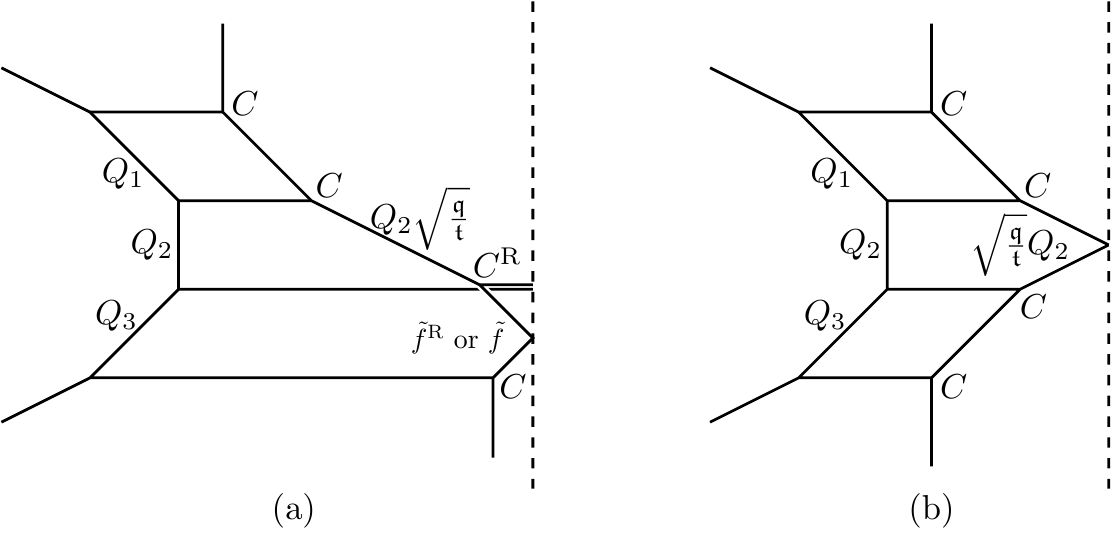}
		\caption{Two 5-brane webs for $D_2(2,2)$ gauge theory which are related by a series of flops.}
		\label{fig:D22,2ss}
	\end{figure}
	From Figure \ref{fig:D22,2}(a) we can ``flop" the lower $C^{\text{R}}$ vertex\footnote{By ``flop", we mean that we deform 5-brane web near an ON-plane such that the shape of the 5-brane web is changed so that the horizontal K\"ahler parameter from the ON-plane to the corresponding vertex becomes its inverse. More precisely, the position on the ON-plane is shifted toward the 5-brane web.  } or both the $C^{\text{R}}$ vertices into the ON-plane and reflect back, then we obtain Figure \ref{fig:D22,2ss}(a) and \ref{fig:D22,2ss}(b) respectively. Correspondingly, the vertices and edge factors also change due to the flopping, we have labelled them in Figure \ref{fig:D22,2ss}. Note that when using \eqref{eq:newedge1} or \eqref{eq:newedge1p} to determine the edge factor of the reflected brane that connects $C$ and $C^{\text{R}}$ in Figure \ref{fig:D22,2ss}(a), if we choose $C$ as the vertex that is reflected by the ON-plane, then we should use $\tilde{f}^{\text{R}}$ as the framing factor function for $\calF$, if we choose $C^{\text{R}}$ as the vertex that is reflected by the ON-plane, then we should use $\tilde{f}$ as the framing factor function for $\calF$. The derivation is just like the previous derivations that we have done in finding out the edge factors, so we omit it here.
	
	\begin{figure}[t]
		\centering
		\includegraphics[scale=0.38]{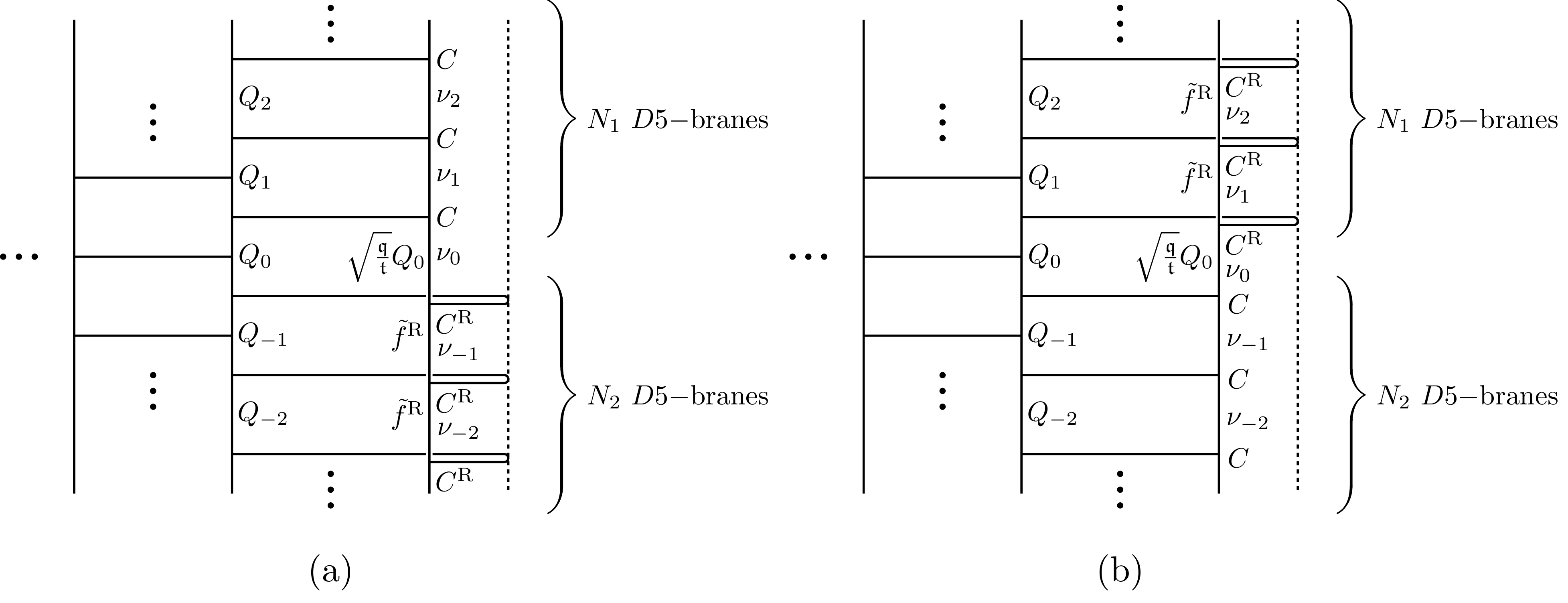}
		\caption{5-brane configurations for general $D$-type quiver gauge theory with an ON-plane located on the right.
		}
		\label{fig:DN1,N2R}
	\end{figure}
	\begin{figure}[t]
		\centering
		\includegraphics[scale=0.38]{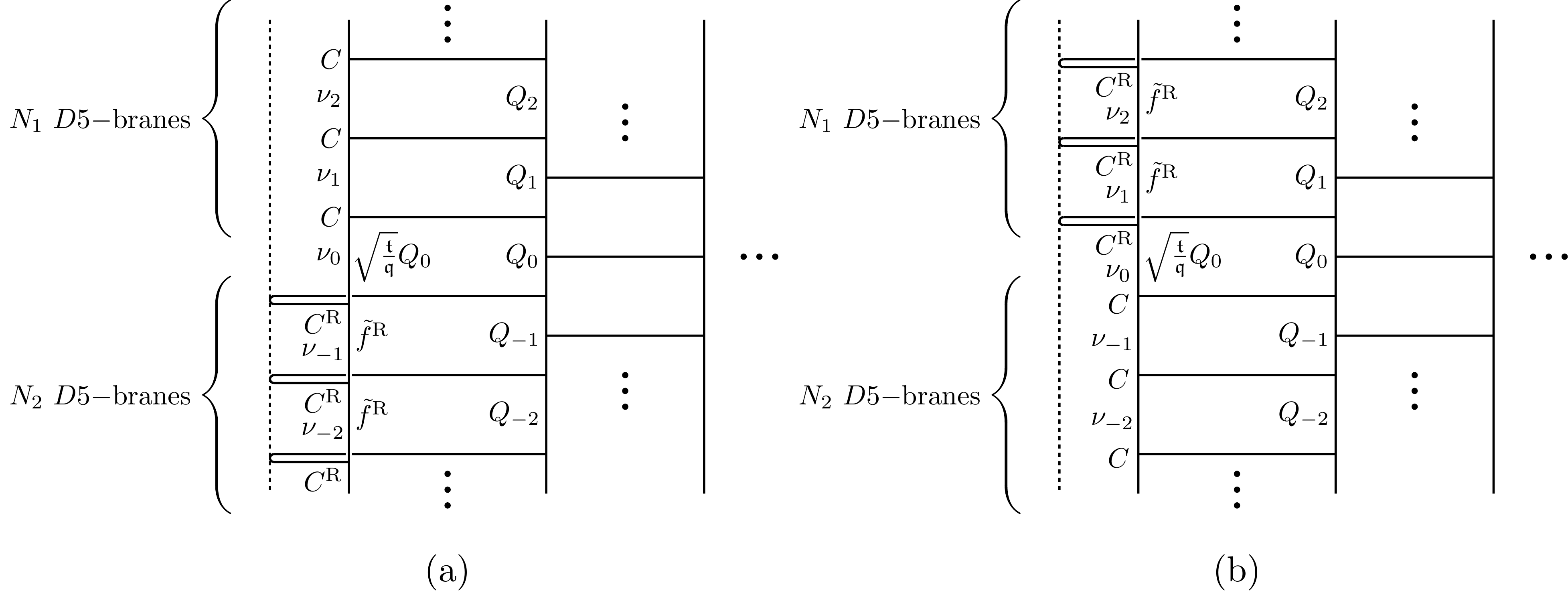}
		\caption{5-brane webs for general $D$-type quiver gauge theory with an ON-plane located on the left.}
		\label{fig:DN1,N2L}
	\end{figure}
	A general $D$-type quiver gauge theory with SU$(N_1)$ and SU$(N_2)$ gauge groups at its bivalent nodes is illustrated in Figure \ref{fig:DN1,N2R}(a), the sub webs of the two SU nodes do not overlap. The analysis for refinement we have done in the $D_2(2,2)$ case can be easily generalized to this case, here we just state the result without proof. Along the strip of NS5-branes next to the ON-plane, we multiply the factor $\sqrt{\fq/\ft}$ to $Q_0$ of $\nu_0$ which is between the two SU nodes, all the vertices and framing factors in the SU$(N_2)$ nodes are $C^{\text{R}}$ and $\tilde{f}^{\text{R}}$. If we flop the vertices $C^{\text{R}}$ into the ON-plane and reflect back, the vertices will become $C$, 
	as 
	illustrated in Figure \ref{fig:D22,2ss}.

\bigskip	
\paragraph{Summary of our proposal on topological vertex with an ON-plane.}
5-brane configurations can be viewed as consisting of vertical strips of NS5-branes glued by horizontal D5-branes with vertex factors and non-preferred direction edge factors living on vertical strips and preferred direction edge factors living on D5-branes. Away from ON-planes, such factors are the same as the usual vertex and edge factors as in the cases without ON-planes. We propose that new vertex and edge factors are only presented on the strips of NS5-branes next to an ON-plane with the following steps:
\begin{enumerate}
	\item [(1)]{\bf Deform 5-brane webs so that bivalent sub webs have no overlapping phase.}
	5-brane web with an ON-plane describes a $D$-type quiver which has an SU gauge theory node at each bivalent leg. A typical 5-brane web with an ON-plane for such quiver may be drawn in a way that two sub webs for each SU gauge theory look overlapping as depicted in Figure \ref{fig:splitconf}(a). %
\begin{figure}[t]
		\centering
		\includegraphics[scale=0.5]{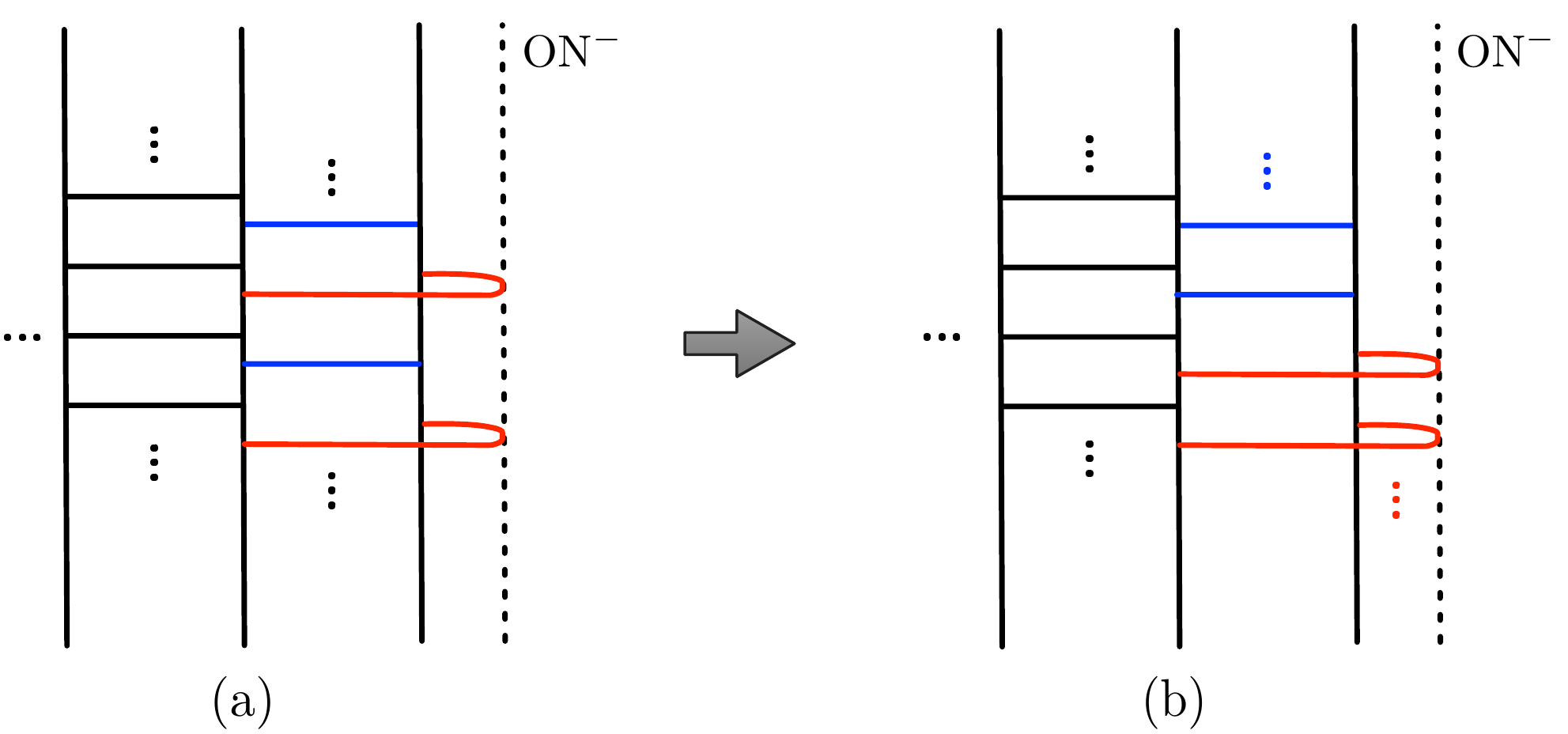}
		\caption{Two phases of 5-brane webs with an ON-plane: (a) Overlapping phase of two bivalent SU gauge groups of a $D$-type quiver (b) Split phase of SU gauge groups of a $D$-type quiver. D5-branes in red and D5-branes in blue correspond to each SU gauge theory of two ends of bivalent node.}
		\label{fig:splitconf}
\end{figure}%
Such overlapping phase can be deformed to yield two sub webs which do not have any overlapping phase as in Figure \ref{fig:splitconf}(b). In order for the topological vertex with an ON-plane to be applied, we require that 5-brane configuration should have no overlapping phase.

\item [(2)] {\bf Apply $C^\text{R}, \tilde{f}^\text{R}$ along the strip next to ON-planes.}\\ 
		If the prescription (1) is satisfied, we claim that there are two kinds of vertices that we need to assign on the strip next to an ON-plane.  
		One is the usual topological vertex factor $C$, the other one is the reflected refined topological vertex factor $C^{\text{R}}$ defined in \eqref{eq:CR}. The framing factor for the edges that connect $C^{\text{R}}$'s is the new framing factor $\tilde{f}^{\text{R}}$ defined in \eqref{eq:fR}. If we draw the web diagram in the standard way like Figure \ref{fig:DN1,N2R} or Figure \ref{fig:DN1,N2L}, then the assignment of the vertex and framing factors on the strip should follow that in Figure \ref{fig:DN1,N2R}(a) or Figure \ref{fig:DN1,N2R}(b) when the ON-plane is on the right-hand side, and should follow that in Figure \ref{fig:DN1,N2L}(a) or Figure  \ref{fig:DN1,N2L}(b) when the ON-plane is on the left-hand side.
		
		\item[(3)] {\bf Dress additional $\ft,\fq$ factors on the edge connecting two SU sub webs.}\\
		$\sqrt{\frac{\fq}{\ft}}$ or $\sqrt{\frac{\ft}{\fq}}$ needs to be multiplied to the K\"ahler parameters of the edges that connect the two SU nodes which is illustrated in Figure \ref{fig:DN1,N2R} and Figure \ref{fig:DN1,N2L}.
		
		\item[(4)]{\bf Introduce the new reflected edge factor to the edge reflected by an ON-plane.}\\
		For edges that are reflected by an ON-plane as in Figure \ref{fig:reflection of ref TV}(a), the corresponding edge factor is given by \eqref{eq:newedge1} when the reflected vertex is associated with the vertex factor of $C_{}(\ft,\fq)$ or $C^\text{R}(\ft,\fq)$. When the reflected vertex is of $C_{}(\fq,\ft)$ or $C^\text{R}(\fq,\ft)$, the corresponding edge factor is given by \eqref{eq:newedge1p}.
\end{enumerate}
We also remark that we can ``flop" the vertices into the ON-planes to obtain diagrams like those given in Figure \ref{fig:D22,2ss}. For such cases, $C$ and $C^{\text{R}}$ become $C^{\text{R}}$ and $C$ respectively after flopping, and the edge factors also change according to the edge factors presented in Figure \ref{fig:D22,2ss}.

\bigskip

\section{Examples}
	\label{sec:SU2withfour}
	In this section, we demonstrate how to apply the refined topological vertex with ON-planes that we proposed in the previous section with instructive examples as well as some rank-2 theories whose ADHM construction is not known. We provide computational details.  
	
\subsection{5d SU(2)+4{\bf F}}
\begin{figure}
	\centering
	\includegraphics[scale=0.4]{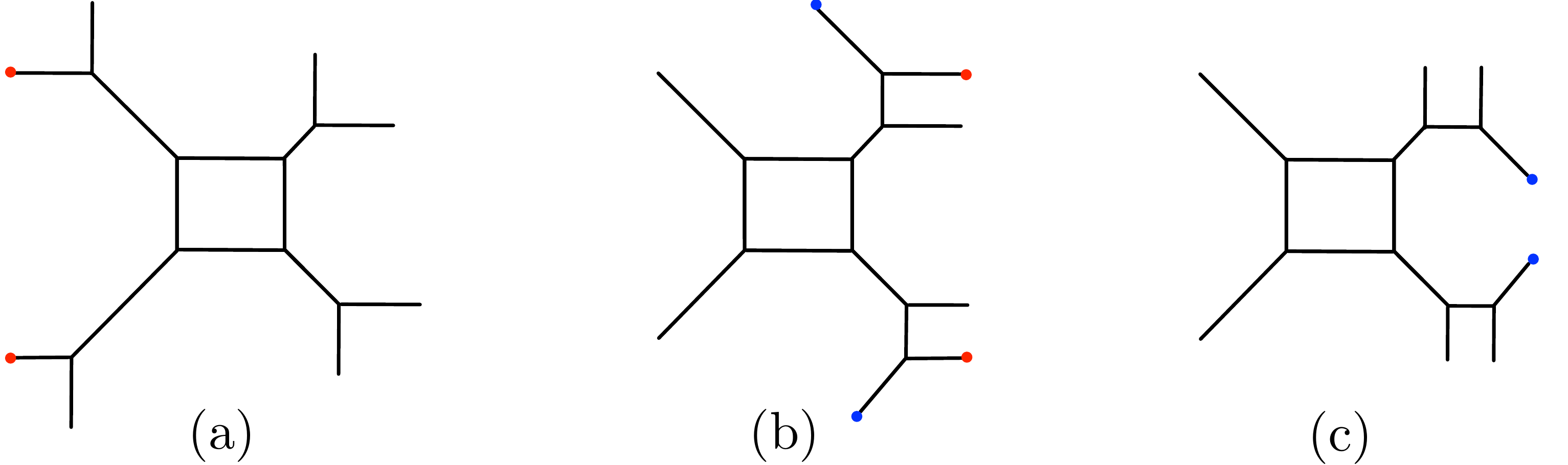}
	\caption{5-brane webs for SU(2)+4{\bf F}. (a) A typical 5-brane web. (b) Performing Hanany-Witten moves to the left. (c) A 5-brane configuration as a $D_3$-type quiver, ``SU(1)"$-$SU(2)$-$``SU(1)", after further Hanany-Witten moves.}
	\label{fig:SU2+4F}
	\end{figure}
\begin{figure}
		\centering
		\includegraphics[scale=1]{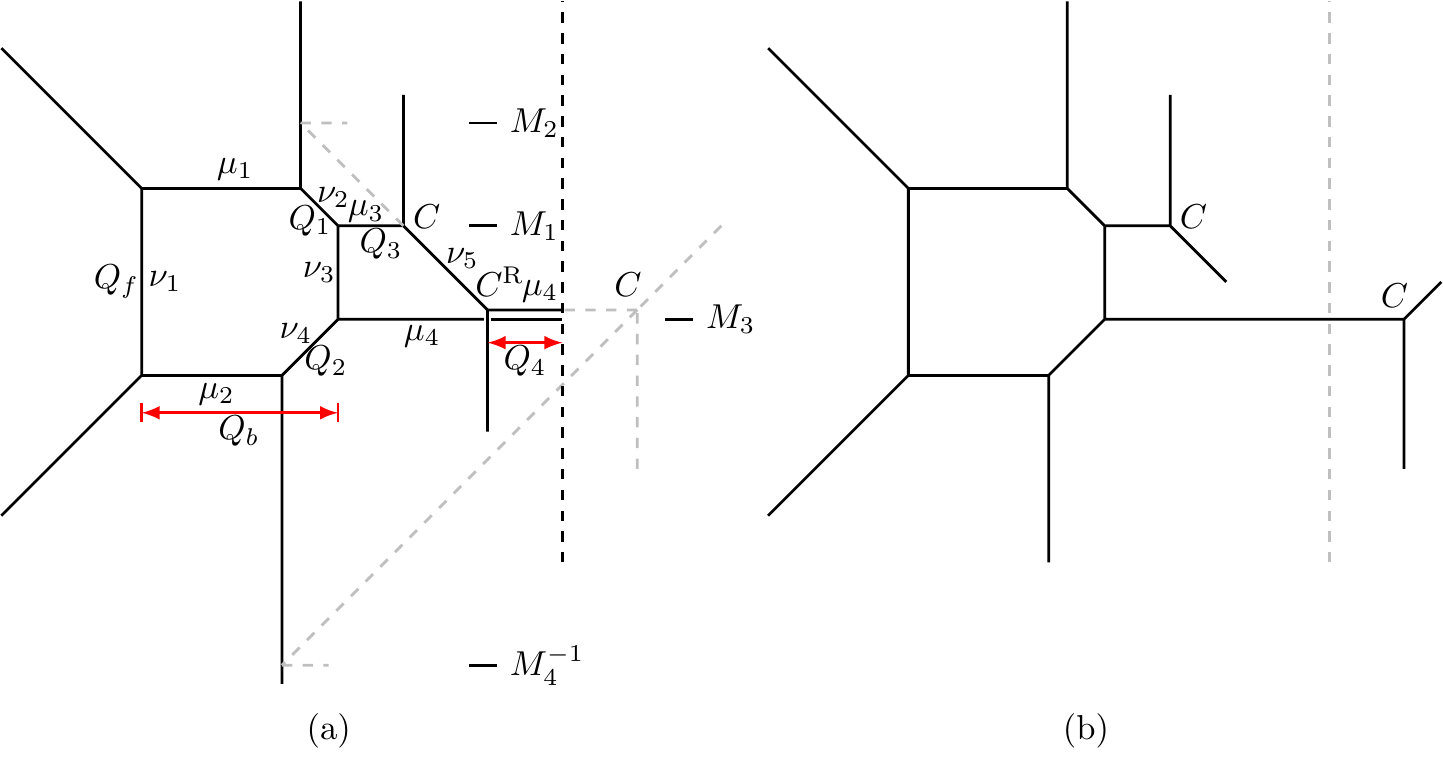}
		\caption{A 5-brane web for SU(2)+4{\bf F} as a $D_3$-quiver, SU(1)$-$SU(2)$-$SU(1). The vertical dotted line represents an ON$^-$-plane.}
		\label{fig:D5 reflection}
\end{figure}

5d $\mathcal{N}=1$ SU$(2)$ gauge theory with four hypermultiplets in the fundamental representation (SU$(2)+4\mathbf{F}$) is a proper instructive example for applying topological vertex with ON-planes as its refined partition function is well known and also this theory can be regarded as a quiver ``SU(1)"$-$SU(2)$-$``SU(1)". A typical 5-brane configuration for SU(2)+4{\bf F} is depicted in Figure \ref{fig:SU2+4F}(a). It is easy to see that Hanany-Witten moves can deform the 5-brane configuration so that it can be viewed as ``SU(1)"$-$SU(2)$-$``SU(1)" as in Figure \ref{fig:SU2+4F}(c), where an ``SU(1)" is made out of one D5-brane suspended between two NS5-branes that does not have Coulomb branch but has the coupling \cite{Hayashi:2015fsa}. One can see that together with the bifundamental degrees of freedom connecting SU(2) and ``SU(1)",  ``SU(1)"$-$SU(2)$-$``SU(1)" captures the same degrees of freedom as those of SU(2)+4{\bf F}. As it can be also seen as a $D_3$-type quiver, the SU(2)+4{\bf F} theory can be described with a 5-brane configuration with an ON-plane as given in Figure \ref{fig:D5 reflection}(a).

Following our proposal with ON-planes presented in the previous section, we assign the vertex factors $C$ and $C^{\text{R}}$ on 5-brane web as shown in Figure \ref{fig:D5 reflection}(a). In particular,  $C^{\text{R}}$ is assigned to the vertex on the strip next to an ON-plane, which can be viewed as a configuration by reflecting and gluing Figure \ref{fig:D5 reflection}(b) which is one of possible 5-brane configurations for SU(2)+4{\bf F} as depicted in Figure \ref{fig:SU2+4F}. It follows from Figure \ref{fig:D5 reflection} that the K\"ahler parameters for the masses of 4 flavors $M_i=e^{-\beta m_i},i=1,2,3,4$ can be readily identified. For later convenience, we have defined the fourth mass with $M_4^{-1}$ in Figure \ref{fig:D5 reflection}(a). 

The partition function for SU(2)+4{\bf F} based on our proposal in the previous section then takes the form: 
\begin{align}
	Z^\text{SU(2)+4{\bf F}}
	&=\sum_{\boldsymbol{\mu,\nu}}C_{\nu_1^t\text{\o}\mu_1^t}(\fq,\ft)C_{\text{\o}\nu_1\mu_2^t}(\fq,\ft)C_{\text{\o}\nu_2^t\mu_1}(\ft,\fq)C_{\nu_3^t\nu_2\mu_3^t}(\fq,\ft)C_{\nu_4^t\nu_3\mu_4^t}(\fq,\ft)\crcr
	&\times C_{\nu_4\text{\o}\mu_2}(\ft,\fq) C_{\text{\o}\nu_5^t\mu_3}(\ft,\fq)C^{\text{R}}_{\text{\o}\nu_5\mu_4^t}(\fq,\ft) (-Q_f)^{|\nu_1|}\tilde{f}_{\nu_1}(\fq,\ft)(-Q_1)^{|\nu_2|}(-\frac{Q_f}{Q_{1,2}})^{|\nu_3|}\crcr
	&\times
	\tilde{f}_{\nu_3}(\fq,\ft)(-Q_2)^{|\nu_4|}
		(-\frac{Q_f}{Q_{1,2}}\sqrt{\frac{\fq}{\ft}})^{|\nu_5|}(-\frac{Q_b}{Q_1})^{|\mu_1|}(-\frac{Q_b}{Q_2})^{|\mu_2|}(-Q_3)^{|\mu_3|}\crcr
	&\times 
		(+\frac{Q_{f,3,4^2}}{Q_{1,2}})^{|\mu_4|}f_{\mu_4^t}(\fq,\ft)\frac{\tilde{Z}_{\mu_4}(\ft,\fq)}{\tilde{Z}_{\mu_4^t}(\fq,\ft)}\ ,
\end{align}
where $\boldsymbol{\mu, \nu}$ means a collective notation for all Young diagrams associated with $\mu_i$ and $\nu_i$, and we have used the shorthand notation $Q_{i,j,k^2}\equiv Q_{i}Q_{j}Q_{k}^2.$ Recall that if an empty set is assigned to the new vertex factor $C^\text{R}$ in the non-preferred direction, then $C^\text{R}$ is equivalent to usual $C$. Hence, in this example, we have $C^{\text{R}}_{{\text{\o}}\nu_5\mu_4^t}(\fq,\ft)=C_{{\text{\o}}\nu_5\mu_4^t}(\fq,\ft)$.  
After simplifying the Young diagram sum $\boldsymbol{\nu}$ along the non-preferred directions, we obtain
	\begin{align}
		Z^\text{SU(2)+4{\bf F}}
		=&~\sum_{\boldsymbol{\mu}}(-Q_3)^{|\mu_3|}\left(-\frac{Q_b}{Q_1}\right)^{|\mu_1|}\left(-\frac{Q_b}{Q_2}\right)^{|\mu_2|}\left(-\frac{Q_{3,4^2,f}}{Q_{1,2}}\right)^{|\mu_4|}\nonumber\\
		\times&~\fq^{\frac12\sum_{i=1}^4||\mu_i||^2}\ft^{\frac12\sum_{i=1}^4||\mu_i^t||^2}\prod_{i=1}^4\tilde{Z}_{\mu_i}(\ft,\fq)\tilde{Z}_{\mu_i^t}(\fq,\ft)\nonumber\\
		\times&~\frac{\calR_{\mu_1^t\mu_3}(Q_1)\calR_{\mu_1^t\mu_4}(\frac{Q_f}{Q_2})\calR_{\mu_3^t\mu_2}(\frac{Q_f}{Q_1})\calR_{\mu_4^t\mu_2}(Q_2)}{\calR_{\mu_1^t\mu_2}(Q_f\sqrt{\frac{\fq}{\ft}})\calR_{\mu_1^t\mu_2}(Q_f\sqrt{\frac{\ft}{\fq}})}\ .
		\label{eq:su2+4F simplified}
	\end{align}

We now express this partition function in terms of the gauge theory parameters. Based on the 5-brane web in Figure \ref{fig:D5 reflection}, one readily finds the map between the K\"ahler parameters and the gauge theory parameters,
	\begin{align}
		&Q_b=A^2u\sqrt{\frac{M_3}{M_1M_2M_4}},\qquad\quad~\,
		Q_f=A^2,\cr
		&Q_1=\frac{A}{M_1},\qquad Q_2=AM_3,\qquad
		Q_3=\frac{M_2}{A},\qquad
		Q_4=\sqrt{\frac{M_3M_4}{M_1M_2}}\ ,
		\label{eq:su2w4ap}
	\end{align}
where $u$ is the instanton factor and $A$ is the Coulomb branch parameter. By substituting \eqref{eq:su2w4ap} into \eqref{eq:su2+4F simplified} and using \eqref{eq:RMN}, we find
	\begin{equation}
		Z^\text{SU(2)+4{\bf F}}=Z^{N_f=4}_{\text{M}}\sum_{\mu_1,\mu_2}Z_{\mu_1\mu_2}^{N_f=4}\sum_{\mu_3}Z_{\mu_1\mu_2\mu_3}(M_1,M_2)\sum_{\mu_4}Z_{\mu_1\mu_2\mu_4}(M_3,M_4)\ ,
		\label{eq:Z4}
	\end{equation}
where
	\begin{align}
		&Z^{N_f=4}_{\text{M}} =\frac{\calM(A^2)\calM(A^2\frac{\ft}{\fq})}{\calM(\frac{A}{M_1}\sqrt{\frac{\ft}{\fq}})\calM(AM_1\sqrt{\frac{\ft}{\fq}})\calM(\frac{A}{M_3}\sqrt{\frac{\ft}{\fq}})\calM(AM_3\sqrt{\frac{\ft}{\fq}})}\ ,
		\label{eq:Z4M}\\
		&Z_{\mu_1\mu_2}^{N_f=4}=\frac{\fq^{\frac{||\mu_1||^2+||\mu_2||^2}{2}}\ft^{\frac{||\mu_1^t||^2+||\mu_2^t||^2}{2}}(-Au)^{|\mu_1|+|\mu_2|}\tilde{Z}_{\mu_1}(\ft,\fq)\tilde{Z}_{\mu_1^t}(\fq,\ft)\tilde{Z}_{\mu_2}(\ft,\fq)\tilde{Z}_{\mu_2^t}(\fq,\ft)}{\calN_{\mu_1\mu_2}(A^2)\calN_{\mu_1\mu_2}(A^2\frac{\ft}{\fq})}\ , \label{eq:ZN4mu12}\\
		&Z_{\mu_1\mu_2\mu_3}(M_1,M_2)=\Big(\sqrt{\frac{M_1}{M_2}}\Big)^{|\mu_1|}\Big(\sqrt{\frac{1}{M_1M_2}}\Big)^{|\mu_2|}\Big(-\frac{M_2}{A}\Big)^{|\mu_3|}\fq^{\frac{||\mu_3||^2}{2}}\ft^{\frac{||\mu_3^t||^2}{2}}\crcr
		&\qquad\qquad\qquad\quad~ \times\tilde{Z}_{\mu_3}(\ft,\fq)\tilde{Z}_{\mu_3^t}(\fq,\ft)\calN_{\mu_1\mu_3}(\tfrac{A}{M_1}\sqrt{\tfrac{\ft}{\fq}})\calN_{\mu_3\mu_2}(AM_1\sqrt{\tfrac{\ft}{\fq}})\ ,
		\label{eq:Zmu123}\\
	&Z_{\mu_1\mu_2\mu_4}(M_3,M_4)=\Big(\sqrt{\frac{M_3}{M_4}}\Big)^{|\mu_1|}\Big(\sqrt{\frac{1}{M_3M_4}}\Big)^{|\mu_2|}\Big(-\frac{M_4}{A}\Big)^{|\mu_4|}\fq^{\frac{||\mu_4||^2}{2}}\ft^{\frac{||\mu_4^t||^2}{2}}\crcr
		&\qquad\qquad\qquad\quad~ \times\tilde{Z}_{\mu_4}(\ft,\fq)\tilde{Z}_{\mu_4^t}(\fq,\ft)\calN_{\mu_1\mu_4}(\tfrac{A}{M_3}\sqrt{\tfrac{\ft}{\fq}})\calN_{\mu_4\mu_2}(AM_3\sqrt{\tfrac{\ft}{\fq}})\ .	\label{eq:Zmu124}
\end{align}

\begin{figure}[t]
		\centering
		\includegraphics[scale=1]{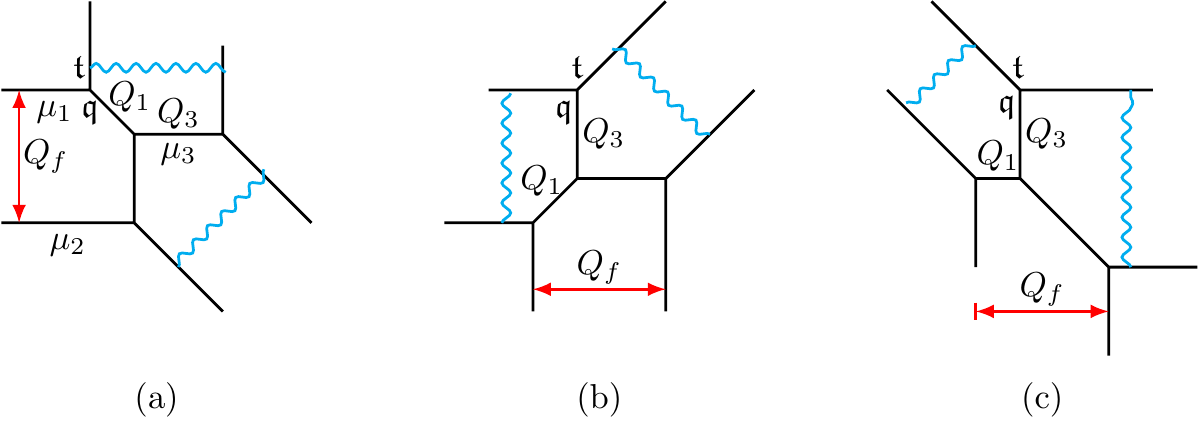}
		\caption{Sources of the extra factors.}
		\label{fig:extra factors}
\end{figure}

Here, we remark some technical points. The contribution of $Z_{\mu_1\mu_2\mu_3}(M_1,M_2)$ to the perturbative part of the partition function is $\sum_{\mu_3}Z_{\text{\o}\text{\o}\mu_3}(M_1,M_2)$, so we need to sum over all the Young diagrams $\mu_3$. This means that one sums over all the partitions associated with the Young diagram $\mu_3$ from zero partition to infinite partitions. In order to compute such an infinite summation, we take the logarithm of $\sum_{\mu_3}Z_{\text{\o}\text{\o}\mu_3}(M_1,M_2)$ and then do a series expansion of it with respect to $M_2$. Practically, one can sum over the Young diagrams $\mu_3$ up to some finite box number as the upper bound such that for terms involving $M_2$, lower order terms do not get corrected even though one increases the upper bound, while higher order terms do get corrected when the upper bound of $|\mu_3|$ increase, that is an artifact of a finite sum which will not be presented when we actually sum all the way to infinity.  
By using this trick, we are able to find the correct lower order expansion of $\sum_{\mu_3}Z_{\text{\o}\text{\o}\mu_3}(M_1,M_2)$, from the expansion we find that it has the following form as the Plethystic exponential,
	\begin{align}
		\sum_{\mu_3}Z_{\text{\o}\text{\o}\mu_3}(M_1,M_2)=&~\text{PE}\bigg[\frac{-M_1M_2\sqrt{\fq\ft}-A^2M_1M_2\sqrt{\fq\ft}+AM_2(M_1^2\fq+\ft)}{AM_1(1-\fq)(1-\ft)}\bigg]\cr
		=&~\frac{\calM(M_1M_2)\calM(\frac{M_2}{M_1}\frac{\ft}{\fq})}{\calM(\frac{M_2}{A}\sqrt{\frac{\ft}{\fq}})\calM(AM_2\sqrt{\frac{\ft}{\fq}})}\ .
		\label{eq:Z00mu3}
	\end{align}
Notice that $\calM(M_1M_2)$ and $\calM(\frac{M_2}{M_1}\frac{\ft}{\fq})$ do not depend on the Coulomb branch parameter $A$, so they are the extra factors. In terms of K\"ahler parameters, they are given as  $\calM(\frac{Q_3Q_f}{Q_1})$ and $\calM(Q_1Q_3\frac{\ft}{\fq})$ respectively. These extra factors can be also seen from 5-brane webs. For instance, $Q_1Q_3$ is the distance between the two external parallel branes in the upper part of Figure \ref{fig:D5 reflection}(a), and hence it contributes to the extra factor $\calM(Q_1Q_3\frac{\ft}{\fq})$. On the other hand, the extra factor $\calM(\frac{Q_3Q_f}{Q_1})$ is not easy to directly see from Figure \ref{fig:D5 reflection}(a). We note that these two extra factors rather can be seen from Figure \ref{fig:extra factors}. 
The function $Z_{\mu_1\mu_2\mu_3}(M_1,M_2)$ is proportional to the topological string partition function of Figure \ref{fig:extra factors}(a). The extra factors $\calM(\frac{Q_3Q_f}{Q_1}),\calM(Q_1Q_3\frac{\ft}{\fq})$ correspond to the two $(p,q)$-strings in blue which connect the parallel external branes. Figure \ref{fig:extra factors}(b) is the S-dual of Figure \ref{fig:extra factors}(a), from Figure \ref{fig:extra factors}(b) it is easy to see the extra factor corresponding to the left vertical string which is $\calM(Q_1Q_3\frac{\ft}{\fq})$. Figure \ref{fig:extra factors}(c) is obtained by an $SL(2,\mathbb{Z})$ transformation of Figure \ref{fig:extra factors}(b), from which we can easily see that the other extra factor $\calM(\frac{Q_3Q_f}{Q_1})$ corresponds to 
the $(p,q)$-string in blue that is vertically expanded on the right side. 

The same analysis applies to $\sum_{\mu_4}Z_{\text{\o}\text{\o}\mu_4}(M_3,M_4)$. Combining these results with \eqref{eq:Z4M}, we finally obtain the full perturbative part\footnote{After necessary flops: $\calM(Q;\ft,\fq)\rightarrow\calM(\frac{1}{Q};\fq,\ft)=\calM(\frac{1}{Q}\frac{\ft}{\fq};\ft,\fq)$.},
	\begin{multline}
		Z^\text{SU(2)+4{\bf F}}_{\text{pert}}
		=\calM(A^2)\calM(A^2\frac{\ft}{\fq})\Bigg[\calM(\frac{A}{M_1}\sqrt{\frac{\ft}{\fq}})\calM(AM_1\sqrt{\frac{\ft}{\fq}})\calM(\frac{A}{M_3}\sqrt{\frac{\ft}{\fq}})\calM(AM_3\sqrt{\frac{\ft}{\fq}})\\\times\calM(\frac{A}{M_2}\sqrt{\frac{\ft}{\fq}})\calM(AM_2\sqrt{\frac{\ft}{\fq}})\calM(\frac{A}{M_4}\sqrt{\frac{\ft}{\fq}})\calM(AM_4\sqrt{\frac{\ft}{\fq}})\Bigg]^{-1}.
	\end{multline}
	Written as Plethystic exponential, 
	\begin{equation}
		Z^\text{SU(2)+4{\bf F}}_{\text{pert}}=\text{PE}\left[\frac{1}{(1-\ft)(1-\fq)}\left((\fq+\ft)A^2-\chi_{\textbf{8v}}^{\text{SO(8)}}\sqrt{\fq\ft}A\right)\right]\ , 
		\label{eq:Zpert4}
	\end{equation}
	where $\chi^{\text{SO(8)}}_{\textbf{8v}}$ is the character for the 8-dimensional vector representation of SO(8). From here on, we use   $\chi^{\text{SO(8)}}_{\textbf{n}}$ to denote the character of the $\mathbf{n}$-dimensional representation of SO(8), defined in Appendix \ref{sec:App}.

	Extracting the perturbative part from \eqref{eq:Z4}, one finds that the instanton part can be expressed as 
	\begin{equation}
		Z^\text{SU(2)+4{\bf F}}_{\text{inst}}=\sum_{\mu_1,\mu_2}Z_{\mu_1\mu_2}^{N_f=4}\Bigg(\frac{\sum_{\mu_3}Z_{\mu_1\mu_2\mu_3}(M_1,M_2)}{\sum_{\mu_3}Z_{\text{\o}\text{\o}\mu_3}(M_1,M_2)}\Bigg)\Bigg(\frac{\sum_{\mu_4}Z_{\mu_1\mu_2\mu_4}(M_3,M_4)}{\sum_{\mu_4}Z_{\text{\o}\text{\o}\mu_4}(M_3,M_4)}\Bigg)\ .
		\label{eq:Z4instunsim}
	\end{equation}
At first sight, it appears that we need to sum over all the Young diagrams associated with $\mu_3,\mu_4$ in order to obtain the coefficients of $u$ at a fixed order because $Z_{\mu_1\mu_2}^{N_f=4}$ is proportional to $u^{|\mu_1|+|\mu_2|}$. However, the sum is truncated at finite order.  Recall that the SU$(2)+4\mathbf{F}$ theory has an SO$(8)$ global symmetry which can be further enhanced to SO(10) \cite{Seiberg:1996bd, Kim:2012gu}, and also notice that the individual factor $\frac{\sum_{\mu_3}...}{\sum_{\mu_3}...}$ in the parenthesis only involves $M_1$ and $M_2$ which is of an SO$(4)=$SU($2$)$\times$SU($2$) symmetry, exchanging $M_1$ and $M_1^{-1}$ as well as $M_2$ and $M_2^{-1}$. 
If we Taylor expand $\frac{\sum_{\mu_3}...}{\sum_{\mu_3}...}$ with respect to $M_2$, the lowest order is $M_2^{-\frac{|\mu_1|+|\mu_2|}{2}}$, and it looks like the highest order could be infinity, but because of the symmetry between $M_2$ and $M_2^{-1}$, the highest order should be $M_2^{\frac{|\mu_1|+|\mu_2|}{2}}$, so higher order terms greater than $\frac{|\mu_1|+|\mu_2|}{2}$ would cancel with one another. We define $\calZ_{\mu_1\mu_2}\equiv\frac{\sum_{\mu_3}...}{\sum_{\mu_3}...}$ and rewrite it as follows,  
	\begin{equation}
		\calZ_{\mu_1\mu_2}(M_1,M_2)\equiv\frac{\sum_{\mu_3}Z_{\mu_1\mu_2\mu_3}(M_1,M_2)}{\sum_{\mu_3}Z_{\text{\o}\text{\o}\mu_3}(M_1,M_2)}=\frac{\sum_{\mu_3}B_{\mu_1\mu_2\mu_3}M_2^{|\mu_3|}}{\sum_{\mu_3}D_{\mu_3}M_2^{|\mu_3|}}M_2^{-\frac{|\mu_1|+|\mu_2|}{2}}\ ,
		\label{eq:defZmu12}
	\end{equation}
where unimportant coefficients are denoted by $B$, $D$ for simplicity. 
	The lowest order in the Taylor expansion of the fractional factor in \eqref{eq:defZmu12} is 1 and the highest order should be $M_2^{|\mu_1|+|\mu_2|}$, we obtain the following formula:
	\begin{equation}
		\frac{\sum_{\mu_3}B_{\mu_1\mu_2\mu_3}M_2^{|\mu_3|}}{\sum_{\mu_3}D_{\mu_3}M_2^{|\mu_3|}}M_2^{-\frac{|\mu_1|+|\mu_2|}{2}}=\left(\frac{\sum_{\mu_3}^{|\mu_3|\leq|\mu_1|+|\mu_2|}B_{\mu_1\mu_2\mu_3}M_2^{|\mu_3|}}{\sum_{\mu_3}^{|\mu_3|\leq|\mu_1|+|\mu_2|}D_{\mu_3}M_2^{|\mu_3|}}\right)^{\leq|\mu_1|+|\mu_2|}M_2^{-\frac{|\mu_1|+|\mu_2|}{2}}\ .
		\label{eq:cutoff}
	\end{equation}
On the right-hand side we sum over Young diagrams 
no greater than $|\mu_1|+|\mu_2|$. The superscript of the big parentheses means that we only keep the terms whose $M_2$ orders are no greater than $|\mu_1|+|\mu_2|$ after Taylor expanding the fraction in the parentheses with respect to $M_2$. In this way\footnote{See also section 4.3.2 in \cite{Hayashi:2013qwa} for a related discussion.}, we can compute the sum over $\mu_3$ exactly by summing over finite number of Young diagrams. 

We note that one may find that even after we obtain the right-hand side of \eqref{eq:cutoff}, the result is still not very compact, although we have a finite polynomial with $M_2$. To obtain a compact expression, we further expand it with respect to the Coulomb branch parameter $A$, again by a naive analysis that the lowest order term should be $A^{-|\mu_1|-|\mu_2|}$ and the highest order term should be $A^{2|\mu_1|+2|\mu_2|}$, it turns out that the actual computation yields that the lowest order term is 1 and the highest order term is $A^{|\mu_1|+|\mu_2|}$, due to drastic cancellations in lower and higher orders.  
We list the results of $\calZ_{\mu_1\mu_2}$'s for a few lower orders. For one-instanton partition function, the corresponding $\calZ_{\mu_1\mu_2}$'s take the form
	\begin{align}
		\calZ_{\{1\},\text{\o}}(M_1,M_2)=&~\frac{M_1+M_2}{\sqrt{M_1M_2}}-\Big(\frac{1}{\sqrt{M_1M_2}}+\sqrt{M_1M_2}\Big)\sqrt{\frac{\fq}{\ft}}A\ ,\nonumber\\
		\calZ_{\text{\o},\{1\}}(M_1,M_2)=&~\Big(\frac{1}{\sqrt{M_1M_2}}+\sqrt{M_1M_2}\Big)-\frac{M_1+M_2}{\sqrt{M_1M_2}}\sqrt{\frac{\ft}{\fq}}A\ .
		\label{eq:ZZ1}
	\end{align}
For two-instanton partition function, the corresponding $\calZ_{\mu_1\mu_2}$'s are given by
	\begin{align}
		\calZ_{\{2\},\text{\o}}(M_1,M_2)=&~1+\fq+\frac{M_1}{M_2}+\frac{M_2}{M_1}-\Big(\frac{1}{M_1}+\frac{1}{M_2}+M_1+M_2\Big)\sqrt{\frac{\ft}{\fq}}(1+\fq)A\nonumber\\
		&+\Big((1+\fq)\frac{\fq}{\ft}+\big(\frac{1}{M_1M_2}+M_1M_2\big)\frac{\fq^2}{\ft}\Big)A^2\ ,\nonumber\\
		\calZ_{\{1,1\},\text{\o}}(M_1,M_2)=&~1+\frac{1}{\ft}+\frac{M_1}{M_2}+\frac{M_2}{M_1}-\Big(\frac{1}{M_1}+\frac{1}{M_2}+M_1+M_2\Big)
		\sqrt\frac{\fq}{\ft}\big(1+\frac1\ft\big)A\nonumber\\
		&+\Big(\frac{\fq+\fq\ft}{\ft^2}+\big(\frac{1}{M_1M_2}+M_1M_2\big)\frac{\fq}{\ft^2}\Big)A^2\ ,\nonumber\\
		\calZ_{\{1\},\{1\}}(M_1,M_2)=&~\frac{1}{M_1}+M_1+\frac{1}{M_2}+M_2-\bigg(\frac{(1+\fq)(1+\ft)}{\sqrt{\fq\ft}}\nonumber\\
		&+\big(\frac{1}{M_1M_2}+M_1M_2\big)\sqrt{\frac{\fq}{\ft}}+\big(\frac{M_1}{M_2}+\frac{M_2}{M_1}\big)\sqrt{\frac{\ft}{\fq}}\,\bigg)A\nonumber\\
		&+\big(\frac{1}{M_1}+M_1+\frac{1}{M_2}+M_2\big)A^2\ ,\nonumber\\
		\calZ_{\text{\o},\{2\}}(M_1,M_2)=&~1+\frac{1}{\fq}+\frac{1}{M_1M_2}+M_1M_2-\big(\frac{1}{M_1}+M_1+\frac{1}{M_2}+M_2\big)
		\sqrt\frac\ft\fq \big(1\!+\!\frac1\fq\big)A\nonumber\\
		&+\Big(\frac{\ft+\fq\ft}{\fq^2}+\big(\frac{M_1}{M_2}+\frac{M_2}{M_1}\big)\frac{\ft}{\fq^2}\Big)A^2\ ,\nonumber\\
		\calZ_{\text{\o},\{1,1\}}(M_1,M_2)=&~1+\ft+\frac{1}{M_1M_2}+M_1M_2-\Big(\frac{1}{M_1}+M_1+\frac{1}{M_2}+M_2\Big)\sqrt{\frac{\ft}{\fq}}(1+\ft)A\nonumber\\
		&+\Big(\frac{\ft}{\fq}+\big(1+\frac{M_1}{M_2}+\frac{M_2}{M_1}\big)\frac{\ft^2}{\fq}\Big)A^2\ .
		\label{eq:ZZ2}
	\end{align}

Now we rewrite the instanton partition function \eqref{eq:Z4instunsim} as
\begin{equation}
		Z^{\text{SU(2)+4$\mathbf{F}$}}_{\text{inst}}=\sum_{\mu_1,\mu_2}Z_{\mu_1\mu_2}^{N_f=4}\calZ_{\mu_1\mu_2}(M_1,M_2)\calZ_{\mu_1\mu_2}(M_3,M_4)\ .
\end{equation}
By substituting the form of $\calZ_{\mu_1\mu_2}(M_1,M_2)\calZ_{\mu_1\mu_2}(M_3,M_4)$ at each order of the instanton, we can obtain the instanton partition function. For instance, the one-instanton partition function takes the following form up to $A^2$,
	\begin{equation}
		Z^{\text{SU(2)+4$\mathbf{F}$}}_{\text{inst}}\!=1+\!\left(\!-\frac{\sqrt{\fq\ft}\chi_{\textbf{8s}}^{\text{SO(8)}}}{(1-\fq)(1-\ft)}A+\frac{(\fq+\ft)\chi_{\textbf{8c}}^{\text{SO(8)}}}{(1-\fq)(1-\ft)}A^2+\calO(A^3)\!\right)u+\calO(u^2),
		\label{eq:Zinst4}
	\end{equation}
	where $\chi^{\text{SO(8)}}_{\bf 8s}$ and $\chi^{\text{SO(8)}}_{\bf 8c}$ are respectively the character for the spinor and conjugate spinor representation of SO(8), defined in Appendix \ref{sec:App}. 
	The perturbative and one-instanton results given in \eqref{eq:Zpert4} and \eqref{eq:Zinst4} perfectly agree with 
	the partition functions computed from the conventional 5-brane web diagram for SU(2)+4{\bf F} without an ON-plane, as expected.   
	The prescription of topological vertex in this example also coincide with the one that were found in \cite{Bourgine:2017rik} because $C^{\text{R}}$ reduces to $C$ in this case.
	
We also remark that there are other shape of 5-brane configurations than those that we discussed earlier. For instance,  Figure \ref{fig:(1,1) reflection} is another 5-brane diagram for SU$(2)+4\mathbf{F}$ which can be obtained by flopping the vertex associated with $C^{\text{R}}$ in Figure \ref{fig:D5 reflection}(a) into the ON-plane and reflecting back. By our proposal, the corresponding topological string partition function is given by
	\begin{figure}
		\centering
		\includegraphics[scale=1]{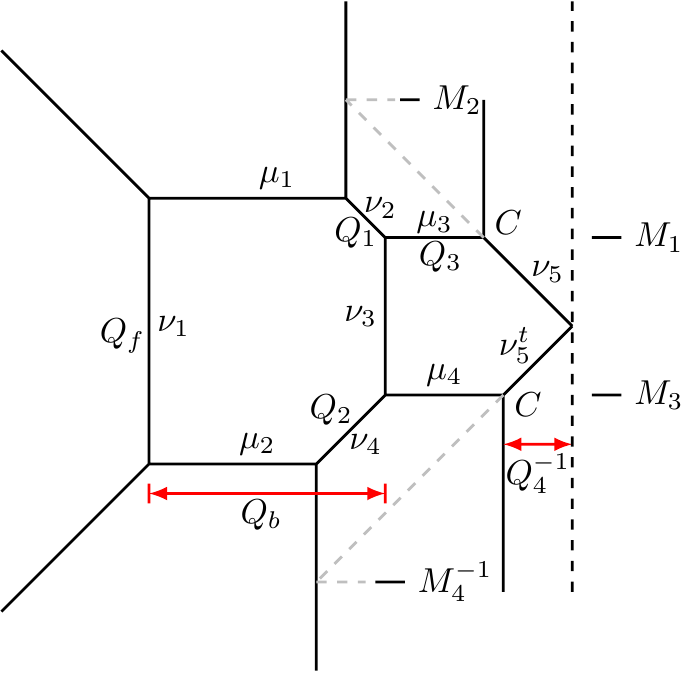}
		\caption{A 5-brane web for SU(2) with 4 flavors by flopping.}
		\label{fig:(1,1) reflection}
	\end{figure} 
\begin{align}
	{Z'}^\text{SU(2)+4{\bf F}}&=\sum_{\boldsymbol{\mu,\nu}}C_{\nu_1^t\text{\o}\mu_1^t}(\fq,\ft)C_{\text{\o}\nu_1\mu_2^t}(\fq,\ft)C_{\text{\o}\nu_2^t\mu_1}(\ft,\fq)C_{\nu_3^t\nu_2\mu_3^t}(\fq,\ft)C_{\nu_4^t\nu_3\mu_4^t}(\fq,\ft)\cr 
	&\times C_{\nu_4\text{\o}\mu_2}(\ft,\fq)C_{\text{\o}\nu_5^t\mu_3}(\ft,\fq)C_{\nu_5^t\text{\o}\mu_4}(\ft,\fq)(-Q_f)^{|\nu_1|}\tilde{f}_{\nu_1}(\fq,\ft)(-Q_1)^{|\nu_2|}\cr 
	&\times(-\frac{Q_f}{Q_{1,2}})^{|\nu_3|}\tilde{f}_{\nu_3}(\fq,\ft)(-Q_2)^{|\nu_4|}(\frac{Q_f}{Q_{1,2}}\sqrt{\frac{\fq}{\ft}})^{|\nu_5|} f_{\nu_5^t}(\ft,\fq)(-\frac{Q_b}{Q_1})^{|\mu_1|}\cr
	&
	\times (-\frac{Q_b}{Q_2})^{|\mu_2|}(-Q_3)^{|\mu_3|}(-\frac{Q_{f,3,4^2}}{Q_{1,2}})^{|\mu_4|}\ .
\end{align}	
After summing over $\boldsymbol{\nu}$, we obtain
	\begin{align}
		{Z'}^{SU(2)+4{\bf F}} =&\sum_{\boldsymbol{\mu}}\left(-\frac{Q_b}{Q_1}\right)^{|\mu_1|}\left(-\frac{Q_b}{Q_2}\right)^{|\mu_2|}(-Q_3)^{|\mu_3|}\left(-\frac{Q_{f,3,4^2}}{Q_{1,2}}\right)^{|\mu_4|}\nonumber\\
		\times&
		\fq^{\frac12\sum_{i=1}^4||\mu_i||^2}\ft^{\frac12\sum_{i=1}^4||\mu_i^t||^2}\  \prod_{i=1}^4\tilde{Z}_{\mu_i}(\ft,\fq)\tilde{Z}_{\mu_i^t}(\fq,\ft)\ \nonumber\\
		\times&\frac{\calR_{\mu_1^t\mu_3}(Q_1)\calR_{\mu_1^t\mu_4}(\frac{Q_f}{Q_2})\calR_{\mu_3^t\mu_2}(\frac{Q_f}{Q_1})\calR_{\mu_4^t\mu_2}(Q_2)}{\calR_{\mu_1^t\mu_2}(Q_f\sqrt{\frac{\fq}{\ft}})\calR_{\mu_1^t\mu_2}(Q_f\sqrt{\frac{\ft}{\fq}})}\ ,
	\end{align}
	which is exactly the same as \eqref{eq:su2+4F simplified}. This confirms that our proposal for the topological vertex with ON-planes also works well for the flopped diagrams.
	\bigskip


\subsection{5d SU(2)+8{\bf F}:
		E-string theory as 5d affine \texorpdfstring{$D_4$}{D4} quiver}\label{sec:D4D4}
\begin{figure}
		\centering
		\includegraphics[scale=1]{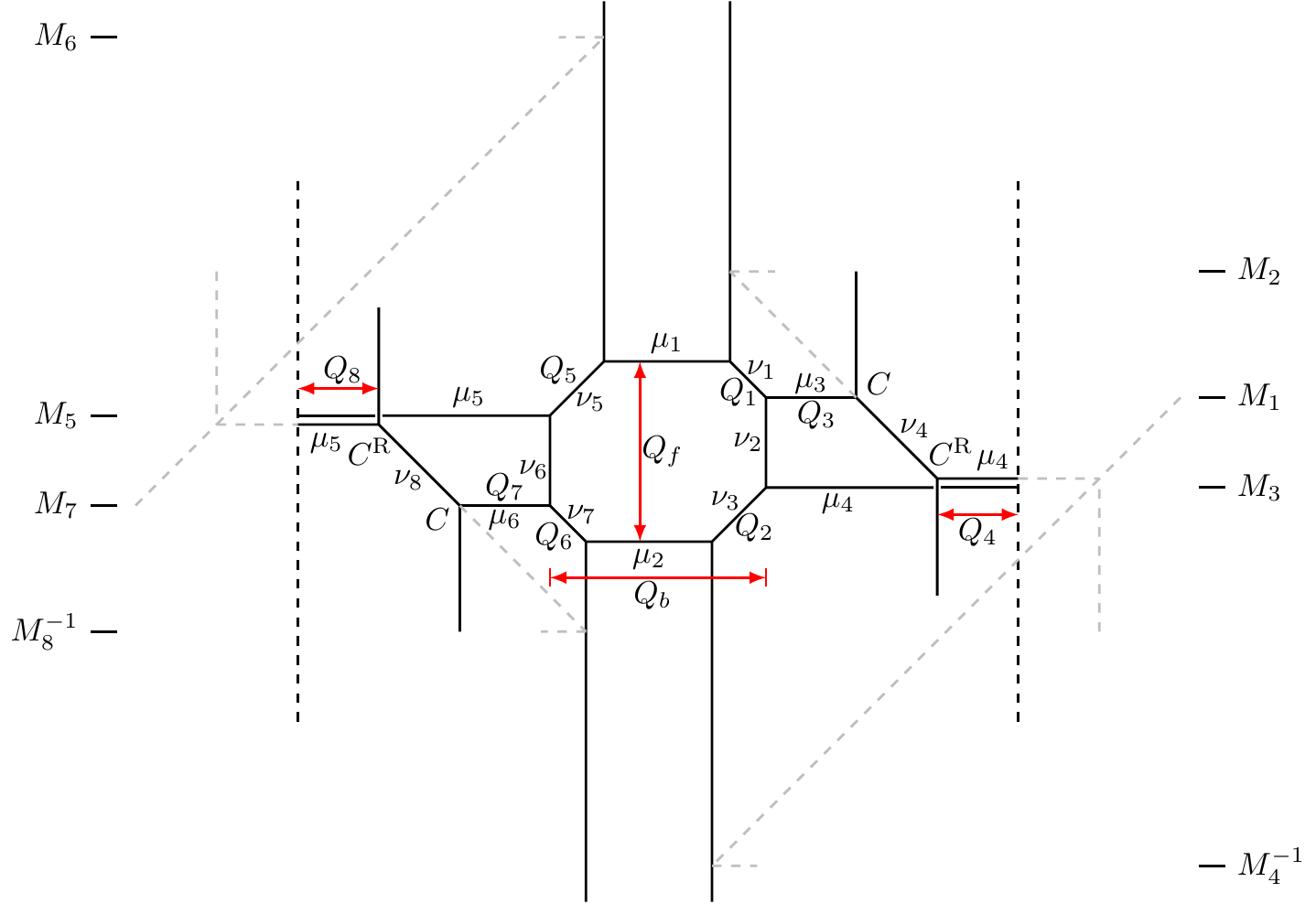}
		\caption{A 5-brane web for E-string theory on a circle as affine $D_4$-quiver. 
					}
		\label{fig:UV E-string theory}
	\end{figure}

In this subsection, we apply our proposal to a 5-brane system with two ON-planes. As shown in the previous subsection, SU(2)+4{\bf F} can be realized with an ON-plane. Now we add four more flavors to the 5-brane configuration which yields a 5-brane web for SU(2)+8{\bf F} whose UV completion 
is realized as the E-string theory. There are several different ways of representing the E-string theory on a circle using Type IIB 5-brane web. As SU$(2)+8\mathbf{F}$, the corresponding 5-brane web is given by a Tao web diagram which does not involve any orientifolds \cite{Kim:2015jba}. As $Sp(1)+8\mathbf{F}$, one can represent it with two O7$^-$-planes~\cite{Hayashi:2015fsa} or with two O5-planes~\cite{Kim:2017jqn}. In particular, as an affine $D_4$-quiver
	\begin{align}
		\begin{tikzpicture}
			\draw[thick](-1.4,.8)--(-0.6,0.2);
			\draw[thick](-1.4,-.8)--(-0.6,-0.2);
			\draw[thick](0.6,0.2)--(1.4,.8);
			\draw[thick](0.6,-0.2)--(1.4,-.8); 
			\node at (0,0){SU$(2)$};
			\node at (-2,1){SU$(1)$};
			\node at (-2,-1){SU$(1)$};
			\node at (2,1){SU$(1)$};
			\node at (2,-1){SU$(1)$};
		\end{tikzpicture}\qquad,
	\end{align}
one can use a 5-brane with two ON-planes as depicted in Figure~\ref{fig:UV E-string theory}. We use this  5-brane web to compute the partition function by our proposal of topological vertex with ON-planes.  As its partition function is non-trivial, this case would be another good example for testing our proposal.

As one can see, Figure \ref{fig:UV E-string theory} is similar to the SU$(2)+4\mathbf{F}$ web diagrams discussed in the previous section, the left and right hand sides of Figure \ref{fig:UV E-string theory} are two copies of the right hand side of the diagram in Figure \ref{fig:D5 reflection}(a). The eight mass parameters $M_i=e^{-\beta m_i},i=1,\cdots,8$ can be accordingly labelled in the 5-brane web diagram, where two $C^{\text{R}}$ are introduced to the vertex nearest to each ON-plane.  
By using our proposal with ON-planes, we write the topological string partition function for SU$(2)$+8{\bf F} as
	\begin{align}
		Z^\text{SU(2)+8{\bf F}}\! &=\sum_{\boldsymbol{\mu,\nu}}C_{\text{\o}\nu_1^t\mu_1}(\ft,\fq)C_{\nu_2^t\nu_1\mu_3^t}(\fq,\ft)C_{\nu_3^t\nu_2\mu_4^t}(\fq,\ft)C_{\nu_3\text{\o}\mu_2}(\ft,\fq)C_{\text{\o}\nu_4^t\mu_3}(\ft,\fq)\cr
		&\times C^{\text{R}}_{\text{\o}\nu_4\mu_4^t}(\fq,\ft) C_{\nu_5^t\text{\o}\mu_1^t}(\fq,\ft)C_{\nu_5\nu_6^t\mu_5}(\ft,\fq)C_{\nu_6\nu_7^t\mu_6}(\ft,\fq)C_{\text{\o}\nu_7\mu_2^t}(\fq,\ft)\crcr
		&\times C^{\text{R}}_{\text{\o}\nu_8^t\mu_5}(\ft,\fq)C_{\text{\o}\nu_8\mu_6^t}(\fq,\ft)(-Q_1)^{|\nu_1|} (-\frac{Q_f}{Q_{1,2}})^{|\nu_2|}\tilde{f}_{\nu_2}(\fq,\ft)
		(-Q_2)^{|\nu_3|}\nonumber\\
		&
		\times (-\frac{Q_f}{Q_{1,2}}\sqrt{\frac{\fq}{\ft}})^{|\nu_4|}(-Q_5)^{|\nu_5|}(-\frac{Q_f}{Q_{5,6}})^{|\nu_6|}\tilde{f}_{\nu_6}^{-1}(\fq,\ft) (-Q_6)^{|\nu_7|}(-\frac{Q_f}{Q_{5,6}}\sqrt{\frac{\ft}{\fq}})^{|\nu_8|}\nonumber\\
		&
		\times(-\frac{Q_b}{Q_{1,5}})^{|\mu_1|}
		f_{\mu_1}(\ft,\fq)(-\frac{Q_b}{Q_{2,6}})^{|\mu_2|}f_{\mu_2}(\ft,\fq)^{-1}(-Q_3)^{|\mu_3|}(+\frac{Q_{3,f,4^2}}{Q_{1,2}})^{|\mu_4|}f_{\mu_4^t}(\fq,\ft)\cr
		&
		\times \frac{\tilde{Z}_{\mu_4}(\ft,\fq)}{\tilde{Z}_{\mu_4^t}(\fq,\ft)}
		(+\frac{Q_{8^2,f,7}}{Q_{5,6}})^{|\mu_5|}f_{\mu_5}(\ft,\fq)\frac{\tilde{Z}_{\mu_5^t}(\fq,\ft)}{\tilde{Z}_{\mu_5}(\ft,\fq)}(-Q_7)^{|\mu_6|}\ .
	\end{align}
After simplifying the sum over $\boldsymbol{\nu}$, we express the partition function as 
\begin{small}
\begin{align}
Z^\text{SU(2)+8{\bf F}}\! =&\sum_{\boldsymbol{\mu}\!}\!\frac{\calR_{\mu_1^t\mu_3\!}(\!Q_1\!)\calR_{\mu_1^t\mu_4\!}(\frac{Q_f}{Q_2}\!)\calR_{\mu_1^t\mu_5\!}(\!Q_5\!)\calR_{\mu_1^t\mu_6\!}(\frac{Q_f}{Q_6}\!)\calR_{\mu_3^t\mu_2\!}(\!\frac{Q_f}{Q_1}\!)\calR_{\mu_4\mu_2\!}(\!Q_2\!)\calR_{\mu_5^t\mu_2\!}(\!\frac{Q_f}{Q_5}\!)\calR_{\mu_6^t\mu_2\!}(\!Q_6\!)\!}{\calR_{\mu_1^t\mu_2}(Q_f\sqrt{\frac{\fq}{\ft}})\calR_{\mu_1^t\mu_2}(Q_f\sqrt{\frac{\ft}{\fq}})}\nonumber \\
&
		\times(-Q_3)^{|\mu_3|}(-Q_7)^{|\mu_6|}\left(\frac{Q_b}{Q_{1,5}}\right)^{|\mu_1|}\left(\frac{Q_b}{Q_{2,6}}\right)^{|\mu_2|}\left(-\frac{Q_{3,4^2,f}}{Q_{1,2}}\right)^{|\mu_4|}\left(-\frac{Q_{7,8^2,f}}{Q_{5,6}}\right)^{|\mu_5|}\nonumber \\
		&\times\fq^{||\mu_2||^2+\frac12\sum_{i=3}^6||\mu_i||^2}\ft^{||\mu_1^t||^2+\frac12\sum_{i=3}^6||\mu_i^t||^2}
		\prod_{i=1}^6\tilde{Z}_{\mu_i}(\ft,\fq)\tilde{Z}_{\mu_i^t}(\fq,\ft)\ . 
		\label{eq:Nfeq8}
\end{align}
\end{small}
From the web diagram in Figure \ref{fig:UV E-string theory}, the map between the K\"ahler parameters and the gauge theory parameters is given as follows:
\begin{align}
		&Q_b=A^2u\sqrt{\frac{M_3M_7}{M_1M_2M_4M_5M_6M_8}},\qquad\quad
		Q_f=A^2,\crcr
		&
		Q_1=\frac{A}{M_1},\qquad
		Q_2=AM_3,\qquad
		Q_3=\frac{M_2}{A},\qquad
		Q_4=\sqrt{\frac{M_3M_4}{M_1M_2}},
		\cr
		&
		Q_5=\frac{A}{M_5},\qquad
		Q_6=AM_7,\qquad
		Q_7=\frac{M_8}{A},\qquad
		Q_8=\sqrt{\frac{M_6M_7}{M_5M_8}}\ ,
		\label{eq:su2with8f}
	\end{align}
where $A$ is the Coulomb branch parameter and $u$ is the instanton factor. After extracting the $\calM$ factors and factorizing the remaining parts, as demonstrated in the previous subsection, we can write the partition function as 
	\begin{align}
		Z^\text{SU(2)+8{\bf F}}=Z^{N_f=8}_{\text{M}}\sum_{\mu_1\mu_2}&Z_{\mu_1\mu_2}^{N_f=8}\sum_{\mu_3}Z_{\mu_1\mu_2\mu_3}(M_1,M_2)\sum_{\mu_4}Z_{\mu_1\mu_2\mu_4}(M_3,M_4)\cr &~~\quad\times\sum_{\mu_5}Z_{\mu_1\mu_2\mu_5}(M_5,M_6)\sum_{\mu_6}Z_{\mu_1\mu_2\mu_6}(M_7,M_8)\ ,
		\label{eq:Z8}
	\end{align}
where
\begin{align}
		Z^{N_f=8}_{\text{M}}=&~\calM(A^2)\calM\big(A^2\frac{\ft}{\fq}\big)\Bigg[\calM(\frac{A}{M_1}\sqrt{\frac{\ft}{\fq}})\calM(AM_1\sqrt{\frac{\ft}{\fq}})\calM(\frac{A}{M_3}\sqrt{\frac{\ft}{\fq}})\calM\big(AM_3\sqrt{\frac{\ft}{\fq}}\big)\crcr
		&~
		\times\calM(\frac{A}{M_5}\sqrt{\frac{\ft}{\fq}})\calM(AM_5\sqrt{\frac{\ft}{\fq}})\calM(\frac{A}{M_7}\sqrt{\frac{\ft}{\fq}})\calM(AM_7\sqrt{\frac{\ft}{\fq}})\Bigg]^{-1}\ ,\\
		Z_{\mu_1\mu_2}^{N_f=8}=&~\frac{\fq^{||\mu_2||^2}\ft^{||\mu_1^t||^2}\tilde{Z}_{\mu_1}(\ft,\fq)\tilde{Z}_{\mu_1^t}(\fq,\ft)\tilde{Z}_{\mu_2}(\ft,\fq)\tilde{Z}_{\mu_2^t}(\fq,\ft)u^{|\mu_1|+|\mu_2|}}{\calN_{\mu_1\mu_2}(A^2)\calN_{\mu_1\mu_2}(A^2\frac{\ft}{\fq})}\ , 
\end{align}
and $Z_{\mu_1\mu_2\mu_i}(M_j,M_k)$ are the same as those defined in \eqref{eq:Zmu123} and \eqref{eq:Zmu124} , 
\begin{align}
    Z_{\mu_1\mu_2\mu_i}(M_j,M_k)=&~\fq^{\frac{||\mu_i||^2}{2}}\ft^{\frac{||\mu_i^t||^2}{2}}Z_{\mu_i}(\ft,\fq)Z_{\mu_i^t}(\fq,\ft)\calN_{\mu_1\mu_i}(\frac{A}{M_j}\sqrt{\frac{\ft}{\fq}})\calN_{\mu_i\mu_2}(AM_j\sqrt{\frac{\ft}{\fq}})\cr
     &\times\left(-\frac{M_k}{A}\right)^{|\mu_i|}\left(\frac{1}{\sqrt{M_jM_k}}\right)^{|\mu_2|}\left(\sqrt{\frac{M_j}{M_k}}\right)^{|\mu_1|}\ .
\end{align}

The perturbative part of the partition function then takes the form
	\begin{align}
		Z^\text{SU(2)+8{\bf F}}_{\text{pert}'}
		=&~Z^{N_f=8}_{\text{M}}\sum_{\mu_3}Z_{\text{\o}\text{\o}\mu_3}(M_1,M_2)\sum_{\mu_4}Z_{\text{\o}\text{\o}\mu_4}(M_3,M_4)\crcr
		&\times \sum_{\mu_5}Z_{\text{\o}\text{\o}\mu_5}(M_5,M_6)\sum_{\mu_6}Z_{\text{\o}\text{\o}\mu_6}(M_7,M_8)\ .
	\end{align}
Plugging in \eqref{eq:Z00mu3}, dropping the extra factors and flopping necessary factors, we obtain the perturbative part in terms of the gauge theory parameters: 
	\begin{align}
		Z^\text{SU(2)+8{\bf F}}_{\text{pert}}=\frac{\calM\big(A^2\big)\calM \big(A^2\frac{\ft}{\fq}\big)}{\prod_{i=1}^8\calM\Big(\frac{A}{M_i}\sqrt{\frac{\ft}{\fq}}\Big)\calM\Big(A M_i\sqrt{\frac{\ft}{\fq}}\, \Big)}\ . 
	\end{align}
	Expressed as the Plethystic exponential, it is given by
	\begin{align}
		Z^\text{SU(2)+8{\bf F}}_{\text{pert}} &=\text{PE}\Big[\frac{1}{(1-\ft)(1-\fq)}\left((\fq+\ft)A^2-\chi_{\bf 16}\sqrt{\fq\ft}A\right)\Big]\ ,
	\end{align}
	where we denote by $\chi_{\text{\bf n}}$ the characters of the $\bf n$-dimensional irreducible representation of SO$(16)$. See also Appendix \ref{sec:App} for their explicit forms.
	
	By extracting the perturbative part from \eqref{eq:Z8}, we find the instanton part has the following structure:
	\begin{multline}
		Z^\text{SU(2)+8{\bf F}}_{\text{inst+extra}}=\sum_{\mu_1,\mu_2}\Bigg(Z_{\mu_1\mu_2}^{N_f=8}\frac{\sum_{\mu_3}Z_{\mu_1\mu_2\mu_3}(M_1,M_2)}{\sum_{\mu_3}Z_{\text{\o}\text{\o}\mu_3}(M_1,M_2)}\frac{\sum_{\mu_4}Z_{\mu_1\mu_2\mu_4}(M_3,M_4)}{\sum_{\mu_4}Z_{\text{\o}\text{\o}\mu_4}(M_3,M_4)}\\\times\frac{\sum_{\mu_5}Z_{\mu_1\mu_2\mu_5}(M_5,M_6)}{\sum_{\mu_5}Z_{\text{\o}\text{\o}\mu_5}(M_5,M_6)}\frac{\sum_{\mu_6}Z_{\mu_1\mu_2\mu_6}(M_7,M_8)}{\sum_{\mu_6}Z_{\text{\o}\text{\o}\mu_6}(M_7,M_8)}\Bigg)\ ,
		\label{eq:inst_extra}
	\end{multline}
where extra factors are included. From Figure \ref{fig:UV E-string theory} we can see that there are four upper and four lower parallel external branes which are also parallel to the two ON-planes, so there will be infinitely many different strings connecting these external parallel branes. They contribute to the extra factors in \eqref{eq:inst_extra} if the corresponding distances between the branes depend on the instanton factor. With the definition in \eqref{eq:defZmu12}, \eqref{eq:inst_extra} is rewritten as
	\begin{equation}
		Z^{\text{SU(2)+8{\bf F}}}_{\text{inst+extra}}
		=\sum_{\mu_1,\mu_2}Z_{\mu_1\mu_2}^{N_f=8}\calZ_{\mu_1\mu_2}(M_1,M_2)\calZ_{\mu_1\mu_2}(M_3,M_4)\calZ_{\mu_1\mu_2}(M_5,M_6)\calZ_{\mu_1\mu_2}(M_7,M_8)\ .
	\end{equation}
	By plugging in the expressions of $\calZ_{\mu_1\mu_2}$'s, for up to two-instanton, \eqref{eq:ZZ1} and \eqref{eq:ZZ2}, we can obtain the instanton partition function which includes the extra factors.

As the Omega deformation parameters capture the left-right spin content $[j_l,j_r]$ of the theory under SU$(2)_l \times SU(2)_r$,
which is \cite{Gopakumar:1998ii,Gopakumar:1998jq, Iqbal:2007ii} defined as
\begin{align}
		[j_{l}, j_r]:=\frac{(-1)^{2j_l +2j_r +1} \Big( (\ft\fq)^{-j_l}+\cdots+ (\ft\fq)^{j_l}\Big)\Big((\frac{\ft}{\fq})^{-j_r} +\cdots  +(\frac{\ft}{\fq})^{j_r}\Big)}{(\ft^{1/2}- \ft^{-1/2})(\fq^{1/2}-\fq^{-1/2})}\ , \label{eq:GVspin}
\end{align}
we express the partition function as the GV invariant with $[j_{l}, j_r]$ and the characters $\chi$ for the flavor symmetry. 
The perturbative part is then re-expressed as
	\begin{align}
		Z^\text{SU(2)+8{\bf F}}_{\text{pert}}
		&=\text{PE}\Big[\,[0,\tfrac12]\, A^2+ \chi_{\bf 16}\,[0,0]A\,\Big]\ . 
	\end{align}
The instanton partition function which includes the extra factors can be written as 
	\begin{equation}
		Z^\text{SU(2)+8{\bf F}}_{\text{inst+extra}}=\text{PE}\Big[\, \sum_{n=1}^{\infty}d_n(A,M_i,\ft,\fq)u^n\,\Big]\ ,
	\end{equation}
	with
	\begin{equation}
		d_n(A,M_i,\ft,\fq)=\sum_{m=0}^{\infty}d_{n,m}(M_i,\ft,\fq)A^m\ ,
	\end{equation}
	where the mass $M_i$ terms are organized as the characters $\chi_{\text{\bf n}}$ of SO(16), whose explicit forms are given in Appendix \ref{sec:App}. 
	As the extra factor is the part that does not depend on the Coulomb branch parameter $A$, the extra factor is expressed as $Z_{\text{extra}}= {\rm PE} \big[ \sum_{n=1}^{\infty}d_{n ,0}u^n\big]$. Discarding this extra factor, we find the instanton part is given by
\begin{align}\label{eq:Estring-instanton-refind}
		Z^{\text{SU(2)+8$\mathbf{F}$}}_{\text{inst}}=\text{PE}\Big[\, \sum_{n,m=1}^{\infty} d_{n,m} A^m u^n\,\Big]\ 	,
	\end{align}
	where, 
	up to $u^3$ and $A^2$, the explicit forms of $d_{n,m}$ with $n\le 3,m\le 2$ are given as follows:  
	\begin{align}\label{eq:fmn for E-string}
		d_{1,1} &= \chi_{\overline{\bf 128}}\,[0,0]\ ,  \\
		d_{1,2} &= \chi_{\bf 128}\,[0,\tfrac12]\ ,\crcr
		d_{2,1}& =\chi_{\bf 16}\,[\tfrac12,\tfrac12]+\big(\chi_{\bf 560}\,+\chi_{\bf 16}\,\big)[0,0]\ ,
		\crcr 
		d_{2,2}&=[1,\tfrac32]+\big(\chi_{\bf 120}\,+1\big)[\tfrac12,1]+[\tfrac12,0]+\big(\chi_{\bf 1820}\,+\chi_{\bf 120}\,+2\big)[0,\tfrac12]\ , \quad
		\crcr
		d_{3,1}& =\chi_{\overline{\bf 128}}\,[\tfrac12,\tfrac12]+\big(\chi_{\overline{\bf 1920}}+\chi_{\overline{\bf 128}}\,\big)[0,0]\ ,
		\crcr
		d_{3,2}&=\chi_{\bf 128} [1,\tfrac32]+\!\big(\chi_{\bf 1920} +\!2 \chi_{\bf 128}\big) [\tfrac12, 
		1] + \!\chi_{\bf 128}  [\tfrac12, 0] +\! \big(\chi_{\bf 13312}\! +\! \chi_{\bf1920}\! + \! 3 \chi_{\bf128}\big) [0, \tfrac12]\ . \nonumber
	\end{align}
The result completely agrees with the known refined partition function \cite{Kim:2014dza, Kim:2015jba, Kim:2020hhh}. We note that there has been several partition function computations for SU(2)+8{\bf F} based the topological vertex \cite{Kim:2015jba, Kim:2017jqn,Kim:2020npz}, but all these attempts are computed in the unrefined limit $\ft=\fq$. The result \eqref{eq:Estring-instanton-refind} is the first example that correctly reproduces the refined result based on topological vertex formalism.

\bigskip
\subsection{5d SU(3) theory at CS level 7}\label{sec:SU(3)CS7}
In this subsection, we consider yet another non-trivial theory of rank-2.   It was proposed in \cite{Hayashi:2018lyv} that pure SU(3) theories at various Chern-Simons (CS) level $\kappa$ can have 5-brane webs with ON-plane(s). Such 5-brane webs can be understood as a Higgsing from D-type quivers. We first consider the example of $\kappa=7$, denoted as SU(3)$_7$. It is obtained by Higgsing the $D_4$-type quiver
\begin{align}\label{eq:quiverSU3_7}
	\begin{tikzpicture}
		\draw[thick](-1.4,0)--(-0.7,0);
		\draw[thick](0.7,0.2)--(1.4,.8);
		\draw[thick](0.7,-0.2)--(1.4,-.8); 
		\node at (0,0){SU$(3)_1$};
		\node at (-2,0){SU$(2)$};
		\node at (2,1){SU$(2)$};
		\node at (2,-1){SU$(2)$};
	\end{tikzpicture}		
\end{align}
\begin{figure}[htbp]
	\centering
	\includegraphics[scale=1]{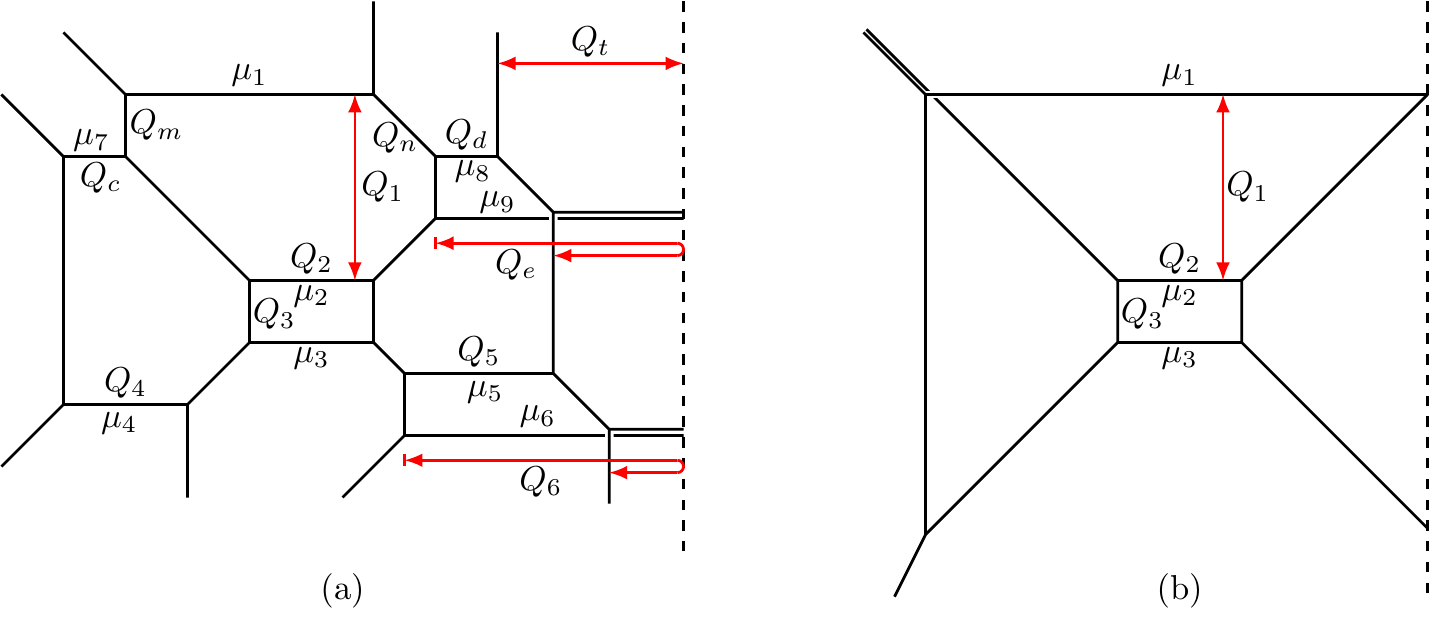}
	\caption{(a) The $D_4$-type quiver theory. (b) SU(3) theory at CS level 7.}
	\label{fig:su3_cs7_al}
\end{figure}where the middle node is an SU(3) theory at the CS level 1, the corresponding web diagram is illustrated in Figure \ref{fig:su3_cs7_al}(a). Here, the Higgsing procedure is as follows: A Higgsing of an SU(2) gives rise to an antisymmetric hypermultiplet ($\mathbf{AS}$) to SU(3), increasing the CS level of SU(3) by $\frac32$~\cite{Hayashi:2018lyv}. As we have three SU(2), we get SU(3)$_{\frac{11}{2}}+3 \mathbf{AS}$ after the Higgsing, in Figure \ref{fig:su3_cs7_al}(a) it corresponds to shrink $Q_m,Q_c,Q_n,Q_d,Q_t,Q_e$. Finally, since an antisymmetric hypermultiplet transforms as $\bar{\mathbf{3}}$, the decoupling of an antisymmetric hypermultiplet further increases the CS level of SU(3) by $\frac12$. Hence, decoupling all the $\mathbf{AS}$ which means flopping $Q_4,Q_5,Q_6$ downward to infinity yields SU(3)$_7$ whose web diagram is illustrated in Figure \ref{fig:su3_cs7_al}(b). 
\begin{figure}[htbp]
	\centering
	\includegraphics[scale=1]{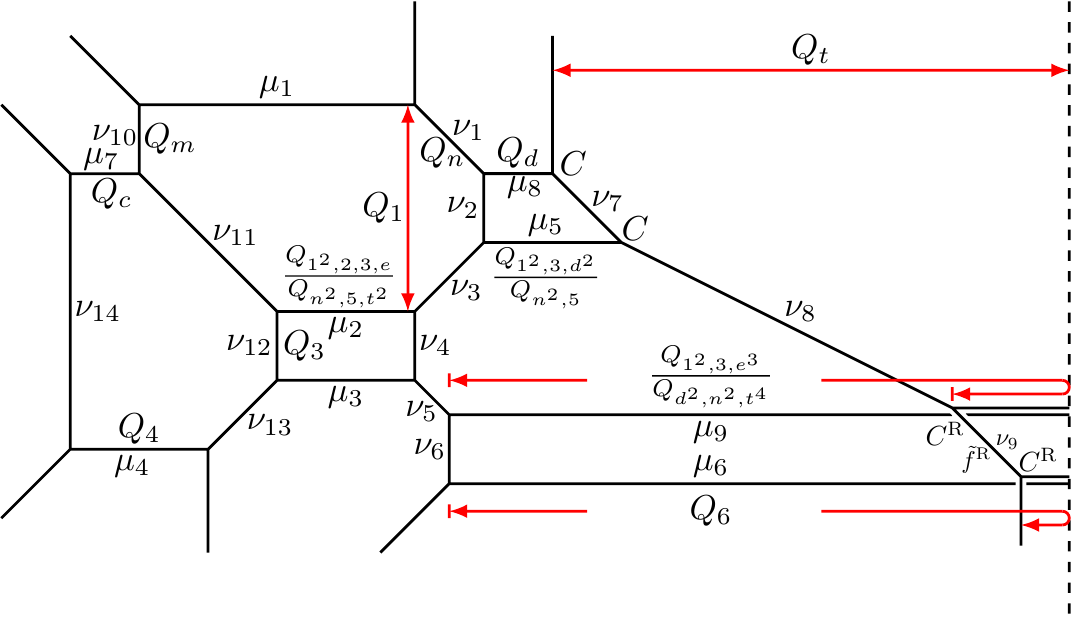}
	\caption{The $D_4$-type quiver theory with separated SU(2) sub webs on the right.}
	\label{fig:SU(3)_7}
\end{figure}

However, if we start from Figure \ref{fig:su3_cs7_al}(a) to compute the topological string partition function, the two SU(2) sub webs on the right side are overlapping. In order to use our ON proposal, we have to swap the positions of the brane $\mu_9$ and $\mu_5$ in the web diagram to make the sub webs of the two SU(2) nodes separated which corresponds to Figure \ref{fig:SU(3)_7}. Due to this swapping, the K\"ahler parameters in Figure \ref{fig:su3_cs7_al}(a) change into the K\"ahler parameters in Figure \ref{fig:SU(3)_7}. Then, by our proposal of topological vertex with ON-planes, the input data of topological vertex for Figure \ref{fig:SU(3)_7} is:
\begin{small}
	\begin{align}
		{Z'}^{D_4}=&~\sum_{\boldsymbol{\mu,\nu}}C_{\text{\o}\nu_1^t\mu_1}(\ft,\fq)C_{\nu_2^t\nu_1\mu_8^t}(\fq,\ft)C_{\nu_3^t\nu_2\mu_5^t}(\fq,\ft)C_{\nu_3\nu_4^t\mu_2}(\ft,\fq)C_{\nu_4\nu_5^t\mu_3}(\ft,\fq)C_{\nu_6^t\nu_5\mu_9^t}(\fq,\ft)C_{\text{\o}\nu_6\mu_6^t}(\fq,\ft)\nonumber\\
		\times& C_{\text{\o}\nu_7^t\mu_8}(\ft,\fq)C_{\nu_7\nu_8^t\mu_5}(\ft,\fq)C^{\text{R}}_{\nu_9^t\nu_8\mu_9^t}(\fq,\ft)C^{\text{R}}_{\text{\o}\nu_9\mu_6^t}(\fq,\ft)C_{\nu_{10}\text{\o}\mu_1^t}(\fq,\ft)C_{\nu_{10}\nu_{11}^t\mu_7}(\ft,\fq)\nonumber\\
		\times& C_{\nu_{12}^t\nu_{11}\mu_2^t}(\fq,\ft)C_{\nu_{13}^t\nu_{12}\mu_3^t}(\fq,\ft)C_{\nu_{13}\text{\o}\mu_4}(\ft,\fq)C_{\nu_{14}\text{\o}\mu_7^t}(\fq,\ft)C_{\text{\o}\nu_{14}\mu_4^t}(\fq,\ft)(-Q_n)^{|\nu_1|}\nonumber\\
		\times&(-\frac{Q_{1^2,3,d}}{Q_{n^2,5}})^{|\nu_2|} \tilde{f}_{\nu_2}(\fq,\ft)(-\frac{Q_{n,5}}{Q_{1,3,d}})^{|\nu_3|}(-Q_3)^{|\nu_4|}\tilde{f}_{\nu_4}(\fq,\ft)^{-1}(-\frac{Q_{d,n,t^2}}{Q_{e,3,1}})^{|\nu_5|}(-\frac{Q_{1^2,3,e^3}}{Q_{d^2,n^2,t^4,6}})^{|\nu_6|}\nonumber\\
		\times&\tilde{f}_{\nu_6}(\fq,\ft)(-\frac{Q_{1^2,3,d}}{Q_{n^2,5}})^{|\nu_7|}\tilde{f}_{\nu_7}(\fq,\ft)^{-1}(-\frac{Q_{n^2,t^2,5}}{Q_{1^2,3,e}}\sqrt{\frac{\fq}{\ft}})^{|\nu_8|}(-\frac{Q_{1^2,3,e^3}}{Q_{d^2,n^2,t^4,6}})^{|\nu_9|}\tilde{f}^{\text{R}}_{\nu_9}(\fq,\ft)\nonumber\\
		\times&(-Q_m)^{|\nu_{10}|}(-\frac{Q_1}{Q_m})^{\nu_{11}} (-Q_3)^{|\nu_{12}|}\tilde{f}_{\nu_{12}}(\fq,\ft)(-\frac{Q_{c,1}}{Q_{m,4}})^{|\nu_{13}|}(-\frac{Q_{c,3,1^2}}{Q_{m^2,4}})^{|\nu_{14}|}\tilde{f}_{\nu_{14}}(\fq,\ft)\nonumber\\
		\times&(-\frac{Q_{1^2,2,e}}{Q_{n^2,t^2,m,d}})^{|\mu_1|}(-\frac{Q_{1^2,2,3,e}}{Q_{n^2,5,t^2}})^{|\mu_2|} f_{\mu_2}(\ft,\fq)^{-1}(-\frac{Q_{1^2,2,3,e}}{Q_{n^2,5,t^2}})^{|\mu_3|}f_{\mu_3}(\ft,\fq)(-Q_d)^{|\mu_8|}\nonumber\\
		\times&(-\frac{Q_{1^2,3,d^2}}{Q_{n^2,5}})^{|\mu_5|}f_{\mu_5}(\ft,\fq)^2(\frac{Q_{1^2,3,e^3}}{Q_{d^2,n^2,t^4}})^{|\mu_9|}f_{\mu_9}(\ft,\fq)^{-2} f_{\mu_9^t}(\fq,\ft)\frac{\tilde{Z}_{\mu_9}(\ft,\fq)}{\tilde{Z}_{\mu_9^t}(\fq,\ft)}Q_6^{|\mu_6|}\nonumber\\
		\times&f_{\mu_6^t}(\fq,\ft)\frac{\tilde{Z}_{\mu_6}(\ft,\fq)}{\tilde{Z}_{\mu_6^t}(\fq,\ft)}(-Q_c)^{|\mu_7|}(-Q_4)^{|\mu_4|}\ .
	\end{align}
\end{small}
After summing over the non-preferred direction Young diagrams, the result is
\begin{equation}
	{Z'}^{D_4}=\sum_{\boldsymbol{\mu}}{Z'}^{D_4}_{\mu_1\mu_2\mu_3}Z^{D_4}_{\mu_1\mu_2\mu_3\mu_7\mu_4}{Z'}^{D_4}_{\mu_1\mu_2\mu_3\mu_8\mu_5}{Z'}^{D_4}_{\mu_1\mu_2\mu_3\mu_9\mu_6}
	\label{eq:Z'su3cs7}
\end{equation}
where
\begin{align}
	{Z'}^{D_4}_{\mu_1\mu_2\mu_3}=&~W_{\mu_1}(\ft,\fq)(-\frac{Q_{1^2,2,e}}{Q_{n^2,t^2,m,d}})^{|\mu_1|}W_{\mu_2}(\ft,\fq)(-\frac{Q_{1^2,2,3,e}}{Q_{n^2,5,t^2}})^{|\mu_2|}f_{\mu_2}(\ft,\fq)^{-1}W_{\mu_3}(\ft,\fq)\nonumber\\
	\times&(-\frac{Q_{1^2,2,3,e}}{Q_{n^2,5,t^2}})^{|\mu_3|}f_{\mu_3}(\ft,\fq)\times\bigg[\calR_{\mu_1^t\mu_2}(Q_1\sqrt{\frac{\fq}{\ft}})\calR_{\mu_1^t\mu_2}(Q_1\sqrt{\frac{\ft}{\fq}})\calR_{\mu_1^t\mu_3}(Q_{1,3}\sqrt{\frac{\fq}{\ft}})\nonumber\\
	\times&\calR_{\mu_1^t\mu_3}(Q_{1,3}\sqrt{\frac{\ft}{\fq}})\calR_{\mu_2^t\mu_3}(Q_3\sqrt{\frac{\fq}{\ft}})\calR_{\mu_2^t\mu_3}(Q_3\sqrt{\frac{\ft}{\fq}})\bigg]^{-1}\ ,\label{eq:Z'123}\\
	Z^{D_4}_{\mu_1\mu_2\mu_3\mu_7\mu_4}=&~W_{\mu_7}(\ft,\fq)(-Q_c)^{|\mu_7|}W_{\mu_4}(\ft,\fq)(-Q_4)^{|\mu_4|}\calR_{\mu_1^t\mu_7}(Q_m)\calR_{\mu_7^t\mu_2}(\frac{Q_1}{Q_m})\nonumber\\
	\times&\calR_{\mu_7^t\mu_3}(\frac{Q_{1,3}}{Q_m})\calR_{\mu_1^t\mu_4}(\frac{Q_{1^2,3,c}}{Q_{m,4}})\calR_{\mu_2^t\mu_4}(\frac{Q_{1,3,c}}{Q_{m,4}})\calR_{\mu_3^t\mu_4}(\frac{Q_{1,c}}{Q_{m,4}})\nonumber\\
	\times&\bigg[\calR_{\mu_7^t\mu_4}(\frac{Q_{c,3,1^2}}{Q_{m^2,4}}\sqrt{\frac{\fq}{\ft}})\calR_{\mu_7^t\mu_4}(\frac{Q_{c,3,1^2}}{Q_{m^2,4}}\sqrt{\frac{\ft}{\fq}})\bigg]^{-1}\ ,\label{eq:Z'74}\\
	{Z'}^{D_4}_{\mu_1\mu_2\mu_3\mu_8\mu_5}=&~W_{\mu_8}(\ft,\fq)(-Q_d)^{|\mu_8|}W_{\mu_5}(\ft,\fq)(-\frac{Q_{1^2,3,d^2}}{Q_{n^2,5}})^{|\mu_5|}f_{\mu_5}(\ft,\fq)^2\calR_{\mu_1^t\mu_8}(Q_n)\nonumber\\
	\times&\calR_{\mu_8^t\mu_2}(\frac{Q_1}{Q_n})\calR_{\mu_8^t\mu_3}(\frac{Q_{1,3}}{Q_n})\calR_{\mu_1^t\mu_5}(\frac{Q_{1^2,3,d}}{Q_{n,5}})\calR_{\mu_5^t\mu_2}(\frac{Q_{n,5}}{Q_{1,3,d}})\calR_{\mu_5^t\mu_3}(\frac{Q_{n,5}}{Q_{1,d}})\nonumber\\
	\times&\bigg[\calR_{\mu_8^t\mu_5}(\frac{Q_{1^2,3,d}}{Q_{n^2,5}}\sqrt{\frac{\fq}{\ft}})\calR_{\mu_8^t\mu_5}(\frac{Q_{1^2,3,d}}{Q_{n^2,5}}\sqrt{\frac{\ft}{\fq}})\bigg]^{-1}\ ,\label{eq:Z'85}\\
	{Z'}^{D_4}_{\mu_1\mu_2\mu_3\mu_9\mu_6}=&~W_{\mu_9}(\ft,\fq)(-\frac{Q_{1^2,3,e^3}}{Q_{d^2,n^2,t^4}})^{|\mu_9|}f_{\mu_9}(\ft,\fq)^{-2}W_{\mu_6}(\ft,\fq)(-Q_6)^{|\mu_6|}\calR_{\mu_1^t\mu_9}(\frac{Q_{d,n,t^2}}{Q_e})\nonumber\\
	\times&\calR_{\mu_2^t\mu_9}(\frac{Q_{d,n,t^2}}{Q_{1,e}})\calR_{\mu_3^t\mu_9}(\frac{Q_{d,n,t^2}}{Q_{1,3,e}})\calR_{\mu_1^t\mu_6}(\frac{Q_{1^2,3,e^2}}{Q_{d,n,t^2,6}})\calR_{\mu_2^t\mu_6}(\frac{Q_{1,3,e^2}}{Q_{d,n,t^2,6}})\nonumber\\
	\times&\calR_{\mu_3^t\mu_6}(\frac{Q_{1,e^2}}{Q_{d,n,t^2,6}})\bigg[\calR_{\mu_9^t\mu_6}(\frac{Q_{1^2,3,e^3}}{Q_{d^2,n^2,t^4,6}}\sqrt{\frac{\fq}{\ft}})\calR_{\mu_9^t\mu_6}(\frac{Q_{1^2,3,e^3}}{Q_{d^2,n^2,t^4,6}}\sqrt{\frac{\ft}{\fq}})\bigg]^{-1}\ .\label{eq:Z'96}
\end{align}
We have defined 
\begin{equation}
	W_{\mu}(\ft,\fq)\equiv\ft^{\frac{||\mu^t||^2}{2}}\fq^{\frac{||\mu||^2}{2}}\tilde{Z}_{\mu^t}(\fq,\ft)\tilde{Z}_{\mu}(\ft,\fq)
\end{equation}
for convenience, this factor is universal for every color brane. 

However, in order to do the Higgsing and decoupling, it is much more convenient to use Figure \ref{fig:su3_cs7_al}(a). So, we transform the result obtained from Figure \ref{fig:SU(3)_7} to the one that corresponds to Figure \ref{fig:su3_cs7_al}(a) by using the following formula \cite{Mitev:2014jza}:
\begin{align}
	\calR_{\mu^t\nu}(Q^{-1};\ft,\fq)\rightarrow(Q^{-1})^{|\mu|+|\nu|}f_{\mu}(\ft,\fq)^{-1}f_{\nu}(\ft,\fq)\calR_{\nu^t\mu}(Q;\ft,\fq)\ ,\label{eq:Rexchange}
\end{align}
which can be derived from 
\begin{align}
    \calN_{\mu\nu}(\sqrt{\frac{\ft}{\fq}}Q^{-1};\ft,\fq)&=(Q^{-1})^{|\mu|+|\nu|}f_{\mu}(\ft,\fq)^{-1}f_{\nu}(\ft,\fq)\calN_{\nu\mu}(\sqrt{\frac{\ft}{\fq}}Q;\ft,\fq)\ ,\\
	\calM(\sqrt{\frac{\ft}{\fq}}Q^{-1};\ft,\fq)&\rightarrow\calM(\sqrt{\frac{\ft}{\fq}}Q;\ft,\fq)\ ,
\end{align}
where $\rightarrow$ means ``equal up to a flop''. We use \eqref{eq:Rexchange} to exchange the order of indices of $\calR_{\mu_5^t\mu_2}$,$\calR_{\mu_5^t\mu_3}$,$\calR_{\mu_2^t\mu_9}$ and $\calR_{\mu_3^t\mu_9}$ which are affected by the swapping of color branes from Figure \ref{fig:SU(3)_7} to Figure \ref{fig:su3_cs7_al}(a). This procedure also generates $Q$ factors and framing factors which will be rearranged into the $Z$'s of \eqref{eq:Z'su3cs7}. We then have
\begin{equation}
	Z^{D_4}=\sum_{\boldsymbol{\mu}}Z^{D_4}_{\mu_1\mu_2\mu_3}Z^{D_4}_{\mu_1\mu_2\mu_3\mu_7\mu_4}Z^{D_4}_{\mu_1\mu_2\mu_3\mu_8\mu_5}Z^{D_4}_{\mu_1\mu_2\mu_3\mu_9\mu_6}
	\ , \label{eq:Zsu3cs7} 
\end{equation}
where
\begin{align}
	Z^{D_4}_{\mu_1\mu_2\mu_3}=&~W_{\mu_1}(\ft,\fq)(-\frac{Q_{1^2,2,e}}{Q_{n^2,t^2,m,d}})^{|\mu_1|}W_{\mu_2}(\ft,\fq)(-Q_2)^{|\mu_2|}f_{\mu_2}(\ft,\fq)^{-1}W_{\mu_3}(\ft,\fq)\nonumber\\
	\times&(-Q_2)^{|\mu_3|}f_{\mu_3}(\ft,\fq)\times\bigg[\calR_{\mu_1^t\mu_2}(Q_1\sqrt{\frac{\fq}{\ft}})\calR_{\mu_1^t\mu_2}(Q_1\sqrt{\frac{\ft}{\fq}})\calR_{\mu_1^t\mu_3}(Q_{1,3}\sqrt{\frac{\fq}{\ft}})\nonumber\\
	\times&\calR_{\mu_1^t\mu_3}(Q_{1,3}\sqrt{\frac{\ft}{\fq}})\calR_{\mu_2^t\mu_3}(Q_3\sqrt{\frac{\fq}{\ft}})\calR_{\mu_2^t\mu_3}(Q_3\sqrt{\frac{\ft}{\fq}})\bigg]^{-1}\ ,\label{eq:Z123}\\
	Z^{D_4}_{\mu_1\mu_2\mu_3\mu_7\mu_4}=~&W_{\mu_7}(\ft,\fq)(-Q_c)^{|\mu_7|}W_{\mu_4}(\ft,\fq)(-Q_4)^{|\mu_4|}\calR_{\mu_1^t\mu_7}(Q_m)\calR_{\mu_7^t\mu_2}(\frac{Q_1}{Q_m})\nonumber\\
	\times&\calR_{\mu_7^t\mu_3}(\frac{Q_{1,3}}{Q_m})\calR_{\mu_1^t\mu_4}(\frac{Q_{1^2,3,c}}{Q_{m,4}})\calR_{\mu_2^t\mu_4}(\frac{Q_{1,3,c}}{Q_{m,4}})\calR_{\mu_3^t\mu_4}(\frac{Q_{1,c}}{Q_{m,4}})\nonumber\\
	\times&\bigg[\calR_{\mu_7^t\mu_4}(\frac{Q_{c,3,1^2}}{Q_{m^2,4}}\sqrt{\frac{\fq}{\ft}})\calR_{\mu_7^t\mu_4}(\frac{Q_{c,3,1^2}}{Q_{m^2,4}}\sqrt{\frac{\ft}{\fq}})\bigg]^{-1}\ ,\label{eq:Z74}\\
	Z^{D_4}_{\mu_1\mu_2\mu_3\mu_8\mu_5}=&~W_{\mu_8}(\ft,\fq)(-Q_d)^{|\mu_8|}W_{\mu_5}(\ft,\fq)(-Q_5)^{|\mu_5|}\calR_{\mu_1^t\mu_8}(Q_n)\nonumber\\
	\times&\calR_{\mu_8^t\mu_2}(\frac{Q_1}{Q_n})\calR_{\mu_8^t\mu_3}(\frac{Q_{1,3}}{Q_n})\calR_{\mu_1^t\mu_5}(\frac{Q_{1^2,3,d}}{Q_{n,5}})\calR_{\mu_2^t\mu_5}(\frac{Q_{1,3,d}}{Q_{n,5}})\calR_{\mu_3^t\mu_5}(\frac{Q_{1,d}}{Q_{n,5}})\nonumber\\
	\times&\bigg[\calR_{\mu_8^t\mu_5}(\frac{Q_{1^2,3,d}}{Q_{n^2,5}}\sqrt{\frac{\fq}{\ft}})\calR_{\mu_8^t\mu_5}(\frac{Q_{1^2,3,d}}{Q_{n^2,5}}\sqrt{\frac{\ft}{\fq}})\bigg]^{-1}\ ,\label{eq:Z85}\\
	Z^{D_4}_{\mu_1\mu_2\mu_3\mu_9\mu_6}=&~W_{\mu_9}(\ft,\fq)(-Q_e)^{|\mu_9|}W_{\mu_6}(\ft,\fq)(-Q_6)^{|\mu_6|}\calR_{\mu_1^t\mu_9}(\frac{Q_{d,n,t^2}}{Q_e})\nonumber\\
	\times&\calR_{\mu_9^t\mu_2}(\frac{Q_{1,e}}{Q_{d,n,t^2}})\calR_{\mu_9^t\mu_3}(\frac{Q_{1,3,e}}{Q_{d,n,t^2}})\calR_{\mu_1^t\mu_6}(\frac{Q_{1^2,3,e^2}}{Q_{d,n,t^2,6}})\calR_{\mu_2^t\mu_6}(\frac{Q_{1,3,e^2}}{Q_{d,n,t^2,6}})\nonumber\\
	\times&\calR_{\mu_3^t\mu_6}(\frac{Q_{1,e^2}}{Q_{d,n,t^2,6}})\bigg[\calR_{\mu_9^t\mu_6}(\frac{Q_{1^2,3,e^3}}{Q_{d^2,n^2,t^4,6}}\sqrt{\frac{\fq}{\ft}})\calR_{\mu_9^t\mu_6}(\frac{Q_{1^2,3,e^3}}{Q_{d^2,n^2,t^4,6}}\sqrt{\frac{\ft}{\fq}})\bigg]^{-1}\ .\label{eq:Z96}
\end{align}
We can see from the above results that the $Q$ factors and framing factors generated through \eqref{eq:Rexchange} compensate the $Q$ factors and framing factors\footnote{Here the framing factors are the $\calF_{\bullet}$ in \eqref{eq:deriveEF} which are framing factors of the unreflected branes.} corresponding to the color branes $\mu_2$,$\mu_3$,$\mu_5$ and $\mu_9$ in Figure \ref{fig:SU(3)_7} to make the recombined $Q$ factors and framing factors correspond to the original unswapped web diagram in Figure \ref{fig:su3_cs7_al}(a). And now, the forms of \eqref{eq:Z85} and \eqref{eq:Z96} are exactly like the form of \eqref{eq:Z74}. Then we take the Higgsing limit which means that $Q_m$,$Q_c$,$Q_n$,$Q_d$,$Q_t$,$Q_e$ in Figure \ref{fig:su3_cs7_al}(a) shrink and take some $\ft,\fq$ related values in the limit. For $Q_m$,$Q_c$,$Q_n$,$Q_d$ their limits are found in \cite{Cheng:2018wll}, which is $\sqrt{\frac{\fq}{\ft}}$ for all of them. As the right hand side of the middle SU(3) node in Figure \ref{fig:su3_cs7_al}(a) has two separate SU(2) nodes, we can find\footnote{By observation from Figure \ref{fig:su3_cs7_al}(a), $Q_d$ is the length of color brane $\mu_8$, $Q_e$ is the length of color brane $\mu_9$, $Q_n$ is the distance between $\mu_1$ and $\mu_8$, $\frac{Q_{n,d,t^2}}{Q_e}$ is the distance between $\mu_1$ and $\mu_9$.} that $Q_d$ in one node corresponds to $Q_e$ in the other node, similarly $Q_n$ in one node corresponds to $\frac{Q_{n,d,t^2}}{Q_e}$ in the other node, so $Q_e\rightarrow\sqrt{\frac{\fq}{\ft}}$ and $Q_t\rightarrow1$ in the Higgsing limit.

In the partition function \eqref{eq:Zsu3cs7} there are three factors which come from the three SU(3)-SU(2) bifundamental contributions:
\begin{equation}
	\calR_{\mu_1^t\mu_7}(Q_m)\calR_{\mu_1^t\mu_8}(Q_n)\calR_{\mu_1^t\mu_{9}}(\frac{Q_{d,n,t^2}}{Q_e})\ .
\end{equation}
After dropping the $\calM$'s from the $\calR$'s, the remaining is
\begin{equation}
	\calN_{\mu_1\mu_7}(Q_m\sqrt{\frac{\ft}{\fq}})\calN_{\mu_1\mu_8}(Q_n\sqrt{\frac{\ft}{\fq}})\calN_{\mu_1\mu_{9}}(\frac{Q_{d,n,t^2}}{Q_e}\sqrt{\frac{\ft}{\fq}})\ .
\end{equation}
Then we plug in the Higgsing limit values for the K\"ahler parameters, we obtain
\begin{equation}
	\calN_{\mu_1\mu_7}(1;\ft,\fq)\calN_{\mu_1\mu_8}(1;\ft,\fq)\calN_{\mu_1\mu_{9}}(1;\ft,\fq) \ .
\end{equation}
By using the following formula which is found in \cite{Cheng:2018wll}:
\begin{equation}
	\calN_{\mu\alpha}(1;\ft,\fq)\neq 0,~\text{only if}~\mu\succcurlyeq\alpha\ ,
\end{equation}
the non-zero contributing Young diagrams are now restricted to be $\mu_{7,8,9}\preccurlyeq\mu_1$.
After taking this Higgsing limit, the partition function $Z^{D_4}$ in \eqref{eq:Zsu3cs7} can now be written as
\begin{equation}
	{Z''}^{D_4}=Z^{D_4}_{\text{M}}\sum_{\boldsymbol{\mu}}\left(Z_{\mu_1\mu_2\mu_3}^{\kappa=7}\prod_{i=7,8,9}Z_{\mu_1\mu_2\mu_3\mu_i}\prod_{j=4,5,6}Z_{\mu_1\mu_2\mu_3\mu_{j+3}\mu_j}^{Q_j}\right)\ ,
	\label{eq:separatePFofCS7}
\end{equation}
where
\begin{equation}
	Z^{D_4}_{\text{M}}=\frac{\calM(Q_1)\calM(Q_3)\calM(Q_{1,3})\calM(Q_3\frac{\ft}{\fq})}{\calM(1)^3\calM(Q_1\frac{\ft}{\fq})^2\calM(Q_{1,3}\frac{\ft}{\fq})^2}\prod_{j=4,5,6}\frac{\calM(\frac{Q_{1^2,3}}{Q_j}\sqrt{\frac{\ft}{\fq}})\calM(\frac{Q_{1^2,3}}{Q_j}\frac{\ft^{3/2}}{\fq^{3/2}})}{\calM(\frac{Q_{1^2,3}}{Q_j}\sqrt{\frac{\ft}{\fq}})\calM(\frac{Q_{1,3}}{Q_j}\sqrt{\frac{\ft}{\fq}})\calM(\frac{Q_1}{Q_j}\sqrt{\frac{\ft}{\fq}})}\ ,
	\label{eq:ZMofhiggsedk7}
\end{equation}
\begin{multline}
	Z_{\mu_1\mu_2\mu_3}^{\kappa=7}=\frac{\fq^{\frac{||\mu_1||^2}{2}+||\mu_2||^2}\ft^{\frac{||\mu_1^t||^2}{2}+||\mu_3^t||^2}\prod_{i=1,2,3}\tilde{Z}_{\mu_i}(\ft,\fq)\tilde{Z}_{\mu_i^t}(\fq,\ft)\left(-Q_{1^2,2}\frac{\ft^{3/2}}{\fq^{3/2}}\right)^{|\mu_1|}Q_2^{|\mu_2|+|\mu_3|}}{\calN_{\mu_1\mu_2}(Q_1)\calN_{\mu_1\mu_2}(Q_1\frac{\ft}{\fq})\calN_{\mu_1\mu_3}(Q_{1,3})\calN_{\mu_1\mu_3}(Q_{1,3}\frac{\ft}{\fq})\calN_{\mu_2\mu_3}(Q_3)\calN_{\mu_2\mu_3}(Q_3\frac{\ft}{\fq})}\ ,
	\label{eq:ZmainofCS7}
\end{multline}

\begin{multline}
	Z_{\mu_1\mu_2\mu_3\mu_i}=\fq^{\frac{||\mu_i||^2}{2}}\ft^{\frac{||\mu_i^t||^2}{2}}\tilde{Z}_{\mu_i}(\ft,\fq)\tilde{Z}_{\mu_i^t}(\fq,\ft)(-\sqrt{\frac{\fq}{\ft}})^{|\mu_i|}\calN_{\mu_1\mu_i}(1)\calN_{\mu_i\mu_2}(Q_1\frac{\ft}{\fq})\calN_{\mu_i\mu_3}(Q_{1,3}\frac{\ft}{\fq})
	\label{eq:Zmu8ofCS7}
\end{multline}
for $i=7,8,9$, and 
\begin{align}
	Z_{\mu_1\mu_2\mu_3\mu_{j+3}\mu_j}^{Q_j}&=\fq^{\frac{||\mu_j||^2}{2}}\ft^{\frac{||\mu_j^t||^2}{2}}\tilde{Z}_{\mu_j}(\ft,\fq)\tilde{Z}_{\mu_j^t}(\fq,\ft)(-Q_j)^{|\mu_j|}\nn\\
	&\times\frac{\calN_{\mu_1\mu_j}(\frac{Q_{1^2,3}}{Q_j}\sqrt{\frac{\ft}{\fq}})\calN_{\mu_2\mu_j}(\frac{Q_{1,3}}{Q_j}\sqrt{\frac{\ft}{\fq}})\calN_{\mu_3\mu_j}(\frac{Q_1}{Q_j}\sqrt{\frac{\ft}{\fq}})}{\calN_{\mu_{j+3}\mu_j}(\frac{Q_{1^2,3}}{Q_j}\sqrt{\frac{\ft}{\fq}})\calN_{\mu_{j+3}\mu_j}(\frac{Q_{1^2,3}}{Q_j}\frac{\ft^{3/2}}{\fq^{3/2}})}
	\label{eq:ZQ4ofCS7}
\end{align}
for $j=4,5,6$\ . We note that from the web diagram in Figure \ref{fig:su3_cs7_al}(b), the map from K\"ahler parameters to gauge theory parameters is given by 
\begin{align}
	Q_1=\frac{A_1^2}{A_2},\qquad
	Q_2=\frac{A_2^2}{A_1^3}u,\qquad
	Q_3=\frac{A_2^2}{A_1},\qquad
	\label{eq:subCS7}
\end{align}
where $A_1$ and $A_2$ are the Coulomb branch parameters, $u$ is the instanton factor.

Now we compute the perturbative part of the partition function which corresponds to $\mu_{1,2,3}=\text{\o}$, and because $\mu_{7,8,9}\preccurlyeq\mu_1$, $\mu_{7,8,9}$ are also empty in this case.
\begin{equation}
	{Z''}^{D_4}_{\text{pert}}=Z_{\text{M}}^{D_4}\sum_{\mu_4}Z_{\text{\o}\text{\o}\text{\o}\text{\o}\mu_4}^{Q_4}\sum_{\mu_5}Z_{\text{\o}\text{\o}\text{\o}\text{\o}\mu_5}^{Q_5}\sum_{\mu_6}Z_{\text{\o}\text{\o}\text{\o}\text{\o}\mu_6}^{Q_6}\ .
	\label{eq:pertofCS7}
\end{equation}
We first sum over $Z_{\text{\o}\text{\o}\text{\o}\text{\o}\mu_4}^{Q_4}$ from zero to some finite Young diagram boxes number of $\mu_4$, and then series expand it with respect to $Q_1$ and $Q_4$. There are both negative power and positive power terms of $Q_4$ in the expansion, but when we increase the upper bound of the boxes number of $\mu_4$, lower positive power terms of $Q_4$ disappear, in the limit that the upper bound of boxes number of $\mu_4$ goes to infinity, all the positive power terms of $Q_4$ will disappear and only negative power terms of $Q_4$ remain which are not changing once the upper bound exceeds some finite number, then we take the limit that $Q_4$ goes to $\infty$, we obtain\begin{equation}
	\calZ_{0}\equiv\lim_{Q_4\rightarrow\infty}\sum_{\mu_4}Z_{\text{\o}\text{\o}\text{\o}\text{\o}\mu_4}^{Q_4}=\text{PE}\left[\frac{Q_1(1+Q_3)\ft}{(1-\fq)(1-\ft)}\right] \ .
	\label{eq:Z7Q4ptlim}
\end{equation}
Because $Z^{Q_{5}},Z^{Q_{6}}$ have the same form as $Z^{Q_4}$, they will have the same limit \eqref{eq:Z7Q4ptlim} when $Q_{5},Q_{6}\rightarrow\infty$. Substitute \eqref{eq:Z7Q4ptlim} into \eqref{eq:pertofCS7}, take the decoupling limit, transform K\"ahler parameters into gauge theory parameters and drop the factors that do not depend on $A_1$ or $A_2$, we obtain the perturbative part of SU$(3)_7$:
\begin{align}
	Z_{\text{pert}}^{\text{SU}(3)_7}&=\frac{\calM(Q_1)\calM(Q_3)\calM(Q_{1,3})\calM(Q_3\frac{\ft}{\fq})}{\calM(Q_1\frac{\ft}{\fq})^2\calM(Q_{1,3}\frac{\ft}{\fq})^2}\calZ_0^3\nonumber\\
	&=\text{PE}\left[\frac{\fq+\ft}{(1-\fq)(1-\ft)}\left(\frac{A_1^2}{A_2}+A_1A_2+\frac{A_2^2}{A_1}\right)\right]\ .
\end{align}

The instanton part of \eqref{eq:separatePFofCS7} is
\begin{equation}
	{Z''}_{\text{inst}}^{D_4}=\sum_{\mu_1,\mu_2,\mu_3}\Bigg(
	Z_{\mu_1\mu_2\mu_3}^{\kappa=7}\prod_{j=4,5,6}\sum_{\mu_{j+3}}\left( Z_{\mu_1\mu_2\mu_3\mu_{j+3}}\frac{\sum_{\mu_j}Z_{\mu_1\mu_2\mu_3\mu_{j+3}\mu_j}^{Q_j}}{\sum_{\mu_j}Z_{\text{\o}\text{\o}\text{\o}\text{\o}\mu_j}^{Q_j}}\right)\Bigg)\ .
\end{equation}
We define
\begin{equation}
	\calZ_{\mu_1\mu_2\mu_3}\equiv\lim_{Q_4\rightarrow\infty}\sum_{\mu_7}\left(Z_{\mu_1\mu_2\mu_3\mu_7}\frac{\sum_{\mu_4}Z_{\mu_1\mu_2\mu_3\mu_7\mu_4}^{Q_4}}{\sum_{\mu_4}Z_{\text{\o}\text{\o}\text{\o}\text{\o}\mu_4}^{Q_4}}\right)\ .
	\label{eq:ZQ4insdef7}
\end{equation}
Because $\mu_7\preccurlyeq\mu_1$, the summation over $\mu_7$ is finite, while the summation over $\mu_4$ is up to infinity. In order to compute $\calZ_{\mu_1\mu_2\mu_3}$, we first sum over $\mu_4$ up to a finite boxes number in the numerator and denominator of \eqref{eq:ZQ4insdef7}, then we expand the whole expression \eqref{eq:ZQ4insdef7} with respect to $Q_1$ and $Q_4$. There are both negative power and positive power terms of $Q_4$ appearing in the expansion, but when we increase the upper bound of the boxes number of $\mu_4$, lower order positive power terms of $Q_4$ disappear, in the limit that the upper bound of boxes number of $\mu_4$ goes to infinity, all the positive power terms of $Q_4$ will disappear and only negative power terms of $Q_4$ remain which are not changing once the upper bound exceeds some finite number. Higher order terms of $Q_1$ also disappear because their coefficients contain only positive power terms of $Q_4$. Then we take the limit that $Q_4$ goes to $\infty$, we will obtain $\calZ_{\mu_1\mu_2\mu_3}$ expanded with respect to $Q_1$ up to finite order. Because $Z^{Q_5},Z^{Q_6}$ have the same form as $Z^{Q_4}$, they will have the same limit \eqref{eq:ZQ4insdef7} when $Q_5,Q_6\rightarrow\infty$.
Because $Z_{\mu_1\mu_2\mu_3}^{\kappa=7}$ is proportional to $Q_2^{|\mu_1|+|\mu_2|+|\mu_3|}$, Young diagram assignments that satisfy $|\mu_1|+|\mu_2|+|\mu_3|=1$ contribute to one-instanton partition function, and the corresponding $\calZ$'s are
\begin{align}
	\calZ_{\{1\},\text{\o},\text{\o}}&=1-\frac{\fq}{\ft}Q_1(1+Q_3)\ ,\nonumber\\
	\calZ_{\text{\o},\{1\},\text{\o}}&=1+Q_1(Q_3-\frac{\ft}{\fq})\ ,\nonumber\\
	\calZ_{\text{\o},\text{\o},\{1\}}&=1+Q_1(1-\frac{\ft}{\fq}Q_3)\ .
	\label{eq:oneinst}
\end{align}
Young diagram assignments that satisfy $|\mu_1|+|\mu_2|+|\mu_3|=2$ contribute to two-instanton partition function, then the corresponding $\calZ$'s are given as follows:
\begin{align}
	\calZ_{\{2\},\text{\o},\text{\o}}&=1-\frac{\fq(1+\fq)}{\ft}Q_1(1+Q_3)+\frac{\fq^3}{\ft^2}Q_1^2(1+Q_3+\fq Q_3+Q_3^2)\ ,\nonumber\\
	\calZ_{\{1,1\},\text{\o},\text{\o}}&=1-\frac{\fq(1+\ft)}{\ft^2}Q_1(1+Q_3)+\frac{\fq^2}{\ft^4}Q_1^2(\ft+Q_3+\ft Q_3+\ft Q_3^2)\ ,\nonumber\\
	\calZ_{\{1\},\{1\},\text{\o}}&=1+Q_1\left(1-\frac{1}{\ft}-\frac{\ft}{\fq}+Q_3-\frac{\fq}{\ft}(\ft+Q_3)\right)+\frac{1}{\ft}Q_1^2(1+Q_3)(\ft-\fq Q_3)\ ,\nonumber\\
	\calZ_{\{1\},\text{\o},\{1\}}&=1+Q_1\left(1-\frac{\fq}{\ft}+Q_3(1-\fq-\frac{1}{\ft}-\frac{\ft}{\fq})\right)-\frac{1}{\ft}Q_1^2(1+Q_3)(\fq-\ft Q_3)\ ,\nonumber\\
	\calZ_{\text{\o},\{1\},\{1\}}&=1+\frac{\fq-\ft}{\fq}Q_1(1+Q_3)+Q_1^2\left(\ft Q_3+\frac{\ft^2}{\fq^2}Q_3-\frac{\ft+(\ft-1)Q_3+\ft Q_3^2}{\fq}\right)\ ,\nonumber\\
	\calZ_{\text{\o},\{2\},\text{\o}}&=1+\frac{1+\fq}{\fq^2}Q_1(\fq^2 Q_3-\ft)+Q_1^2\left(Q_3^2-\frac{(1+\fq)\ft}{\fq^2}Q_3+\frac{\ft^2}{\fq^3}\right)\ ,\nonumber\\
	\calZ_{\text{\o},\{1,1\},\text{\o}}&=1+\frac{1+\ft}{\fq\ft}Q_1(\fq Q_3-\ft^2)+Q_1^2\left(\frac{\ft^3}{\fq^2}+Q_3^2-\frac{(\ft+\ft^2)Q_3}{\fq}\right)\ ,\nonumber\\
	\calZ_{\text{\o},\text{\o},\{2\}}&=1+\frac{1+\fq}{\fq^2}Q_1(\fq^2-\ft Q_3)+Q_1^2\left(1-\frac{\ft}{\fq^2}Q_3-\frac{\ft}{\fq}Q_3+\frac{\ft^2}{\fq^3}Q_3^2\right)\ ,\nonumber\\
	\calZ_{\text{\o},\text{\o},\{1,1\}}&=1+\frac{1+\ft}{\fq\ft}Q_1(\fq-\ft^2 Q_3)+Q_1^2\left(1+\frac{\ft^3}{\fq^2}Q_3^2-\frac{\ft(1+\ft)Q_3}{\fq}\right)\ .
	\label{eq:twoinst}
\end{align}
So the instanton partition function of SU$(3)_7$ is
\begin{equation}
	Z_{\text{inst}}^{\text{SU}(3)_7}=\sum_{\mu_1,\mu_2,\mu_3}Z_{\mu_1\mu_2\mu_3}^{\kappa=7}\calZ_{\mu_1\mu_2\mu_3}^3\ ,
	\label{eq:instSU(3)_7}
\end{equation}
where $Z_{\mu_1\mu_2\mu_3}^{\kappa=7}$ is defined in \eqref{eq:ZmainofCS7}. Summing over the Young diagrams, one can express the instanton partition function as an expansion of $Q_2$. Here we write it as a PE:
\begin{align}
	Z_{\text{inst}}^{\text{SU}(3)_7}&
	=1+\sum_{n=1}Z_n Q_2^n
	= \text{PE}\left[\sum_{n=1}f^{\kappa=7}_n(Q_1,Q_3;\ft,\fq) \, Q_2^n\right]\ ,
	\label{eq:finPE7}
\end{align}
where $Z_n$ is the $n$-instanton partition function up to an overall factor $(A^2_2/A^3_1)^n$ due to \eqref{eq:subCS7}.  
By substituting \eqref{eq:ZmainofCS7} and \eqref{eq:oneinst} into \eqref{eq:instSU(3)_7}, we can obtain the one-instanton partition function, then by \eqref{eq:finPE7}, we have the exact form of the one-instanton part:
\begin{align}
	f_1^{\kappa=7}
	=\frac{\ft}{(1-\fq)(1-\ft)\fq}\Bigg(&\frac{Q_1^2(\fq Q_1(1+Q_3)-\ft)^3}{(1-Q_1)(1-Q_1Q_3)(\fq Q_1-\ft)(\fq Q_1Q_3-\ft)}\cr
	&+\frac{(\fq+\fq Q_1Q_3-Q_1 \ft)^3}{(1-Q_1)(1-Q_3)(\fq Q_3-\ft)(Q_1\ft-\fq)}\crcr
	&+\frac{(\fq+\fq Q_1-Q_1 Q_3 \ft)^3}{(1-Q_3)(1-Q_1Q_3)(\fq-Q_3\ft)(\fq-Q_1Q_3\ft)}\Bigg).\label{eq:f1k=7PE}
\end{align}
To compare this with the known result from the blowup \cite{Kim:2020hhh}, we expand \eqref{eq:f1k=7PE} with respect to $Q_1$ and $Q_3$, which gives
\begin{small}
\begin{align}
	f_1^{\kappa=7}\!
	&=\frac{\fq+\ft}{(\fq-1) (\ft-1)}\bigg[1+Q_1+(\frac{\fq}{\ft}+\frac{\ft}{\fq})Q_3+Q_1^2+(\frac{\fq}{\ft}+\frac{\ft}{\fq}+1)Q_1Q_3+(\frac{\fq^2}{\ft^2}+\frac{\ft^2}{\fq^2}+1)Q_3^2\nn\\
	&+\!Q_1^3+(\frac{\fq}{\ft}\!+\frac{\ft}{\fq}\!+1)Q_1^2Q_3\!+(\frac{\fq^2}{\ft^2}+\!\frac{\ft^2}{\fq^2}+\!\frac{\fq}{\ft}+\!\frac{\ft}{\fq}+1)Q_1Q_3^2+\!(\frac{\fq^2}{\ft^2}+\!\frac{\ft^2}{\fq^2}+\!\frac{\fq}{\ft}+\!\frac{\ft}{\fq}+\!2)Q_1^2Q_3^2\nn\\
	&+\!(\frac{\fq^3}{\ft^3}+\!\frac{\ft^3}{\fq^3}+\!\frac{\fq}{\ft}+\!\frac{\ft}{\fq})Q_3^3+\!(\frac{\fq}{\ft}+\!\frac{\ft}{\fq}\!+1)Q_1^3Q_3+\!(\frac{\fq^3}{\ft^3}\!+\frac{\fq^2}{\ft^2}\!+\frac{\ft^3}{\fq^3}\!+\frac{\ft^2}{\fq^2}\!+\frac{\fq}{\ft}\!+\frac{\ft}{\fq}\!+1)Q_1Q_3^3\nn\\
	&+(\frac{\fq^2}{\ft^2}+\frac{\ft^2}{\fq^2}+\frac{\fq}{\ft}+\frac{\ft}{\fq}+2)Q_1^3Q_3^2+(\frac{\fq^3}{\ft^3}+\frac{\fq^2}{\ft^2}+\frac{\ft^3}{\fq^3}+\frac{\ft^2}{\fq^2}+\frac{2 \fq}{\ft}+\frac{2 \ft}{\fq}+1)Q_1^2Q_3^3\nn\\
	&+(\frac{\fq^3}{\ft^3}+\frac{\fq^2}{\ft^2}+\frac{\ft^3}{\fq^3}+\frac{\ft^2}{\fq^2}+\frac{2 \fq}{\ft}+\frac{2 \ft}{\fq}+2)Q_1^3Q_3^3\bigg] + \mathcal{O}(Q_1^4, Q_3^4)\ ,  
\end{align}
\end{small}
\noindent where $\mathcal{O}(Q_1^4, Q_3^4)$ denotes terms either of order $Q_1^4$ or higher or of $Q_3^4$ or higher. 
We express this in terms of the spin state $[j_l, j_r]$ defined in \eqref{eq:GVspin}, 
\begin{align}
    f_1^{\kappa=7}\! =&~ [0, \tfrac12] + [0,\tfrac32] Q_3+[0,\tfrac52] Q_3^2+ [0, \tfrac12] Q_1 +\Big([0,\tfrac12]+[0,\tfrac32]\Big) Q_1Q_3\crcr
    &+ \Big( [0,\tfrac32]+[0,\tfrac52]\Big) Q_1 Q_3^2+ [0,\tfrac12] Q_1^2 + \Big( [0,\tfrac12] +[0,\tfrac32]\Big) Q_1^2 Q_3    \crcr
    &+ \Big( [0,\tfrac12] +[0,\tfrac32]+[0,\tfrac52]\Big) Q_1^2 Q_3^2\crcr
&  + [0, \tfrac72]Q_3^3
   + \Big([0, \tfrac52]+[0, \tfrac72]\Big)Q_1 Q_3^3
   + \Big([0, \tfrac32]+[0, \tfrac52]+[0, \tfrac72]\Big)Q_1^2Q_3^3 \crcr
& + [0, \tfrac12]Q_1^3+
   \Big([0, \tfrac12]+[0, \tfrac32]\Big)Q_1^3 Q_3 +\Big([0, \tfrac12]+[0, \tfrac32]+[0, \tfrac52]\Big)Q_1^3 Q_3^2
   \cr
   &
   + 
   \Big([0, \tfrac12]+[0, \tfrac32]+[0, \tfrac52]+[0, \tfrac72]\Big)Q_1^3 Q_3^3 +\mathcal{O}(Q_1^4, Q_3^4)
   \ . \label{eq:f1k=7GV}
\end{align}
This agrees with the blowup result in \cite{Kim:2020hhh}.

By substituting \eqref{eq:twoinst} into \eqref{eq:instSU(3)_7}, we can obtain the two-instanton partition function. As a PE form \eqref{eq:finPE7}, we can obtain the exact form $f_2^{\kappa=7}$.\footnote{As the exact form $f_2^{\kappa=7}$ is rather long, we put the exact form into a Mathematica file 
attached as an ancillary file.}
Here, we expand $f_2^{\kappa=7}$ to order $Q_1^2Q_3^2$: 
\begin{small}
\begin{align}
	f_2^{\kappa=7}\!=&~\frac{1}{(1-\fq)(1-\ft)}\bigg[(\frac{\fq^3}{\ft^2}+\frac{\ft^3}{\fq^2}+\frac{\fq^2}{\ft}+\frac{\ft^2}{\fq}+\fq+\ft)Q_3+(\frac{\fq^3}{\ft^2}+\frac{\ft^3}{\fq^2}+\frac{2 \fq^2}{\ft}+\frac{2 \ft^2}{\fq}+2 \fq\nn\\
	&+2 \ft)Q_1Q_3+(\frac{\fq^5}{\ft^3}+\frac{\fq^4}{\ft^2}+\frac{\fq^4}{\ft^3}+\frac{\fq^4}{\ft^4}+\frac{2 \fq^3}{\ft^2}+\frac{\fq^3}{\ft^3}+\frac{\fq^2}{\ft^2}+\frac{\ft^5}{\fq^3}+\frac{\ft^4}{\fq^2}+\frac{\ft^4}{\fq^3}+\frac{\ft^4}{\fq^4}+\frac{2 \ft^3}{\fq^2}\nn\\
	&+\frac{\ft^3}{\fq^3}+\frac{\ft^2}{\fq^2}+\frac{\fq^3}{\ft}+\frac{2 \fq^2}{\ft}+\fq^2+\frac{\ft^3}{\fq}+\frac{2 \ft^2}{\fq}+\fq \ft+\frac{\fq}{\ft}+\frac{\ft}{\fq}+2 \fq+\ft^2+2 \ft+1)Q_3^2\nn\\
	&+(\frac{\fq^3}{\ft^2}+\frac{\ft^3}{\fq^2}+\frac{2 \fq^2}{\ft}+\frac{2 \ft^2}{\fq}+3 \fq+3 \ft)Q_1^2Q_3+(\frac{\fq^5}{\ft^3}+\frac{2 \fq^4}{\ft^2}+\frac{2 \fq^4}{\ft^3}+\frac{\fq^4}{\ft^4}+\frac{5 \fq^3}{\ft^2}+\frac{2 \fq^3}{\ft^3}\nn\\
	&+\frac{2 \fq^2}{\ft^2}+\frac{\ft^5}{\fq^3}+\frac{2 \ft^4}{\fq^2}+\frac{2 \ft^4}{\fq^3}+\frac{\ft^4}{\fq^4}+\frac{5 \ft^3}{\fq^2}+\frac{2 \ft^3}{\fq^3}+\frac{2 \ft^2}{\fq^2}+\frac{2 \fq^3}{\ft}+\frac{6 \fq^2}{\ft}+2 \fq^2+\frac{2 \ft^3}{\fq}+\frac{6 \ft^2}{\fq}\nn\\
	&+2 \fq \ft+\frac{2 \fq}{\ft}+\frac{2 \ft}{\fq}+6 \fq+2 \ft^2+6 \ft+2)Q_1Q_3^2+(\frac{\fq^5}{\ft^3}+\frac{2 \fq^4}{\ft^2}+\frac{2 \fq^4}{\ft^3}+\frac{\fq^4}{\ft^4}+\frac{6 \fq^3}{\ft^2}+\frac{2 \fq^3}{\ft^3}\nn\\
	&+\frac{3 \fq^2}{\ft^2}+\frac{\ft^5}{\fq^3}+\frac{2 \ft^4}{\fq^2}+\frac{2 \ft^4}{\fq^3}+\frac{\ft^4}{\fq^4}+\frac{6 \ft^3}{\fq^2}+\frac{2 \ft^3}{\fq^3}+\frac{3 \ft^2}{\fq^2}+\frac{3 \fq^3}{\ft}+\frac{9 \fq^2}{\ft}+3 \fq^2+\frac{3 \ft^3}{\fq}+\frac{9 \ft^2}{\fq}\nn\\
	&+3 \fq \ft+\frac{3 \fq}{\ft}+\frac{3 \ft}{\fq}+10 \fq+3 \ft^2+10 \ft+3)Q_1^2Q_3^2\bigg]+ \mathcal{O}(Q_1^3, Q_3^3)\ .
\end{align}
\end{small}
In terms of the spin state, it is given by
\begin{align}\label{eq:ff2k=7GVspin} 
 f_2^{\kappa=7}\!&=[0,\tfrac52]Q_3+\Big([0,\tfrac52]+[0,\tfrac72]+[\tfrac12,4] \Big)Q_3^2+\Big([0,\tfrac32]+[0,\tfrac52]\Big)Q_1Q_3\\
  &+\Big([0,\tfrac32]+3[0,\tfrac52]+2[0,\tfrac72] +[\tfrac12,3]+[\tfrac12,4]\Big)Q_1Q_3^2+\Big([0,\tfrac12]+[0,\tfrac32]+[0,\tfrac52] \Big)Q_1^2Q_3\crcr
 & +\Big([0,\tfrac12]+3[0,\tfrac32]+4[0,\tfrac52]+2[0,\tfrac72]+[\tfrac12,2] +[\tfrac12,3]+[\tfrac12,4]
 \Big)Q_1^2Q_3^2+\!\mathcal{O}(Q_1^3, Q_3^3),\nonumber 
\end{align}
which agrees with the two-instanton part from the blowup result \cite{Kim:2020hhh}.

\subsection{5d SU(3) theory at CS level 9}\label{sec:SU3CS9}
We consider 5d SU(3) theory at CS level 9 without flavor, denoted as SU(3)$_9$, which is a KK theory arising from 6d SU(3) theory on a circle with a $\mathbb{Z}_2$ twist \cite{Jefferson:2017ahm,Jefferson:2018irk,Razamat:2018gro}. 
Its refined partition function is computed in \cite{Kim:2020hhh} via bootstrapping the BPS spectrum based on the blowup equation \cite{Nakajima:2003pg,Nakajima:2005fg,Gottsche:2006bm}. From the perspective of topological vertex, the partition function for 6d SU(3) theory with a $\mathbb{Z}_2$-twist is obtained in \cite{Kim:2021cua} based on 5-brane web with two $\widetilde{\text{O5}}$-planes, and that for the 5d SU(3)$_9$ is also obtained in the S-dual frame \cite{Hayashi:2020hhb}. However these partition functions based on topological vertex are limited to the unrefined case where $\epsilon_1+\epsilon_2=0$.\footnote{In \cite{Kim:2021cua}, the refined elliptic genus for 6d SU(3) theory with a $\mathbb{Z}_2$ twist is also computed via the ADHM method.} We here, however, compute the refined Nekrasov partition function for 5d SU(3)$_9$ theory, using our proposal for the refined topological vertex with ON-planes. The corresponding 5-brane web  requires two ON-planes and can be obtained from the Higgsing of the following affine $D_4$-type quiver \cite{Hayashi:2018lyv},
\begin{align}\label{eq:D4quiverSU3}
	\begin{tikzpicture}
		\draw[thick](-1.4,.8)--(-0.7,0.2);
		\draw[thick](-1.4,-.8)--(-0.7,-0.2);
		\draw[thick](0.7,0.2)--(1.4,.8);
		\draw[thick](0.7,-0.2)--(1.4,-.8); 
		\node at (0,0){SU$(3)_1$};
		\node at (-2,1){SU$(2)$};
		\node at (-2,-1){SU$(2)$};
		\node at (2,1){SU$(2)$};
		\node at (2,-1){SU$(2)$};
	\end{tikzpicture}		
\end{align}
\begin{figure}[htbp]
	\centering
	\includegraphics[scale=1]{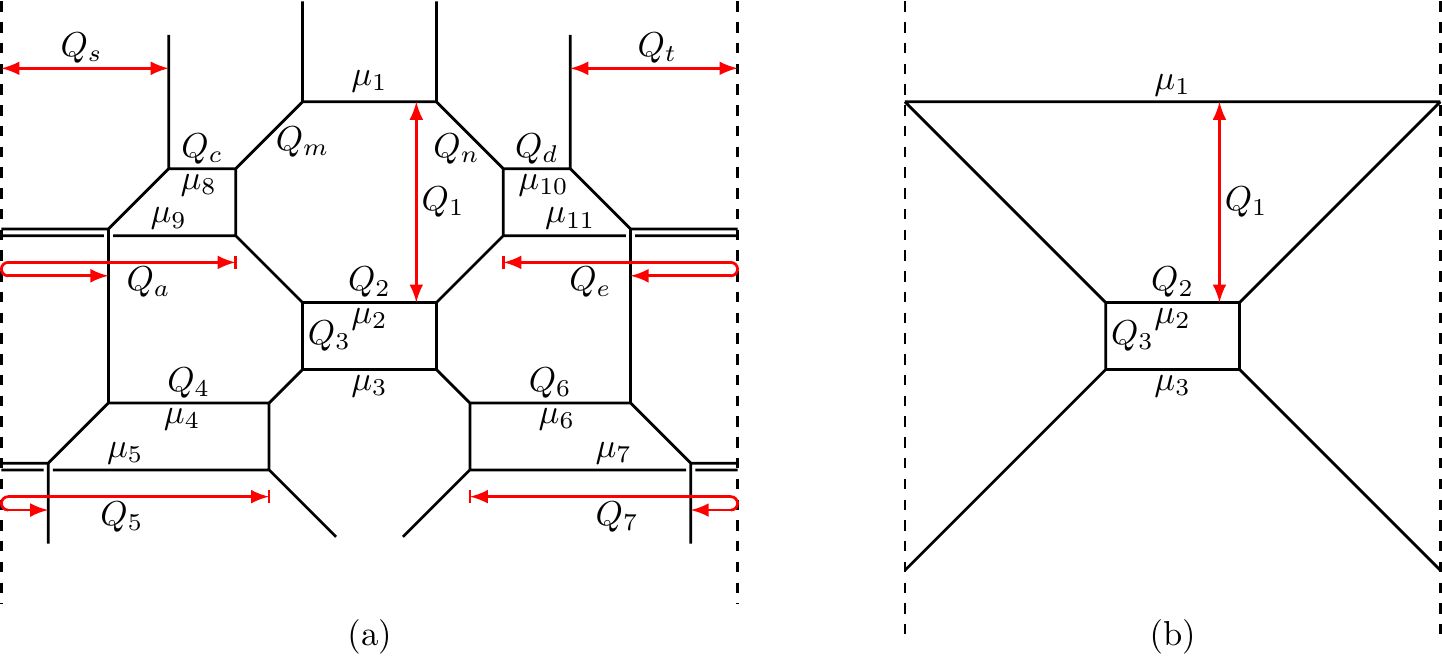}
	\caption{(a) A 5-brane web for the affine $D_4$-type quiver with two ON-planes on the left and right. (b) SU(3) with CS 9 obtained by Higgsing and decoupling.}
	\label{fig:D4quiver SU3}
\end{figure}
\noindent Here, the middle node is the SU(3) theory at the CS level 1 and the corresponding web diagram is depicted in Figure \ref{fig:D4quiver SU3}(a). 
The Higgsing procedure is similar to the SU$(3)_7$ case discussed in the previous section: A Higgsing of an SU(2) gives rise to an antisymmetric hypermultiplet ($\mathbf{AS}$) to SU(3), increasing the CS level of SU(3) by $\frac32$~\cite{Hayashi:2018lyv}. As we have four SU(2), we get SU(3)$_7+4 \mathbf{AS}$ after the Higgsing. Finally, since an antisymmetric hypermultiplet transforms as $\bar{\mathbf{3}}$, the decoupling of an antisymmetric hypermultiplet further increases the CS level of SU(3) by $\frac12$. Hence, decoupling all the $\mathbf{AS}$ yields SU(3)$_9$ whose web diagram is shown in Figure \ref{fig:D4quiver SU3}(b).

As in the previous section where we have computed the partition function of SU$(3)_7$ theory, we start from the web diagram in Figure \ref{fig:D4quiver SU3}(a). The two SU(2) sub webs on the right and the two SU(2) sub webs on the left are both overlapping, in order to compute by our proposal for topological vertex with ON-planes, we need to swap the positions of relevant branes to make the SU(2) sub webs separated. From the Higgsing and decoupling processes of obtaining the SU$(3)_7$ theory, we know that in the Higgsing limit of Figure \ref{fig:D4quiver SU3}(a), $Q_m,Q_c,Q_a,Q_n,Q_d,Q_e\rightarrow\sqrt{\frac{\fq}{\ft}}$ and $Q_s,Q_t\rightarrow1$, in the decoupling limit, $Q_4,Q_5,Q_6,Q_7\rightarrow\infty$. As the procedures are exactly like the procedures done in the previous section on SU$(3)_7$, we omit the detailed computation process and just list the results. 
From the web diagram in Figure \ref{fig:D4quiver SU3}(b), the K\"ahler parameters are expressed in terms of the gauge theory parameters as 
\begin{align}
	Q_1=\frac{A_1^2}{A_2},\qquad
	Q_2=\frac{A_2^2}{A_1^4}u,\qquad
	Q_3=\frac{A_2^2}{A_1},\qquad
	\label{eq:subCS9}
\end{align}
where $A_1$ and $A_2$ are the Coulomb branch parameters and $u$ is the instanton factor. Notice that $Q_1^2 Q_2=u$ and hence any terms with $(Q_1^2 Q_2)^n$ do not explicitly depend on the Coulomb branch parameters and they are the extra factor, which will be modded out when expressing the partition function.

The perturbative part of the partition function is
\begin{align}
	Z_{\text{pert}}^{\text{SU}(3)_9}&=\frac{\calM(Q_1)\calM(Q_3)\calM(Q_{1,3})\calM(Q_3\frac{\ft}{\fq})}{\calM(Q_1\frac{\ft}{\fq})^3\calM(Q_{1,3}\frac{\ft}{\fq})^3}\calZ_0^4\nonumber\\
	&=\text{PE}\left[\frac{\fq+\ft}{(1-\fq)(1-\ft)}\left(\frac{A_1^2}{A_2}+A_1A_2+\frac{A_2^2}{A_1}\right)\right]\ .
\end{align}
The instanton part of the partition function with extra factors included is
\begin{equation}
	Z_{\text{inst+extra}}^{\text{SU}(3)_9}=\sum_{\mu_1,\mu_2,\mu_3} Z_{\mu_1\mu_2\mu_3}^{\kappa=9}\calZ_{\mu_1\mu_2\mu_3}^4\ ,
	\label{eq:instSU(3)_9}
\end{equation}
where
\begin{multline}
	Z_{\mu_1\mu_2\mu_3}^{\kappa=9}\equiv\frac{\fq^{||\mu_2||^2}\ft^{||\mu_1^t||^2+||\mu_3^t||^2}\prod_{i=1,2,3}\tilde{Z}_{\mu_i}(\ft,\fq)\tilde{Z}_{\mu_i^t}(\fq,\ft)\left({Q_{1^2,2}}\frac{\ft^2}{\fq^2}\right)^{|\mu_1|}Q_2^{|\mu_2|+|\mu_3|}}{\calN_{\mu_1\mu_2}(Q_1)\calN_{\mu_1\mu_2}(Q_1\frac{\ft}{\fq})\calN_{\mu_1\mu_3}(Q_{1,3})\calN_{\mu_1\mu_3}(Q_{1,3}\frac{\ft}{\fq})\calN_{\mu_2\mu_3}(Q_3)\calN_{\mu_2\mu_3}(Q_3\frac{\ft}{\fq})}\ ,
	\label{eq:ZmainofCS9}
\end{multline}
and $\calZ_{\mu_1\mu_2\mu_3}$ is the same one that appears in \eqref{eq:instSU(3)_7}. We now expand this instanton part with respect to $Q_2$ which is the K\"ahler parameter proportional to the instanton factor $u$. As discussed in the previous section,  $Z_{\mu_1\mu_2\mu_3}^{\kappa=9}$ is proportional to  $Q_2^{|\mu_1|+|\mu_2|+|\mu_3|}$ (or  $u^{|\mu_1|+|\mu_2|+|\mu_3|}$) and leads to the  instanton expansion. We then organize the sum of the $n$-instanton part $Z_n$ as a PE: 
\begin{align}
	Z_{\text{inst+extra}}^{\text{SU}(3)_9}&=1+\sum_{n=1}Z_n Q_2^n
	= \text{PE}\left[\sum_{n=1}f_n^{\kappa=9}(Q_1,Q_3;\ft,\fq) \, Q_2^n\right]\ .
	\label{eq:finPE}
\end{align}
By summing over the Young diagram assignments in \eqref{eq:instSU(3)_9} with the corresponding $\calZ$'s defined in \eqref{eq:ZQ4insdef7}, order by order in $Q_2$, one can obtain $f_n$. We list a few 
$f_n$ below. In particular, the exact one-instanton part $f_1$ is given as follows:
\begin{align}
	f_1^{\kappa=9}
	=\frac{\ft}{(1-\fq)(1-\ft)\fq^2}&\bigg(\frac{Q_1^2\left(-\fq Q_1(1+Q_3)+\ft\right)^4}{(1-Q_1)(1-Q_1Q_3)(\fq Q_1-\ft)(\fq Q_1Q_3-\ft)}\cr
	&+\frac{(\fq+\fq Q_1Q_3-Q_1\ft)^4}{(1-Q_1)(1-Q_3)(\fq Q_3-\ft)(\ft Q_1-\fq)}\cr
	&+\frac{(\fq+\fq Q_1-\ft Q_1Q_3)^4}{(1-Q_3)(1-Q_1Q_3)(\ft Q_3-\fq)(\ft Q_1Q_3-\fq)}\bigg) .
	\label{eq:f1ex}
\end{align}
To compare this with the result from the blowup \cite{Kim:2020hhh}, we expand $f_1$ with respect to $Q_1,Q_3$, which gives 
\begin{align}
	f_1^{\kappa=9}&=\frac{\fq+\ft}{(1-\fq)(1-\ft)}\bigg[1+Q_1+\Big(\frac{\fq}{\ft}+\frac{\ft}{\fq}\Big)Q_3+\Big(1+\frac{\fq^2}{\ft^2}+\frac{\ft^2}{\fq^2}\Big)Q_3^2\cr
	&\!+\!\Big(1+\frac{\fq}{\ft}+\frac{\ft}{\fq}\Big)Q_1Q_3
	+\Big(1+\frac{\fq^2}{\ft^2}+\frac{\fq}{\ft}+\frac{\ft}{\fq}+\frac{\ft^2}{\fq^2}\Big)Q_1Q_3^2+\Big(1+2\frac{\ft}{\fq}+\frac{\ft^2}{\fq^2}\Big)Q_1^2\cr
	&\!+\!\Big(1+\frac{\fq}{\ft}+\frac{\ft}{\fq}\Big)Q_1^2Q_3+\Big(2+\frac{\fq^2}{\ft^2}+\frac{\fq}{\ft}+\frac{\ft}{\fq}+\frac{\ft^2}{\fq^2}\Big)Q_1^2Q_3^2\crcr
    &+Q_1^3+
   \Big(\frac{\fq}{\ft}+\frac{\ft}{\fq}+\frac{\fq^3}{\ft^3}+\frac{\ft^3}{\fq^3}\Big)Q_3^3+
   \Big(1+\frac{\fq}{\ft}+\frac{\ft}{\fq}\Big)Q_1^3 Q_3\cr
   &
   +
   \Big(1+\frac{\fq^3}{\ft^3}+\frac{\fq^2}{\ft^2}+\frac{\fq}{\ft}+\frac{\ft}{\fq}+\frac{\ft^2}{\fq^2}+\frac{\ft^3}{\fq^3}\Big)Q_3^3 Q_1 +\Big(2+\frac{\fq^2}{\ft^2}+\frac{\fq}{\ft}+\frac{\ft}{\fq}+\frac{\ft^2}{\fq^2}\Big)Q_1^3 Q_3^2\cr
  & +
   \Big(1+\frac{\fq^3}{\ft^3}+\frac{\fq^2}{\ft^2}+\frac{2
   \fq}{\ft}+\frac{2 \ft}{\fq}+\frac{\ft^2}{\fq^2}+\frac{\ft^3}{\fq^3}\Big)Q_3^3 Q_1^2\cr
   &
   + 
   \Big(2+\frac{\fq^3}{\ft^3}+\frac{\fq^2}{\ft^2}+\frac{2
   \fq}{\ft}+\frac{2 \ft}{\fq}+\frac{\ft^2}{\fq^2}+\frac{\ft^3}{\fq^3}\Big)Q_1^3 Q_3^3	\bigg]
   +\mathcal{O}(Q_1^4, Q_3^4)
\ . \label{eq:f1}   
\end{align}
As $Q_1^2Q_2=u$ does not depend on the Coulomb branch parameters, the term proportional to $Q_1^2$ in \eqref{eq:f1} contributes to the extra factor. Separating out such extra factor, we rewrite \eqref{eq:f1} in terms of the spin state $[j_l, j_r]$ defined in \eqref{eq:GVspin}, 
\begin{align}
    f_1^{\kappa=9}&= [0, \tfrac12] + [0,\tfrac32] Q_3+[0,\tfrac52] Q_3^2+ [0, \tfrac12] Q_1 +\Big([0,\tfrac12]+[0,\tfrac32]\Big) Q_1Q_3\crcr
    &+ \Big( [0,\tfrac32] +[0,\tfrac52]\Big) Q_1 Q_3^2+ \Big( [0,\tfrac12] +[0,\tfrac32]\Big) Q_1^2 Q_3
    +\Big( [0,\tfrac12]+[0,\tfrac32] +[0,\tfrac52]\Big) Q_1^2 Q_3^2\crcr
&  + [0, \tfrac72]Q_3^3
   + \Big([0, \tfrac52]+[0, \tfrac72]\Big)Q_1 Q_3^3
   + \Big([0, \tfrac32]+[0, \tfrac52]+[0, \tfrac72]\Big)Q_1^2Q_3^3 \crcr
& + [0, \tfrac12]Q_1^3+
   \Big([0, \tfrac12]+[0, \tfrac32]\Big)Q_1^3 Q_3 +\Big([0, \tfrac12]+[0, \tfrac32]+[0, \tfrac52]\Big)Q_1^3 Q_3^2
   \cr
   &
   + 
   \Big([0, \tfrac12]+[0, \tfrac32]+[0, \tfrac52]+[0, \tfrac72]\Big)Q_1^3 Q_3^3 +\mathcal{O}(Q_1^4, Q_3^4)+
     (\text{extra factor})
   \ .\label{eq:f1GV}
\end{align}
This agrees with the blowup result in \cite{Kim:2020hhh} where the terms up to $Q_1^2Q_3^2$ are presented at the first order of $Q_2$.

The contribution at order $Q_2^2$ leads to two-instanton partition function. Summing over Young diagrams in \eqref{eq:instSU(3)_9} with the corresponding $\calZ$'s in \eqref{eq:twoinst} and combining \eqref{eq:finPE} and \eqref{eq:f1ex}, we can obtain the exact form of $f_2$. As the exact form of $f_2$ is rather long\footnote{The exact form of $f_2$ for the SU(3)$_9$ theory is presented in a Mathematica file attached as an ancillary file.}, we expand the form with respect to $Q_1,Q_3$ up to $Q_1^2Q_3^2$ to compare it with the known result, as follows: 
\begin{small}
\begin{align}
 f_2^{\kappa=9}&=\frac{1}{(1-\fq)(1-\ft)}\bigg[\Big(\fq+\frac{\fq^3}{\ft^2}+\frac{\fq^2}{\ft}+\ft+\frac{\ft^2}{\fq}+\frac{\ft^3}{\fq^2}\Big)Q_3+\Big(1+2\fq+\fq^2+\frac{\fq^4}{\ft^4}+\frac{\fq^3}{\ft^3}\cr
 &+\frac{\fq^4}{\ft^3}+\frac{\fq^5}{\ft^3}+\frac{\fq^2}{\ft^2}+2\frac{\fq^3}{\ft^2}+\frac{\fq^4}{\ft^2}+\frac{\fq}{\ft}+2\frac{\fq^2}{\ft}+\frac{\fq^3}{\ft}+2\ft+\frac{\ft}{\fq}+\fq\ft+\ft^2+\frac{\ft^2}{\fq^2}+2\frac{\ft^2}{\fq}+\frac{\ft^3}{\fq^3}\cr
 &+2\frac{\ft^3}{\fq^2}+\frac{\ft^3}{\fq}+\frac{\ft^4}{\fq^4}+\frac{\ft^4}{\fq^3}+\frac{\ft^4}{\fq^2}+\frac{\ft^5}{\fq^3}\Big)Q_3^2+\Big(2\fq+\frac{\fq^3}{\ft^2}+2\frac{\fq^2}{\ft}+2\ft+2\frac{\ft^2}{\fq}+\frac{\ft^3}{\fq^2}\Big)Q_1Q_3\cr
 &+\Big(2\!+\!6\fq +2\fq^2+\frac{\fq^4}{\ft^4}+2\frac{\fq^3}{\ft^3}+2\frac{\fq^4}{\ft^3}+\frac{\fq^5}{\ft^3}+2\frac{\fq^2}{\ft^2}+5\frac{\fq^3}{\ft^2}+2\frac{\fq^4}{\ft^2}+2\frac{\fq}{\ft}+6\frac{\fq^2}{\ft}+2\frac{\fq^3}{\ft\!}\cr
 &+\!6\ft+\!2\frac{\ft}{\fq}+\!2\fq\ft+\!2\ft^2+\!2\frac{\ft^2}{\fq^2}+\!6\frac{\ft^2}{\fq}+\!2\frac{\ft^3}{\fq^3}+\!5\frac{\ft^3}{\fq^2}+\!2\frac{\ft^3}{\fq}+\!\frac{\ft^4}{\fq^4}+\!2\frac{\ft^4}{\fq^3}+\!2\frac{\ft^4}{\fq^2}+\!\frac{\ft^5}{\fq^3}\Big)Q_1Q_3^2
 \cr
 &
 +\Big(3\fq+\frac{\fq^3}{\ft^2}+2\frac{\fq^2}{\ft}+3\ft+2\frac{\ft^2}{\fq}+\frac{\ft^3}{\fq^2}\Big)Q_1^2Q_3+\Big(3+10\fq+3\fq^2+\frac{\fq^4}{\ft^4}+2\frac{\fq^3}{\ft^3}+2\frac{\fq^4}{\ft^3}\cr
 &
 +\frac{\fq^5}{\ft^3}+3\frac{\fq^2}{\ft^2}+6\frac{\fq^3}{\ft^2}+2\frac{\fq^4}{\ft^2}+3\frac{\fq}{\ft}+9\frac{\fq^2}{\ft}+3\frac{\fq^3}{\ft}+10\ft+3\frac{\ft}{\fq}+3\fq\ft+3\ft^2+3\frac{\ft^2}{\fq^2}+9\frac{\ft^2}{\fq}\cr
 &
 +2\frac{\ft^3}{\fq^3}+6\frac{\ft^3}{\fq^2}+3\frac{\ft^3}{\fq}+\frac{\ft^4}{\fq^4}+2\frac{\ft^4}{\fq^3}+2\frac{\ft^4}{\fq^2}+\frac{\ft^5}{\fq^3}\Big)Q_1^2Q_3^2\bigg]+\mathcal{O}(Q_1^3, Q_3^3)\ ,
	\label{eq:f2}
\end{align}
\end{small}
which can be rewritten as 
\begin{align}
 f_2^{\kappa=9}&=[0,\tfrac52]Q_3+\Big([0,\tfrac52]+[0,\tfrac72]+[\tfrac12,4] \Big)Q_3^2+\Big([0,\tfrac32]+[0,\tfrac52]\Big)Q_1Q_3\label{eq:f2GV} \\
  &+\Big([0,\tfrac32]+3[0,\tfrac52]+2[0,\tfrac72] +[\tfrac12,3]+[\tfrac12,4]\Big)Q_1Q_3^2+\Big([0,\tfrac12]+[0,\tfrac32]+[0,\tfrac52] \Big)Q_1^2Q_3\crcr
 & +\Big([0,\tfrac12]+3[0,\tfrac32]+4[0,\tfrac52]+2[0,\tfrac72]+[\tfrac12,2] +[\tfrac12,3]+[\tfrac12,4]
 \Big)Q_1^2Q_3^2+\mathcal{O}(Q_1^3, Q_3^3)\ ,
	\nonumber
\end{align}
which agrees with the two-instanton part from the blowup result \cite{Kim:2020hhh}. 
Notice that there is no extra factor in \eqref{eq:f2GV}, which is because the extra factor at order $Q_2^2$ comes from terms proportional to $Q_1^4$.

The instanton partition function at higher order $n\ge 3$ can be obtained in the same way. It is to repeat the same computation with $|\mu_1|+|\mu_2|+|\mu_3|=n$. One can also express the partition function in terms of the gauge theory parameters by substituting the K\"ahler parameters with the Coulomb branch parameters and instanton factor in \eqref{eq:subCS9}, after dropping the extra factors.

\bigskip
\section{Conclusion}\label{sec:conclusion}
In this paper, we generalized refined topological vertex formalism so that it is applicable for 5-brane webs with ON-planes. 5-brane system with an ON-plane describes a D-type quiver gauge theory. In order for a 5-brane web with an ON-plane to correctly account for a D-type quiver, there should be no bifundamental contributions between two gauge theories belonging to bivalent nodes of a D-type quiver. To this end, we proposed a new vertex factor $C^{\text{R}}$ for the reflected vertices over an ON-plane and also new associated edge factors. Such new vertex and edge factors ensures the resulting partition functions correctly capture the BPS spectrum of a D-type quiver gauge theory. It is also worthy of noting that one can use only this new vertex factor and edge factor for usual 5-brane webs to obtain the same partition function obtained with the conventional vertex and edge factors as shown in Appendix \ref{app:newfactors}. This means that there are two different types of vertex and edge factors, yielding the same partition function. It may also support our proposal that introducing the new vertex and edge factors associated with the 5-brane web reflected over an ON-plane would be a natural procedure to distinguish original domain from the reflected domain of 5-brane system with an ON-plane.

Through the Higgsing, 5-brane configurations with  ON-plane(s) can give rise to 5-brane webs for SU(3) gauge theories at higher Chern-Simons level. We computed the refined partition function for SU(3)$_7$ and SU(3)$_9$ theories and confirmed that the results perfectly agree with the blowup computation \cite{Kim:2020hhh}. We also checked our proposal against  6d E-string theory on a circle, which requires two ON-planes, which also agree with other known results. These are the first examples that the refined partition functions are obtained based on 5-brane webs with ON-plane. We also present the exact form of one- and two-instanton partition functions of these theories. For theories of higher rank gauge groups or of more complicated matter that can be described by a 5-brane web with an ON-plane \cite{Hayashi:2015vhy}, one can apply our new vertex and edge factors to obtain their partition functions in a straightforward way.

From the perspective of the S-duality, 
5-brane configuration with an O5-plane can be understood as an S-dual configuration of that with an ON-plane. It is natural to generalize our proposal to 5-brane systems with an O5-plane, which describes 5d SO/Sp gauge theories. Though our new $C^{\text{R}}$ is compatible with O5-plane, the preferred direction is assigned along the edges associated with W-bosons, and the resulting partition function become an expansion of Coulomb branch parameters, rather than an instanton expansion. A naive attempt for refining topological vertex with an O5-plane seems more challenging because those branes which are reflected over an O5-plane are of the same $\Omega$ background parameters along the same edge. Hence, it is an interesting direction to pursue to  generalize the refined topological vertex so that it is applicable to 5-brane system with an O5-plane. These would shed some light on the partition functions from newly obtained 5-brane webs which involve $G_2$ gauge theories, SO gauge theories with spinor matter, and Sp or SU gauge theories of hypermultiplets in the rank-3 antisymmetric representations.  

It would be also interesting to study the relation between our proposal for a $D$-type quiver theory and algebraic  constructions based on Ding-Iohara-Miki algebra where the presence of an ON-plane is discussed either from the point of view of the reflection states \cite{Bourgine:2017rik} or as the introduction of a new twisted vertex \cite{Kimura:2019gon}.
\bigskip

\acknowledgments
 We thank Jean-Emile Bourgine, Shi Cheng, Hirotaka Hayashi, Songling He, Hee-Cheol Kim, Minsung Kim, Kimyeong Lee, Xiaobin Li, Satoshi Nawata, Yuji Sugimoto, Ryo Suzuki, Futoshi Yagi, and Ruidong Zhu for useful discussions. We are also grateful to Shi Cheng and Songling He for providing useful Mathematica codes and Futoshi Yagi and Yuji Sugimoto for their comments on the draft. SSK thanks APCTP, Fundan university, KIAS, POSTECH and YMSC Tsinghua university for the hospitality during his visit where part of this work is done. SSK is partially supported by the Fundamental Research Funds for the Central Universities 2682021ZTPY043. XYW is partially supported by NSFC grant No. 11950410490.
\bigskip

\appendix

\section{Characters}\label{sec:App}
The characters for the fundamental weights of SO(8) $\chi_{\bf dim}^{\rm SO(8)}$ are given as follows:
\begin{align}
	\chi_{\bf8v}^{\text{SO(8)}}&=\sum_{i=1}^8X_i^2,\nonumber\\
	\chi_{\bf28}^{\text{SO(8)}}&=\sum_{i=1}^8\sum_{j>i}^8X_i^2X_j^2,\nonumber\\
	%
	\chi_{\text{\bf8s}}^{\text{SO(8)}}&=\frac{1}{2}\left(\prod_{i=1}^4\left(X_i+X_{i+4}\right)+\prod_{i=1}^4\left(X_i-X_{i+4}\right)\right),\nonumber\\
	\chi_{\text{\bf8c}}^{\text{SO(8)}}&=\frac{1}{2}\left(\prod_{i=1}^4\left(X_i+X_{i+4}\right)-\prod_{i=1}^4\left(X_i-X_{i+4}\right)\right),
\end{align}
where $X_i\in\big\{M_1^{1/2},\dots,M_4^{1/2},M_1^{-1/2},\dots,M_4^{-1/2}\big\}$ with the flavor mass fagacities $M_i=e^{-\beta m_i}$.

The characters for the fundamental weights of SO(16) $\chi_{\bf dim}$ are given as follows:
\begin{align}
	\chi_{\bf 16}\,&=\sum_{i=1}^{16}Y_i^2,\nonumber\\
	\chi_{\bf120}\,&=\sum_{i=1}^{16}\sum_{j>i}^{16}Y_i^2Y_j^2,\nonumber\\
	%
	\chi_{\bf560}\,&=\sum_{i=1}^{16}\sum_{j>i}^{16}\sum_{k>j}^{16}Y_i^2Y_j^2Y_k^2,\nonumber\\
	%
	\chi_{\bf 1820}\,&=\sum_{i=1}^{16}\sum_{j>i}^{16}\sum_{k>j}^{16}\sum_{l>k}^{16}Y_i^2Y_j^2Y_k^2Y_l^2,\nonumber\\
	%
	\chi_{\bf 4368}\,&=\sum_{i=1}^{16}\sum_{j>i}^{16}\sum_{k>j}^{16}\sum_{l>k}^{16}\sum_{m>l}^{16}Y_i^2Y_j^2Y_k^2Y_l^2Y_m^2,\nonumber\\
	%
	\chi_{\bf 8008}\,&=\sum_{i=1}^{16}\sum_{j>i}^{16}\sum_{k>j}^{16}\sum_{l>k}^{16}\sum_{m>l}^{16}\sum_{n>m}^{16}Y_i^2Y_j^2Y_k^2Y_l^2Y_m^2Y_n^2,\nonumber\\
	\chi_{\bf 128}\,&=\frac{1}{2}\left(\prod_{i=1}^8\left(Y_i+Y_{i+8}\right)+\prod_{i=1}^8\left(Y_i-Y_{i+8}\right)\right),\nonumber\\
	\chi_{\overline{\bf 128}}\,&=\frac{1}{2}\left(\prod_{i=1}^8\left(Y_i+Y_{i+8}\right)-\prod_{i=1}^8\left(Y_i-Y_{i+8}\right)\right)\ ,
\end{align}
where $Y_i\in\big\{M_1^{1/2},\dots,M_8^{1/2},M_1^{-1/2},\dots,M_8^{-1/2}\big\}$.

The characters for higher dimensional irreducible representations can be expressed in terms of the product of the characters of these fundamental weights. For instance, among those that appear in \eqref{eq:fmn for E-string}, 
\begin{align}
	&\chi_{\overline{\bf 1920}}+\chi_{\overline{\bf 128}} = \chi_{\bf 16}\cdot\chi_{\bf 128}\ ,
	\crcr
	&\chi_{\bf 1920}+\chi_{\bf 128}= \chi_{\bf 16}\cdot\chi_{\overline{\bf 128}}\ ,	 \crcr
	&\chi_{\bf 13312} + \chi_{\bf1920} +  \chi_{\bf128}= \chi_{\bf 120}\cdot\chi_{\bf 128}\ .
\end{align}

\section{Reflected 5-brane web with  \texorpdfstring{$C^{\text{R}}$}{CR} and  \texorpdfstring{$\tilde{f}^{\text{R}}$}{fR}.}\label{app:newfactors}

In the main text, we have introduced new vertex factor $C^{\text{R}}$ and framing factor $\tilde{f}^{\text{R}}$ which account for vertex and framing factors that are reflected over an ON-plane. For a 5-brane web with an ON-plane, one needs to choose the fundamental region, as there is reflected 5-brane configuration over the ON-plane. Because of the equivalence between the original 5-brane web and the reflected image, the partition function based on either the original 5-brane web or the reflected one should be equivalent.  In other words, even though one performs the topological vertex computation based on reflected web diagram with $C^\text{R}$ and $\tilde{f}^{\text{R}}$ factors, the resulting partition function should not make any difference. The new factors $C^\text{R}, \tilde{f}^{\text{R}}$ and conventional $C, \tilde{f}$ are on an equal footing.

\begin{figure}[htbp]
	\centering
	\includegraphics[scale=1]{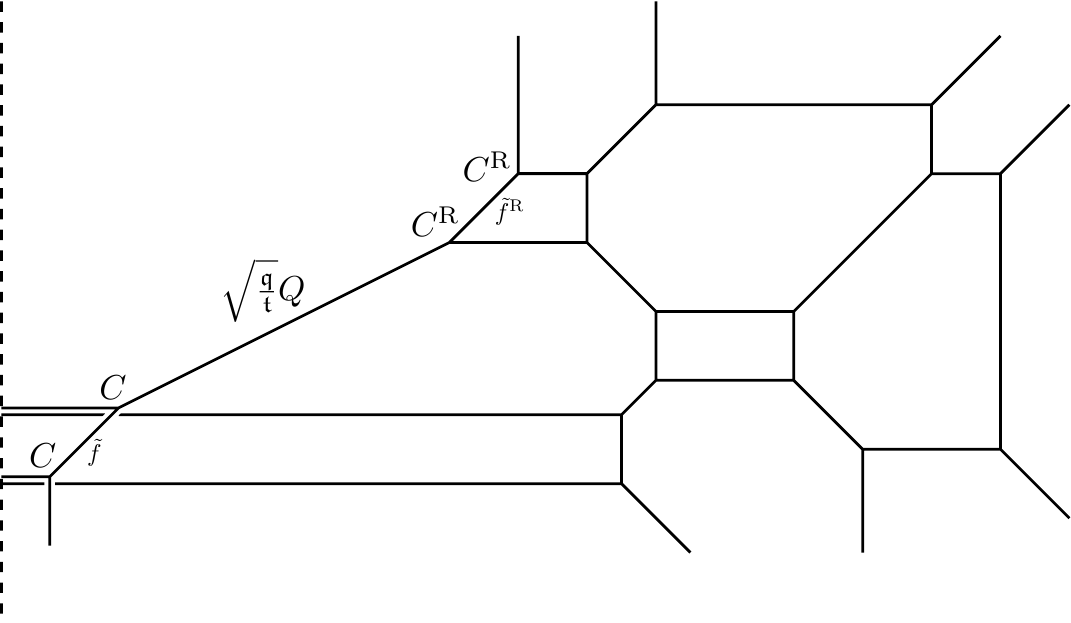}
	\caption{Reflected 5-brane configuration for 
	a $D_4$-type quiver theory discussed in Figure \ref{fig:SU(3)_7}.
	}
	\label{fig:mirror}
\end{figure}

As a concrete example, let us consider a $D_4$-type quiver with SU(3)${}_1$ at the center node connected to tree SU(2) nodes, which is discussed in subsection \ref{sec:SU(3)CS7}. The corresponding 5-brane configuration is presented in Figure \ref{fig:SU(3)_7}. The corresponding reflected 5-brane configuration can be easily depicted as given in Figure \ref{fig:mirror}, where the role of  $C,\tilde{f}$ and $C^{\text{R}},\tilde{f}^{\text{R}}$ exchanged, compared to the original 5-brane web in Figure \ref{fig:SU(3)_7}. In particular, $C,\tilde{f}$ only appear on the strip next to an ON-plane on the left. Note that the preferred direction framing factor $f$ in the reflected web diagram is the same as that in the unreflected web diagram as is mentioned in the main text. One can check that the partition function based on this reflected 5-brane web is the same as the one discussed in subsection \ref{sec:SU(3)CS7}. 

\begin{figure}[htbp]
	\centering
	\includegraphics{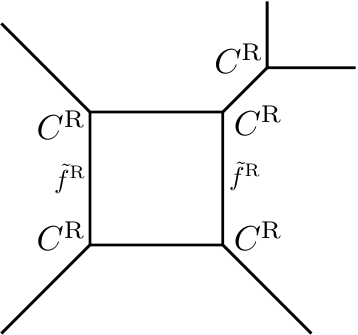}
	\caption{Web diagram of SU$(2)+1\mathbf{F}$ denoted by $C^{\text{R}}$ and $\tilde{f}^{\text{R}}$.}
	\label{fig:su21fmir}
\end{figure}
It naturally follows that even for a 5-brane web without an ON-plane, one can use the new factors $C^\text{R}$ and $\tilde{f}^\text{R}$, instead of using the conventional factors $C$ and $\tilde{f}$. For instance, consider a 5-brane web for SU(2)+1{\bf F} given in Figure \ref{fig:su21fmir}. Here in the figure, we have assigned the new vertex and framing factors $C^\text{R}, \tilde{f}^\text{R}$. Note that four vertex factors out of the five $C^{\text{R}}$'s in fact reduce to the usual vertex factor $C$ because one of their legs in non-preferred direction is external, and hence effectively we have only one $C^\text{R}$ factor presented. One can check that the partition function computation with this $C^\text{R}$ and $\tilde f^\text{R}$ is straightforward and the result is the same as the partition function obtained with  $C$ and $\tilde{f}$.

\bigskip

\bibliographystyle{JHEP}
\bibliography{ref}
\end{document}